
\magnification=\magstep1
\input amssym.def
\input amssym

\def\m@th{\mathsurround=0pt}
\def\mymatrix#1{\null\,\vcenter{\normalbaselines\m@th
	\ialign{\hfil$\scriptstyle ##$\hfil&&\quad\hfil
	$\scriptstyle ##$\hfil\crcr
	\mathstrut\crcr\noalign{\kern-\baselineskip}
	#1\crcr\mathstrut\crcr\noalign{\kern-\baselineskip}}}\,}
\def\tr{\mathop{\rm Tr}\nolimits}
\def\str{\mathop{\rm STr}\nolimits}
\def\sdet{\mathop{\rm SDet}\nolimits}
\def\CN{\mathop{\Bbb C}\nolimits}
\def\RN{\mathop{\Bbb R}\nolimits}
\def\NN{\mathop{\Bbb N}\nolimits}
\def\IN{\mathop{\Bbb Z}\nolimits}
\def\half{{\textstyle {1\over 2}}}

\font\bigbf=cmbx10 at 13pt
\font\bigrm=cmr10 at 14.4pt
\font\medbf=cmbx10 at 12pt
\font\medrm=cmr10 at 12pt

\input harvmac.tex

\lref\qhe{
K.~v.~Klitzing, G.~Dorda, and M.~Pepper, Phys.~Rev.~Lett.~{\bf 45}
(1980) 494; D.C.~Tsui, H.L.~Stoermer, and A.C.~Goddard,
Phys.~Rev.~Lett.~{\bf 48} (1982) 1559.}
\lref\aa{
T.~Ando and H.~Aoki, Sol.~State~Commun.~{\bf 38} (1981) 1079.}
\lref\top{
Q.~Niu, D.J.~Thouless, and Y.S.~Wu, Phys.~Rev.~B{\bf 31} (1985) 3372;
Y.~Avron and R.~Seiler, Phys.~Rev.~Lett.~{\bf 54} (1985) 259.}
\lref\butt{
M.~B\"uttiker, Phys.~Rev.~B{\bf 38} (1988) 9375.}
\lref\expqp{
H.P.~Wei, D.C.~Tsui, M.~Palaanen, and A.A.M.~Pruisken,
Phys.~Rev.~Lett.~{\bf 61} (1988) 1294; S.~Koch, R.~Haug, 
K.~v.~Klitzing, and K.~Ploog, Phys.~Rev.~Lett.~{\bf 67} (1991) 883.}
\lref\cc{
J.T.~Chalker and P.D.~Coddington, J.~Phys.~C{\bf 21} (1988) 2665.}
\lref\hk{
B.~Huckestein and B.~Kramer, Phys.~Rev.~Lett.~{\bf 64} (1990) 1437;
B.~Mieck, Europhys.~Lett.~{\bf 13} (1990) 453.}
\lref\cft{
A.A.~Belavin, A.M.~Polyakov and A.B.~Zamolodchikov, Nucl.~Phys.~B{\bf 241}
(1984) 333; J.~Cardy, in: {\it Phase Transitions and Critical Phenomena} 
vol.~11, eds. C.~Domb and J.L.~Lebowitz (1987).}
\lref\pru{
A.A.M.~Pruisken, Nucl.~Phys.~B{\bf 235} (1984) 277.}
\lref\llp{
H.~Levine, S.B.~Libby, and A.A.M.~Pruisken, 
Phys.~Rev.~Lett.~{\bf 51} (1983) 1915;
Nucl. Phys. B{\bf 240} (1984) 30; 49; 71.}
\lref\prutwo{
A.A.M.~Pruisken, Nucl.~Phys.~B{\bf 285} (1987) 719;
B{\bf 290} (1987) 61.}
\lref\wz{
H.A.~Weidenm\"uller and M.R.~Zirnbauer,
Nucl.~Phys.~B{\bf 305} (1988) 339.}
\lref\hhb{
Y.~Huo, R.E.~Hetzel, and R.N.~Bhatt, Phys.~Rev.~Lett.~{\bf 70} (1993) 481.}
\lref\affone{
I.~Affleck, Phys.~Rev.~Lett.~{\bf 56} (1986) 746.}
\lref\affhal{
I.~Affleck and F.D.M.~Haldane, Phys.~Rev.~B{\bf 36} (1987) 5291.}
\lref\sr{
R.~Shankar and N.~Read, Nucl.~Phys.~B{\bf 336} (1990) 457.}
\lref\scalth{
A.A.M.~Pruisken, Phys.~Rev.~Lett.~{\bf 61} (1988) 1297.}
\lref\afftwo{
I.~Affleck, Nucl.~Phys.~B{\bf 257} (1985) 397; B{\bf 265} (1986) 409; 
B{\bf 305} (1988) 582.}
\lref\haldane{
F.D.M.~Haldane, Phys.~Lett.~93A (1983) 464; Phys.~Rev.~Lett.~{\bf 50}
(1983) 1153; J.~Appl.~Phys.~(USA) {\bf 57} (1985) 3359.}
\lref\vz{
J.J.M.~Verbaarschot and M.R.~Zirnbauer, J.~Phys.~A{\bf 17}
(1985) 1093.}
\lref\ludwig{
A.A.W.~Ludwig, M.P.A.~Fisher, R.~Shankar and G.~Grinstein,
Princeton University Preprint 9/1993.}
\lref\efetov{
K.B.~Efetov, Adv.~in~Phys.~{\bf 32} (1983) 53.}
\lref\weidenm{
H.A.~Weidenm\"uller, Nucl.~Phys.~B{\bf 290} (1987) 87.}
\lref\bs{
H.~Baranger and A.D.~Stone, Phys.~Rev.~B{\bf 40} (1989) 8169.}
\lref\lr{
P.A.~Lee and T.V.~Ramakrishnan, Rev.~Mod.~Phys.~{\bf 57} (1985) 287.}
\lref\chaone{
J.T. Chalker, J.~Phys.~C{\bf 21} (1988) L119.}
\lref\chatwo{
J.T. Chalker, Physica A{\bf 167} (1990) 253.}
\lref\wegone{
F.~Wegner, in: {\it Localisation and Metal Insulator Transitions},
eds. H.~Fritzsche and D.~Adler, Institute for Amorphous Studies
Series (Plenum, New York, 1985).}
\lref\wegtwo{
F.~Wegner, Z.~Phys.~B{\bf 25} (1976) 327.}
\lref\multifrac{
U.~Fastenrath, M.~Janssen, and W.~Pook, Physica A{\bf 191} (1992) 401;
B.~Huckestein and L.~Schweitzer, Physica A{\bf 191} (1992) 406;
Phys.~Rev.~Lett.~{\bf 72} (1994) 713.}
\lref\sw{
L.~Sch\"afer and F.~Wegner, Z.~Phys.~{\bf B38} (1980) 113.}
\lref\mrznpa{
M.R.~Zirnbauer, Nucl.~Phys.~A{\bf 560} (1993) 95.}
\lref\weident{
H.A.~Weidenm\"uller, Physica A{\bf 167} (1990) 28.}
\lref\mrzpha{
M.R.~Zirnbauer, Physica A{\bf 167} (1990) 132.}
\lref\huckestein{
B.~Huckestein, Phys.~Rev.~Lett.~{\bf 72} (1994) 1080.}
\lref\fastenrath{
U.~Fastenrath, Physica A{\bf 189} (1992) 27.}
\lref\wegnert{
F.~Wegner, Z.~Phys.~B{\bf 35} (1979) 207.}
\lref\alk{
B.L.~Al'tshuler, V.I.~Kravtsov, and I.~Lerner, in: {\it Mesoscopic
Phenomena in Solids}, eds. B.L.~Al'tshuler, R.A.~Lee, and R.A.~Webb,
North Holland, Amsterdam 1991.}
\lref\dewitt{
B.~DeWitt, {\it Supermanifolds}, Cambridge University Press,
Cambridge 1984.}
\lref\helgason{
S.~Helgason, {\it Differential geometry, Lie groups, and symmetric 
spaces}, Academic Press, New York 1978; {\it Groups and geometric 
analysis}, Academic Press, Orlando 1984.}
\lref\egh{
T.~Eguchi, P.~Gilkey, and A.J.~Hanson, Phys.~Rep.~{\bf 66},
(1980) 213.}
\lref\berry{
M.V.~Berry, Proc.~Roy.~Soc. (London) A{\bf 392} (1984) 45;
B.~Simon, Phys.~Rev.~Lett.~{\bf 51} (1983) 2167.}
\lref\berezin{
F.A.~Berezin, {\it Introduction to Superanalysis}, D.~Reidel
Publishing Co., Doordrecht, Holland 1987.}
\lref\kn{
S.~Kobayashi and K.~Nomizu, {\it Foundations of Differential
Geometry}, vol.~I, John Wiley \& Sons, New York 1963.}
\lref\fradkin{
E.~Fradkin, {\it Field theories of condensed matter systems}, 
Addison-Wesley, Redwood City 1991.}
\lref\huffmann{
A.~H\"uffmann, Ph.D.~Thesis, University of Cologne (1992);
J.~Phys.~A (in press).}
\lref\mrzcmp{
M.R.~Zirnbauer, Commun.~Math.~Phys.~{\bf 141} (1991) 503.}
\lref\mmz{
M.R.~Zirnbauer, Phys.~Rev.~Lett.~{\bf 69} (1992) 1584;
A.D.~Mirlin, A.~M\"uller-Groeling, and M.R.~Zirnbauer,
Ann.~Phys. (in press).}
\lref\stone{
M.~Stone, Nucl.~Phys.~B{\bf 314} (1989) 557.}
\lref\bv{
N.L.~Balazs and A.~Voros, Phys.~Rep.~{\bf 143} (1986) 109.}
\lref\hr{
F.D.M.~Haldane, Phys.~Rev.~Lett.~{\bf 51} (1983) 605; 
F.D.M.~Haldane and E.~Rezayi, Phys.~Rev.~Lett.~{\bf 54} (1985) 237.}
\lref\mrznpb{
M.R.~Zirnbauer, Nucl.~Phys.~B{\bf 265} (1986) 375.}
\lref\bundschuh{
R.~Bundschuh, Diploma thesis, University of Cologne (1993).}
\lref\altland{
A.~Altland, Z.~Phys.~B{\bf 86} (1991) 105.}
\lref\khmel{
D.E.~Khmel'nitskii, JETP Lett.~{\bf 38} (1983) 454.}
\lref\nread{
The symmetries of the superspin chain were first communicated
to me by N.~Read (Aspen 1993).}
\lref\lc{
D.K.K.~Lee and J.T.~Chalker, Phys.~Rev.~Lett.~{\bf 72} (1994) 1510.}
\lref\dhlee{
D.H.~Lee, {\it Network models of quantum percolation and their
field-theory representation}, cond-mat/9404011.}
\lref\lsm{
E.~Lieb, T.~Schultz, and D.~Mattis, Ann.~Phys.~{\bf 16} (1961) 407.}
\lref\al{
I.~Affleck and E.H.~Lieb, Lett.~Math.~Phys.~{\bf 12} (1986) 57.}
\lref\abram{
M.~Abramowitz and I.A.~Stegun, {\it Handbook of Mathematical
Functions}, National Bureau of Standards, Washington 1964.}
\lref\lwk{
D.H.~Lee, Z.~Wang, and S.A.~Kivelson, Phys.~Rev.~Lett.~{\bf 70}
(1993) 4130.}
\lref\trugman{
S.A.~Trugman, Phys.~Rev.~B{\bf 27} (1983) 7539.}
\lref\baxter{
R.J.~Baxter, {\it Exactly Solved Models in Statistical Mechanics},
Academic Press, London 1982.}
\lref\drinfeld{
V.G.~Drinfel'd, Soviet~Math.~Dokl.~{\bf 32} (1985) 254.}
\lref\krs{
P.P.~Kulish, N.~Yu.~Reshetikhin, and E.K.~Sklyanin,
Lett.~Math.~Phys.~{\bf 5} (1981) 393.}
\lref\hs{
F.D.M.~Haldane, Phys.~Rev.~Lett.~{\bf 60} (1988) 635; B.S.~Shastry,
Phys.~Rev.~Lett.~{\bf 60} (1988) 639.}
\lref\rothstein{
M.J.~Rothstein, Trans.~Am.~Math.~Soc.~{\bf 299} (1987) 387.}

\baselineskip=16pt
\vskip 20pt
\centerline{\bigrm Towards a Theory of the}
\centerline{\bigrm Integer Quantum Hall Transition:}
\centerline{\bigrm From the Nonlinear Sigma Model}
\centerline{\bigrm to Superspin Chains}

\vskip 25pt
\centerline{\medrm Martin R. Zirnbauer}
\centerline{\medrm Institut f\"ur Theoretische Physik}
\centerline{\medrm Universit\"at zu K\"oln, 50937 K\"oln, Germany}
\vskip 30pt

\baselineskip=12pt
\centerline{ABSTRACT}\bigskip

\hskip 0.7cm \vbox{\hsize 11cm \noindent
A careful study of the supersymmetric version of Pruisken's nonlinear 
$\sigma$ model for the integer quantum Hall effect is presented. The 
lattice regularized model is cast in Hamiltonian form by taking the
anisotropic limit and interpreting the topological density as an alternating 
sum of Wess-Zumino terms. It is argued that the relevant large-scale physics 
of the model is preserved by projection of the quantum Hamiltonian on its 
sector of degenerate strong-coupling ground states. For values of the Hall 
conductivity close to $e^2/2h$ (mod $e^2/h$), where a delocalization 
transition occurs, this yields the Hamiltonian of a quantum superspin 
chain which is closely related to an anisotropic version of the 
Chalker-Coddington model. The relation implies that the ratio of magnetic 
length over potential correlation length is an irrelevant parameter at the 
transition. The superspin chain resembles a 1$d$ isotropic antiferromagnet 
with spin 1/2. It has an alternating structure which however permits an 
invariance under translation by one site. The conductance coefficients of 
a quantum Hall system with $N$ small contacts translate into $N$-superspin 
correlation functions which are governed by conformal invariance. The 
superspin formalism provides a framework for studying the crossover from 
classical to quantum percolation. It does not however encompass the 
frequency-dependent correlations of wave amplitudes at criticality.}\medskip

\noindent PACS numbers: 71.30, 72.10, 11.10, 75.10J \medskip
\noindent KEYWORDS: integer quantum Hall effect; metal-insulator transition;
nonlinear $\sigma$ model; supersymmetry; spin chains
\vfill\eject

\baselineskip 18pt
\noindent{\bigbf Contents}\bigskip

\noindent{\medbf 1 Introduction}\medskip

\noindent{\medbf 2 Conformal invariance of critical disordered electron
systems}\smallskip

\hskip 0cm 2.1 Finite frequency breaks conformal invariance\smallskip
\hskip 0cm 2.2 Conformal invariance and d.c. conductance\medskip

\noindent {\medbf 3 Hamiltonian limit of the nonlinear $\sigma$ model}
\smallskip

\hskip 0cm 3.1 Review of Pruisken's model (supersymmetric version)\smallskip
\hskip 0cm 3.2 Gauge fixing\smallskip
\hskip 0cm 3.3 Quantum Hamiltonian for $\sigma_{xy} = 0$\smallskip
\hskip 0cm 3.4 Topological density and Berry phase\smallskip
\hskip 0cm 3.5 Quantum Hamiltonian for $\sigma_{xy} \not= 0$\medskip

\noindent{\medbf 4 Strong-coupling ground states}\smallskip

\hskip 0cm 4.1 Principal remarks\smallskip
\hskip 0cm 4.2 Induced representation\smallskip
\hskip 0cm 4.3 Choice of gauge, and coordinate presentation\smallskip
\hskip 0cm 4.4 Holomorphic sections\smallskip
\hskip 0cm 4.5 Zero modes and coherent states\smallskip
\hskip 0cm 4.6 Edge dynamics and Hall conductance\medskip

\noindent{\medbf 5 Quantum spin chain}\smallskip

\hskip 0cm 5.1 Mapping on a spin chain\smallskip
\hskip 0cm 5.2 Solution of the two-spin problem\medskip

\noindent{\medbf 6 Summary and outlook}\medskip

\noindent{\medbf Appendices}
\vfill\eject

\baselineskip=12pt
\vsize 19cm
\noindent{\bigbf 1 \ Introduction}\medskip

\noindent When a two-dimensional $(2d)$ electron gas is placed in a strong 
magnetic field $B$, the Hall conductance $\sigma_H$ is observed to be a 
rational multiple of $e^2 / h$ near certain magic values of the density \qhe. 
While Coulomb interactions are essential for the occurrence of fractional 
quantization, it is generally agreed that the observation of integer 
values $\sigma_H = n e^2 /h$ $(n \in \IN)$ in low-mobility samples can 
be understood from a model of noninteracting electrons with potential 
disorder. It was first proposed in \aa\ that such a model undergoes a 
sequence of localization-delocalization transitions as the Fermi-energy 
$E_F \sim 1/B$ is varied. Given the localization of all states in the 
Landau band tails, quantization of the Hall conductance has been explained 
by various theories, of which we mention the topological argument \top\ and 
the scattering-theoretic approach \butt. What is less well understood is 
the critical phenomenon which occurs at energies $E_c$ close to a Landau 
level center where some electron states become delocalized. The picture 
that has emerged from experiments \expqp\ and numerical simulations for 
noninteracting systems \refs{\cc,\hk} is that of some kind of second-order 
phase transition with a divergent correlation length $\xi \sim 
|E-E_c|^{-\nu}$, $\nu = 2.3 \pm 0.1$.

We begin with some general statements concerning conformal invariance 
at the integer quantum Hall delocalization transition and other 
metal-insulator transitions in disordered electron systems, in Sect.~2. 
Two ways of regularizing the electron Green's functions are distinguished. 
The system may either be open with absorbing (i.e. conducting) boundary 
conditions at the leads or, alternatively, it can be taken to be closed 
but perturbed by an external field with finite frequency. Both an external 
field and absorption at a boundary cut off the long-time correlations and 
in this sense act as regulators. The introduction of a frequency $\omega$ 
sets a length scale, $L_\omega$, which breaks conformal
invariance. Moreover, the amplitude of critical Green's functions diverges 
for $L_\omega \to \infty$ (or $\omega \to 0$). An interpretation of these 
Green's functions in the framework of conformal field theory \cft\ must 
involve primary fields with scaling dimensions that are {\it negative}. 
The situation is somewhat different when the first option of regularizing 
by absorption at a boundary is utilized. We suggest the novel procedure of 
keeping the size and relative distance of the absorbing contacts finite while 
sending the system size to infinity. In this limit the d.c. conductance, or 
current-current correlation function, decays algebraically with increasing 
distance between the contacts, and we expect it to expand in conformal 
fields that have {\it positive} scaling dimensions only. 

To promote the reasoning of Sect.~2 to a quantitative theory with
the possibility of getting exact analytical solutions, we must
turn to field-theoretic techniques. An effective field theory 
for the integer quantum Hall effect was first formulated by Pruisken 
and collaborators \refs{\pru,\llp}. The field theory is of the 
nonlinear $\sigma$ model type and has the action
\eqn\action{
	S[Q] = {1\over 8} \int_{\cal M} dx dy \left( \sigma_{xx}^0 
	\tr (\partial Q)^2 - \sigma_{xy}^0 
	\tr Q [ \partial_x Q , \partial_y Q ] \right) .}
The matrix-valued fields $Q$ map the configuration space ${\cal 
M}$ of the $2d$ electron gas into a coset space ${\rm U}(2n)/
{\rm U}(n)\times {\rm U}(n)$ with $n = 0$ due to the use of the
replica trick. By the derivation of the model, the bare coupling
constants $\sigma_{xx}^0$ and $\sigma_{xy}^0$ are identified as
the longitudinal and Hall conductivities, respectively, calculated 
in the self-consistent Born approximation and measured in units of $e^2 
/ h$. The field theory with action \action\ attributes to $\sigma_{xy}^0$ 
the meaning of a topological angle $\theta = 2\pi\sigma_{xy}^0$ 
multiplying the topological density $\tr Q[\partial_x Q,\partial_y Q]/8$. 
When ${\cal M}$ has no boundary, the topological density integrates
to $2\pi i$ times an integer and the physics of the model \action\ is 
periodic in $\sigma_{xy}^0$ with period 1.

Pruisken's work on the model \action\ focussed mainly on the semiclassical 
instanton approximation \refs{\prutwo,\wz}, from which he 
proposed a two-parameter renormalization group flow with one (or 
possibly two) unstable fixed point(s) at $\sigma_{xy}^0 = 1/2 \ 
({\rm mod} \ 1)$ and some $\sigma_{xx}^0 = \sigma_{xx}^*$. This 
(or one of these) fixed point(s) is conjectured to be the integer 
quantum Hall delocalization fixed point seen in numerical and real 
experiments. While Pruisken's conjecture has great intuitive appeal 
and is consistent with experiments, it has not led to quantitative 
results, beyond predicting the critical point to be isolated -- with 
no continuous symmetry spontaneously broken and Goldstone bosons 
absent. In particular, the observed universality $\sigma_{xx}^* = 
0.5\pm 0.1$ \hhb\ was not foreseen, and a computation of critical 
indices is still lacking. This is not surprising since the critical 
region is beyond the reach of the instanton approximation, which is 
valid for $\sigma_{xx}^0 \gg 1$ while the integer quantum Hall 
(IQHE) transition occurs at a critical value $\sigma_{xx}^*$ of 
order unity. Another, more serious difficulty arises from the 
constraints imposed by conformal invariance. The field theory 
\action\ at $n = 0$ and $\sigma_{xy}^0 = 1/2 \ ({\rm mod} \ 1)$ must 
be critical with a divergent correlation length -- or else it 
would not describe the IQHE transition. We therefore expect it to 
flow under renormalization to a conformally invariant field theory 
\cft. Now, as was pointed out by Affleck \affone, conformal invariance 
of a $2d$ field theory with continuous symmetry implies that there
exist {\it two} sets of conserved currents, one for the right-moving 
and another one for the left-moving charges. Unfortunately, the 
nonlinear $\sigma$-model \action\ has just {\it one} manifestly conserved 
current, which is its Noether current. (Presumably, in order for the 
required separation into left and right movers to become visible, one 
has to restrict the $\sigma$ model to its low-energy limit.) What 
happens is well understood for the case of one replica ($n = 1$). 
Affleck and Haldane \affhal\ have argued convincingly that the 
${\rm U}(2)/{\rm U}(1)\times {\rm U}(1)$ nonlinear $\sigma$ model 
(also known as the O(3) model) at $\theta = \pi$ is {\it not 
scale-invariant} but renormalizes to the ${\rm SU(2)}$ $k = 1$ 
WZW model -- a conformal field theory with central charge $c = 1$
and two independent conserved currents $J(z)$ and $\bar J(\bar z)$.
This conclusion was confirmed by Shankar and Read \sr. By analogy,
we expect Pruisken's model at $\theta = \pi$ and $n = 0$ to flow to 
some conformal field theory of the Wess-Zumino type. {\it We are thus 
led to reject Pruisken's assertions about the fixed-point structure of 
the model \action\ at $n = 0$.} This model has neither one nor two nontrivial 
fixed points at $\theta = \pi$, but has no such fixed points at all! (Note 
that we are {\it not} calling into question the two-parameter scaling
theory \scalth, which is arguably correct when proper meaning is given 
to the parameters $\sigma_{xx}$ and $\sigma_{xy}$. What we are saying 
is that these parameters have no interpretation as running coupling 
constants of the model \action.)

The lesson learned is that, if our goal is to approach the IQHE fixed 
point by analytic means, we had better abandon the critical (at $\theta
= \pi$) but scale-dependent field theory \action. This logical step was 
taken by Affleck \afftwo\ who exploited the relation of Pruisken's ${\rm U}
(2n)/{\rm U}(n) \times {\rm U}(n)$ replica field theory to certain 
${\rm SU}(2n)$-invariant quantum ``spin'' chains. Such a relation 
first appeared in the work of Haldane \haldane\ who suggested that 
antiferromagnetic spin chains with ${\rm SU}(2)$-invariance and large 
spin $S$ map on the ${\rm U}(2)/{\rm U}(1)\times {\rm U}(1)$ nonlinear 
$\sigma$ model with topological angle $\theta = 2\pi S$. Applying an 
inverse of this mapping \sr\ to Pruisken's model for $n = 1$, one gets an 
interpretation of the Hall conductivity $\sigma_{xy}^0 = 1/2$ (mod 1) as the 
spin $S = 1/2$ of a 1$d$ isotropic antiferromagnet. Even more intriguingly, 
the hypothesis of a sequence of localization-delocalization transitions 
occurring at $\sigma_{xy}^0 = \pm 1/2, \pm 3/2, \pm 5/2$ etc. translates 
into the accepted statement that a $1d$ isotropic $S = 1/2$ antiferromagnet,
while being in a massive dimerized phase for a staggered site-site
interaction, becomes critical as the staggering is turned off.

Affleck's approach, though well conceived, ultimately failed. An early 
prediction ($\nu = 1/2$, gotten from a further mapping of ${\rm SU}(2n)$ 
spin chains onto ${\rm SU}(2n)$ WZW models) was later withdrawn on the 
grounds that the phase transition at $\theta = \pi$ is likely to be of 
{\it first order} for $n \ge 2$, making it difficult if not impossible to 
extrapolate to the replica limit $n = 0$. This did not come as a surprise. 
Problems with the replica $\sigma$ models of localization theory occur 
not only for the integer quantum Hall transition, but are notorious and 
show up whenever one tries to make calculations in a nonperturbative regime 
where observables are sensitive to the global topology of the field space. 
(The first example of a discrepancy with exact results was reported
in \vz.) Here then is a summary of how I view the situation: i) The 
nonlinear $\sigma$ model \action\ does not have the right symmetry structure 
to accommodate conformal invariance and is therefore a bad starting point 
for extracting the critical properties. ii) Replica field theory has 
failed. The replica limit is not smooth and one does not know how to 
extrapolate to $n = 0$.

Clearly, a new analytical approach to the IQHE transition is very much 
called for. Such an approach was recently proposed by Ludwig et al. \ludwig\ 
who introduced a class of model systems which undergo a phase transition 
with a jump in the Hall conductivity even in the absence of disorder. This 
transition is in the Ising universality class and has $\nu = 1$. Adding 
various types of disorder to the pure system, Ludwig et al. found a rich 
array of fixed points and lines, but they were unable to control the 
perturbations that drive the system to the generic IQHE fixed point of 
interest. It remains to be seen whether any quantitative results for the 
proper IQHE transition will be obtained from this approach. In the present 
paper we will follow a more conservative line of attack. It is well known 
that, in addition to the replica trick, there exists another (and actually 
superior) method for calculating disorder averages of products of Green's 
functions for noninteracting electron systems. This is the supersymmetry 
technique pioneered by Efetov \efetov, and it was used in \weidenm\ for the 
derivation of a supersymmetric analog of Pruisken's model \action. This 
latter model, which enjoys a mathematically sound foundation, is the 
object we shall study in the present paper. 

The supersymmetric model was rejected by Affleck on the grounds that 
its fermions do not satisfy the spin-statistics theorem. While this 
is indeed so, I fail to see why it should disqualify the model. To 
be sure, the supersymmetric model was {\it not} constructed so as to
satisfy the narrow requirements of canonical quantum field theory! 
Rather, it arose from a careful study of noninteracting disordered 
electron systems, mapping the problem of calculating ensemble averages 
of products of single-electron Green's functions on the statistical 
mechanics problem of computing the Green's functions of a supersymmetric
field theory. When viewed as a statistical mechanics model, the
supersymmetric model makes perfect sense. In fact, it possesses a 
transfer matrix that passes for a well-defined object even by the 
most rigorous of mathematical standards. It is therefore natural to 
go back to Affleck's perceptive ideas \afftwo\ and try and get them 
straightened out by using supersymmetry. This is what we shall attempt 
to do in this paper. Of course, things will be rough going at times 
because we'll have to make do {\it without} the luxuries of canonical 
quantum field theory (Hilbert space, normalizable states, canonical 
quantization, unitary representations etc.) Nevertheless, as we shall 
see, the first half of Affleck's programme -- mapping of the nonlinear 
$\sigma$ model on a quantum spin chain -- can be executed. The second 
step -- mapping of the spin chain on a field theory of the Wess-Zumino 
type and computation of the critical indices -- is currently under 
investigation.

In Sect.~3 we carefully pass from the 2$d$ Euclidean supersymmetric 
field theory to its 1$d$ quantum Hamiltonian. Our treatment was 
inspired by a pedagogical paper of Shankar and Read \sr\ on the 
O(3) nonlinear $\sigma$ model. Setting up the transfer matrix on a 
mesh of lattice points and taking the anisotropic limit, we obtain 
a 1$d$ lattice Hamiltonian which is a sum of two terms: a ``kinetic 
energy'' governing the field dynamics on individual lattice sites and 
a ``potential energy'' which couples fields on neighboring sites and enforces 
local order. The kinetic energy operator for a site located on one of the 
boundaries of the quantum Hall system is a supersymmetric generalization of 
the Hamiltonian of a charged particle moving in the field of a (fictitious)
magnetic monopole, with the monopole charge being the integer $m$ closest to 
$2\sigma_{xy}^0$. We call this operator the ``monopole Hamiltonian'' for 
short. The value of the monopole charge is quantized by the condition that 
the Dirac string singularities of a corresponding magnetic vector potential 
be unobservable. For half-integral values of $\sigma_{xy}^0$, the single-site 
dynamics in the bulk of the system is governed by the monopole Hamiltonian 
with charge $m = \pm 1$, and the potential energy, which is staggered in 
general, becomes translationally invariant.

The ground state of the monopole Hamiltonian is constructed in Sect.~4. 
The outcome of this calculation can be understood from the simple analogy 
to the Landau problem of a single 2$d$ electron in a magnetic field. In 
a suitable gauge, the lowest Landau level is spanned by holomorphic
functions of the complex coordinate $z = x + iy$, multiplied by a
Gaussian. Reversal of the magnetic field turns holomorphic functions
into antiholomorphic ones. Similarly, the lowest level of the monopole 
Hamiltonian of the $\sigma$ model is a space $V$ ($V^*$) of holomorphic
(antiholomorphic) functions for monopole charge $m > 0$ ($m < 0$). 
The Gaussian gets replaced by something else since the field space
is not flat but curved.

The significance of the integer $m$ becomes particularly clear in 
Sect.~4.6. In previous work on Pruisken's nonlinear $\sigma$ model 
\prutwo, only the {\it local} (longitudinal and Hall) conductivities 
were calculated. However, it is clear that these are {\it insufficient} 
for a complete characterization of transport in a quantum Hall system 
-- as in any mesoscopic conductor -- at $T = 0$. What is needed is the 
entire set of {\it conductance coefficients} \bs. Computing these 
for a circular geometry with $N$ probes, we obtain a result which 
agrees with what is expected from phenomenological theory \butt: the 
nonvanishing conductance coefficients for even $m$ are given by
$|m|$ -- the number of ``edge states''.

The sign of the monopole charge $m = \pm 1$ alternates on sites in 
the bulk, resulting in an alternating sequence of single-site ground 
state spaces $V$, $V^*$, $V$, $V^*$ etc. Our calculation of Sect.~4.6 
suggests that a state in $V \otimes V^* \otimes V \otimes V^* \otimes 
...$ can be thought of as representing a snapshot of the evolution in 
``time'' (i.e. along a quasi-1$d$ strip) of the distribution of 
squared transfer matrix elements in an anisotropic version of the 
Chalker-Coddington model \cc. Guided by this interpretation, 
we argue that the relevant large-scale physics of the $\sigma$ model 
is captured by simply projecting the site-site interaction on the 
space of strong-coupling ground states $V \otimes V^* \otimes V \otimes 
V^* \otimes ...$ . The result of doing such a projection is a quantum 
(super-)spin chain Hamiltonian specified in detail in Sect.~5.1. The 
possibilities for obtaining exact solutions of this or related 
Hamiltonians are under current investigation.

We conclude this introduction with some comments on the style of 
presentation. This paper is intended to be fairly self-contained
and complete, which explains its length. In Sects.~3-4, familiarity 
with supersymmetry is needed only for some calculational details, 
many of which have been relegated to several appendices. The methods 
used in this paper are robust and apply to any symmetric space {\bf 
G/K}, whether it is super or not. Indeed, everything we do is easily 
transcribed to the replica $\sigma$ model by simply replacing supertrace 
by ordinary trace, superdeterminant by ordinary determinant etc. and 
omitting the sign factors from superparity. We take advantage of this 
fact by illustrating many of the concepts at the familiar example of a 
two-sphere ${\rm S}^2 \simeq {\rm U}(2)/{\rm U}(1)\times{\rm U}(1)$.

The material of Sect.~2 is largely independent of the rest of
the paper and will not be used until Sect.~5.1.\bigskip

\noindent{\bigbf 2 \ Conformal Invariance of Critical Disordered 
Electron Systems}\medskip

\noindent Conformal invariance, which occurs in models of field theory 
and statistical mechanics at a critical point \cft, is an especially 
powerful symmetry in two space dimensions, the group of conformal
transformations being in this case infinite-dimensional. Conformal 
invariance connects correlation functions in different geometries
with each other; for example, the conformal mapping $z \mapsto \ln 
z$ relates the correlation functions in the plane to those on a 
cylinder. Relations of this kind are being widely used for the 
computation, numerical and otherwise, of critical indices.

This second section of the paper is concerned with the question 
whether conformal field theory has applications to critical wave 
propagation in 2$d$ disordered media. It is by now well known \lr\ 
that the interference of waves multiply scattered by strong disorder
leads to localization, if the coherence volume is large enough.
The points in parameter space where the transition from diffusive
motion to localization occurs, are characterized by a diverging
correlation length. It is therefore reasonable to ask, and the
question has been asked before \chaone, whether conformal field theory
makes statements about the critical correlations at such a point. 
To be specific, let us consider in the sequel the motion of a single
electron in a disordered solid, although the reasoning used may have 
generalizations to other systems. Then there are (at least) two 
universality classes where a critical point occurs in two dimensions:
(i) systems in a strong magnetic field (quantum Hall systems), and 
(ii) time-reversal invariant systems with spin-orbit scattering. 

For a system of class (i) or (ii), we take $\langle \cdot \rangle$ to 
mean the average over some statistical ensemble of Hamiltonians $H$ 
which is invariant under the group of Euclidean motions of the infinite 
plane. Let $\delta(E) = \pi^{-1} \varepsilon / (\varepsilon^2 + E^2)$ be a 
regularization of the $\delta$-distribution, and let $S_E(x,y)$ denote the 
matrix element of $\delta(E-H)$ between the eigenstates of the position 
operator at $x$ and $y$. With these definitions we consider the averages
\eqn\av{
	\langle S_E(x,x)^p S_{E+\hbar\omega}(y,y)^q \rangle \quad 
	(p,q\in\NN) \quad {\rm and} \quad \langle S_E(x,y) 
	S_{E+\hbar\omega}(y,x) \rangle .}
The first of these measures the correlations of the local density of 
states, while the Fourier transform of the second one yields the dynamic 
structure factor, which can be parametrized in terms of a wave-vector and 
frequency-dependent diffusion constant \chatwo. By the spectral expansion 
$S_E(x,y) = \sum \psi_i(x) \delta(E-E_i) \overline{\psi_i(y)}$, the 
correlation functions \av\ can expressed in terms of the eigenfunctions 
of $H$ with energies $E$ and $E+\hbar\omega$. Introducing the symbol 
$\langle\langle\cdot\rangle\rangle$ to to denote ensemble averaging in 
combination with energy averaging by the regularized $\delta$-distribution, 
we can write the correlators \av\ in the short-hand form
\eqn\KAB{
	K_{AB}(x-y;E,\omega) = \langle\langle A(x) B(y) \rangle\rangle}
where $A(x)$ ($B(y)$) is some polynomial in the eigenfunctions and 
their complex conjugates evaluated at the point $x$ ($y$). $K_{AB}$ 
is translationally and rotationally invariant, by the Euclidean 
invariance of the probability measure defining $\langle\cdot\rangle$. 

At a critical point, $K_{AB}(x-y)$ is expected to show algebraic
decay $\sim |x-y|^{-\eta}$ with some positive exponent $\eta$. When 
the infinite plane is replaced by a cylinder with circumference
$L$, finite-size scaling theory states that $K_{AB}$ should decay
exponentially $\sim \exp(-|x-y|/\lambda L)$. One might now think that,
by the usual hypothesis of conformal invariance at a critical point,
$\eta$ and $\lambda$ are related by 
\eqn\scalrel{
	\eta = 1/\lambda\pi .}
This however is false and we will review the reason why in Sect.~2.1, 
by giving a simplified presentation of the work of Wegner \wegone. The 
lesson learnt from this will prompt us to point out in Sect.~2.2 a 
different class of correlation functions where the relation \scalrel\
does hold.\bigskip

\noindent{\medbf 2.1 Finite frequency breaks conformal invariance}\medskip

\noindent The message of this subsection is easy to state. In addition 
to the correlation length $\xi$ there exists a {\it second} length scale, 
$L_\omega$, which is set by the frequency $\omega$. The correlation 
functions $K_{AB}$, Eq.~\KAB, retain their dependence on $L_\omega$ at 
the critical point, even in the limit $L_\omega \to \infty$. The precise 
form of these correlation functions has been calculated by Wegner \wegone\ 
in some cases. We will now repeat Wegner's basic argument. 

Let $L_\omega$ be defined as the linear size of a system with level 
spacing $\hbar\omega$: $L_\omega = (\hbar\omega\rho)^{-1/2}$, where $\rho$ 
is the density of states per unit of energy and area. At the critical 
energy $E_c$ where the correlation length $\xi \sim |E-E_c|^{-\nu}$ 
diverges, $K_{AB}$ takes the form
\eqn\defkab{
	K_{AB}(z,E_c,\omega) = k_{AB}(|z|,L_\omega) \qquad (z = x - y).}
A necessary condition in order for arguments from conformal invariance to 
apply is that $k_{AB}$ should be ignorant of the length scale $L_\omega$. 
For $L_\omega$ large and fixed, let us therefore specialize to the limit 
$|z| \ll L_\omega$ where we expect power law behavior
	$$
	k_{AB}(|z|,L_\omega) \sim c_0 |z|^{-\eta_{AB}}
	$$
with some exponent $\eta_{AB}$. 

We can obtain more information about $\eta_{AB}$ and the dependence of
$c_0$ on $L_\omega$ by using the real-space renormalization group (RG)
transformation for disordered single-particle systems introduced by
Wegner \wegtwo. To keep the technicalities to a minimum, we assume that 
$A$ and $B$ are scaling fields ({\it i.e.} eigenfields of the RG 
transformation) with scaling dimensions $\Delta_A$ and $\Delta_B$, 
respectively. A change of cutoff scale $a \mapsto b_1 a$ then results in
	$$
	A(x) \mapsto b_1^{-\Delta_A} A(x/b_1), \quad
	B(y) \mapsto b_1^{-\Delta_B} B(y/b_1) .
	$$
Choosing $b_1  = |x-y|/a = |z|/a$, we get
	$$
	k_{AB}(|z|,L_\omega) = (|z|/a)^{-(\Delta_A  + \Delta_B)} 
	k_{AB}(a,L_\omega a/|z|) .
	$$
Note the dependence of the right-hand side on $|z|$ through both the 
algebraic prefactor and the second argument of $k_{AB}$, which is 
$L_\omega a/|z|$. To extract the full $|z|$-dependence of the correlation 
function \defkab, we must therefore perform a {\it second} RG transformation
$b_1 a \mapsto b_2 b_1 a$, taking $L_\omega a / |z|$ to some fixed
value $L_\omega a / (|z|b_2) = L_0$. In calculating the effect of this
transformation, we may no longer renormalize $A$ and $B$ separately,
since the RG transformation $a \mapsto b_1 a  = |z|$ has already brought
the points at which $A$ and $B$ are evaluated within a distance of one 
unit of the cutoff scale. Instead, we must renormalize the local object
$A(x/b_1)B(y/b_1) \simeq (AB)(x/b_1)$. The product $AB$ will not in 
general be a scaling field, but it can be expanded in terms of such fields. 
If $\Delta_{AB}$ denotes the scaling dimension of the most relevant field 
occurring in this expansion, $(AB)(x)$ is renormalized under $a \mapsto 
b_2 a$ by
	$$
	(AB)(x) \mapsto b_2^{-\Delta_{AB}} (AB)(x/b_2)~+~
	{\rm subdominant~terms}.
	$$
Taking $b_2 = L_\omega a / (L_0 |z|) = {\rm const} \times L_\omega /|z|$,
we finally obtain:
\eqn\asymp{
	k_{AB}(|z|,L_\omega) \sim |z|^{-\Delta_{A}-\Delta_{B}
	+\Delta_{AB}} L_\omega^{-\Delta_{AB}} ,}
which expresses the dependence of $k_{AB}$ on both $|z|$ and $L_\omega$ 
by three scaling dimensions $\Delta_A$, $\Delta_B$, and $\Delta_{AB}$. 
On physical grounds, the correlation function $k_{AB}$ must neither 
increase with distance $|z|$ nor decrease with $L_\omega$, $\omega \sim 
L_\omega^{-2}$ being a distance in energy space. Therefore, $\Delta_{AB} 
\le 0$ and $\Delta_{AB} \le \Delta_A + \Delta_B$. For the case of potential 
scattering (orthogonal universality class), Wegner has shown the scaling 
dimensions $\Delta_A$ and $\Delta_B$ for the first choice in \av\ to be 
zero or negative in $2+\epsilon$ dimensions. (Moreover, the scaling 
dimensions decrease without bound as the powers $p$ and $q$ increase.) 
We expect this to be true in general. Thus, the scaling dimensions 
$\Delta_O$ ($O = A, B$, or $AB$) are subject to the inequalities
	$$
	\Delta_O \le 0 , \qquad	
	-\Delta_{AB} \ge -\Delta_A-\Delta_B .
	$$
The second of these can be understood independently of \asymp\ 
by observing that the limit $L_\omega \to \infty$ restricts 
the contributions to $k_{AB}$ to those from a {\it single}
eigenfunction $\psi$ with energy close to $E_c$. The inequality
is a manifestation of the existence of large amplitude fluctuations 
of such an eigenfunction, indicating its multifractal nature \multifrac.

We now return to the question concerning the relation of $\eta_{AB} 
= \Delta_A + \Delta_B - \Delta_{AB}$ to the Lyapunov exponent for 
exponential decay of $k_{AB}$ on the cylinder. The derivation \cft\ of 
relation \scalrel\ assumes that $A$ and $B$ are (spinless) primary fields 
which are dual to each other and belong to a unitary representation 
of the Virasoro algebra, so that $\Delta_A = \Delta_B = \Delta > 0$ 
and $\langle A(x) B(y) \rangle \sim |x-y|^{-2\Delta}$. Clearly, these 
assumptions are violated in the present context. We are therefore 
ill-advised \chaone\ to expect a relation of the form \scalrel\ to hold. (Of 
course, this does not yet rule out the possibility of some generalization 
of \scalrel\ still being valid.)

Could there exist some sort of sensible, nonunitary conformal field 
theory that has an infinite number of primary fields with negative scaling 
dimensions to accommodate the correlation functions \asymp? Whatever 
the answer to this question may turn out to be, it will be constrained 
by the observation \wegone\ that $\langle \langle A(x) \rangle\rangle$ for 
$A(x) = S_E(x,x)^p$ and $p \ge 2$ {\it diverges} as the regularizing 
imaginary frequency $i\varepsilon\to 0$. This divergence is suggestive 
of a field theory with {\it unbroken noncompact symmetry} -- just as 
$\langle A \rangle = 0$ for an order parameter field $A$ would 
signal unbroken compact symmetry. In fact, ever since the seminal work 
of Sch\"afer and Wegner \sw\ and of Efetov \efetov\ it has been understood 
that the problem of calculating disorder averages of powers of wave 
amplitudes maps on field theories of the nonlinear $\sigma$ model type 
with noncompact symmetry. We will study such a field theory in Sects.~3-4. 
The nonpositivity of the dimensions of a subset of its scaling fields
turns out to be intimately related to the fact that these fields are 
{\it improper}, i.e. their application to the vacuum produces states 
that are {\it non-normalizable} in the absence of external frequency
as a regulator for non-compactness. \bigskip

\noindent{\medbf 2.2 Conformal invariance and d.c. conductance}\medskip

\noindent We consider again a $2d$ noninteracting disordered electron 
gas, either in a strong magnetic field or with spin-orbit scattering. 
The configuration space ${\cal M}$ for one electron may be finite or 
infinite. An important difference to earlier is that we now envisage 
the $N$-terminal geometry shown in Figure 1. There are $N$ contact areas 
where electrons are exchanged between ${\cal M}$ and the leads. Applying 
a small voltage (or, more precisely, a small shift of chemical potential) 
to lead $q$ we get a small current flowing in
lead $p$. The constant of proportionality is called the d.c. conductance, 
or conductance coefficient, $g_{pq}$. From linear response theory one can 
derive a formula for $g_{pq}$ as a current-current correlation function 
\bs: $g_{pq} \sim {\rm Tr}(v_p G^+ v_q G^-)$ in symbolic notation where
$v_p$ is the component of the velocity operator normal to the contact area
$C_p$, and the Green's functions $G^\pm$ are those of the {\it total} system 
(including the leads) evaluated at the Fermi energy $E_F$. It turns out that 
an alternative formulation on reduced configuration space ${\cal M}$ (without 
the leads) is possible. If $\Gamma$ is a certain nonlocal kernel determined 
by the lead geometry, it has been shown \mrznpa\ that
\eqn\gpq{
	g_{pq} = \int_{C_p} dw \int_{C_p} dx \int_{C_q} dy 
	\int_{C_q} dz \ \Gamma(w,x) G^+(x,y) \Gamma(y,z) G^-(z,w) .}
$G^\pm$ are now the Green's functions of the {\it isolated} system
subject to certain boundary conditions on the contact areas. 
Qualitatively, $G^\pm = (E_F-H\big|_{\cal M} \pm i\Gamma)^{-1}$.
The exact formulation has been given in \mrznpa. (Note that Eq.~(8)
of \mrznpa\ is misprinted: $G_c^-$ should read $G_c^+$ instead.)
Formula \gpq\ is exact for noninteracting electrons at zero temperature
and for a magnetic field of arbitrary strength. We are measuring 
conductance in its natural units $e^2 / h$. 

The energy argument of the reduced Green's functions $G^\pm$ in \gpq\ 
is a real number $E = E_F$. Why are we permitted to set to zero the 
infinitesimal imaginary part $\pm i\varepsilon$ which is usually added to 
energy to make Green's functions well-defined? The answer is that the 
boundary conditions satisfied by the reduced Green's functions spell 
out the ``openness'' of the system, i.e. the possibility for electrons 
to leave ${\cal M}$ at a contact area $C_r$ and enter the lead $r$. 
The damping caused by the escape of probability from ${\cal M}$ shifts 
the spectrum of $H\big|_{\cal M}$ off the real energy axis and makes the 
limit $\varepsilon\to 0$ exist. This is the crucial point: conductance -- 
as opposed to the Green's functions for an isolated system -- {\it does 
not require any regularizing real or imaginary frequency in order to be 
well-defined.} 

Let us now take the disorder average $\langle g_{pq} \rangle$ and set 
energy right to its critical value $E_c$ where the correlation length 
diverges. When the spatial extension of the contact areas is comparable 
to the system size, as is the case in most experimental geometries, the 
average conductance will become some scale-invariant number $\langle g_{pq}
\rangle = g_{pq}^*$. We find it more interesting however to consider the 
opposite case of {\it small} contacts, see Figure 1. A procedure that 
makes perfect sense is to take the thermodynamic limit of infinite system
size while keeping the relative distance and size of the contacts {\it 
fixed}. We expect the conductance coefficients to be well-behaved and 
finite in this limit. With system size having gone to infinity, there
do not remain any scales over which to measure correlations. (Of course, 
one can still assign some length scale to the ``strength'' or ``decay 
width'' \weident\ of each contact. However, since the contacts are no more 
than a local perturbation on the system, they do not set any scale in 
the sense of scaling theory. Note that, in spite of this, they remain
effective as regulators for conductance even in the thermodynamic limit.) 
With all scales absent, the average conductance $\langle g_{pq}\rangle$, 
and all moments of conductance, must {\it decay algebraically} with 
increasing distance between the contacts. (This will remain true even 
for large contacts, provided that their size is kept fixed as the 
distance between them is sent to infinity.)

Within the framework of the effective field theory described in Sect.~3, 
we can rephrase our assertion as follows. If the linear extension of the 
contacts is less than or comparable to the microscopic cutoff scale of 
the field theory, which is set either by the magnetic length or by the 
elastic mean free path, the contacts can be regarded as point-like. The 
conductance coefficients $\langle g_{pq}\rangle$ for $N$ terminals then 
translate into $N$-point functions
\eqn\Opqr{
	\langle O_p(x_p) O_q(x_q) \prod_r O_r(x_r) \rangle ,}
with {\it local} operators $O_i(x)$, of the field theory. On physical
grounds, these $N$-point functions must decay algebraically on {\it all} 
but the shortest distances when $E = E_c$. There aren't any length scales 
left and, moreover, the disease of divergent amplitudes is now absent. 
It is therefore reasonable to expect that the $N$-point functions \Opqr\ 
can be expanded in terms of scaling fields with {\it positive 
dimensions} of an underlying conformal field theory. (As a matter of fact, 
we shall see in Sect.~5.1 that the $N$-point functions \Opqr\ restrict to
dynamical $N$-spin correlation functions of a superspin chain Hamiltonian
for quantum Hall systems.) More precisely, because of the noncompact 
symmetry of the field theory, we expect a {\it continuum} \mrzpha\ of primary 
fields -- with an associated continuum of scaling dimensions -- to occur 
in the expansion of $O_i(x)$. 

In summary, what we are saying is the following. Critical disordered 
electron systems possess a number of interesting correlation functions 
whose definition involves two (or more) energies $E_+$ and $E_-$. The 
effect of frequency $\omega = (E_+ - E_-)/\hbar$ is similar to the 
effect of an external magnetic field on a critical spin system: it 
sets a length scale and breaks conformal invariance. While conformal 
invariance is restored for the spin system when the external field is 
turned off, the same is not guaranteed to happen for the disordered 
electron system. Many of the correlation functions of interest {\it 
diverge} as $\omega \to 0$. For example, the dynamic structure factor 
does. (This is a consequence of critical long-time correlations at the 
metal-insulator transition.) Any consistent conformal field theory 
interpretation of these correlation functions will have to confront 
the task of defining operator product expansion for an infinite set of 
primary fields whose spectrum of negative dimensions is unbounded from 
below. The situation is much better for correlation functions that 
can be defined at a single energy $E = E_c$. An example is the d.c. 
conductance -- or current-current correlation function -- of an open 
system. While conformal invariance will govern the geometric variation 
of conductance for {\it any} contact size, things become particularly
simple for point contacts: in this case the conductance coefficients
for an $N$-terminal geometry map on $N$-point functions for {\it local}
fields of an effective field theory. These $N$-point functions are 
expected to have a conformal field theory representation involving 
primary fields with positive scaling dimensions only.

I have included the material of Sects.~2.1-2 in this paper to 
motivate why the quest for the conformal limit of the 2$d$ nonlinear 
$\sigma$ model holds more promise than just an explanation of the 
critical indices $\nu = 2.3 \pm 0.1$ \refs{\cc,\hk}, $y_{\rm irr} = 
- 0.40 \pm 0.04$ \huckestein\
for the quantum Hall effect and $\nu = 2.75 \pm 0.15$ for systems 
with spin-orbit scattering \fastenrath. Conformal field theory will also 
predict interesting algebraic decay of the conductance and its 
moments for an $N$-terminal geometry with small contacts.\bigskip

\noindent{\bigbf 3 \ Hamiltonian limit of the nonlinear $\sigma$ model} 
\medskip

\noindent In the remainder of this paper we will make a fresh 
field-theoretic attack on the integer quantum Hall effect, starting from 
a supersymmetric version \weidenm\ of Pruisken's nonlinear $\sigma$
model \pru. Most of what we do in Sects.~3-5 is general and can 
be transcribed to systems with spin-orbit scattering without 
difficulty. I hope to return to such systems in a future publication.

We start out by introducing the nonlinear $\sigma$ model from a gauge
theory point of view in Sects.~3.1-2. We are not aiming at a complete 
solution of this model here. All we care about is its low-energy limit,
describing the long wave length physics of integer quantum Hall systems.
This is important to keep in mind, since restriction to the low-energy 
domain gives us much freedom to manipulate the field theory. We use 
this freedom in Sects.~3.3-5, where a careful derivation of the quantum 
Hamiltonian of the $\sigma$ model is presented. (The well-known connection 
between Feynman's path integral and Schr\"odinger wave mechanics allows 
us to relate the $2d$ classical field theory to a $1d$ quantum field 
theory.) The possibility of such a Hamiltonian formulation was first 
discussed briefly in [10]. Our approach follows the work of Shankar 
and Read \sr\ on the O(3) nonlinear $\sigma$ model with topological 
angle $\theta = \pi$ to some extent. The essential idea is to rewrite 
the topological density as an alternating sum of Wess-Zumino terms.
Gauge ambiguity of the Wess-Zumino term quantizes the values of the 
topological coupling constant for which such a rewriting can be done.
The O(3) model is somewhat special in that its field space is a 
two-sphere and the topological density has an intuitive meaning as 
(the pullback of) the solid-angle two-form. The simplifications 
resulting from this allow Shankar and Read to take some short cuts 
that are not available to us here. \bigskip

\noindent{\medbf 3.1 Review of Pruisken's model 
(supersymmetric version)}\medskip

\noindent An effective field theory for noninteracting electrons moving 
in two dimensions under the influence of a strong magnetic field and a 
Gaussian white noise potential, was first derived by Pruisken \pru. He 
pointed out that a strong magnetic field has the effect of adding a 
so-called topological term to the Lagrangian of the nonlinear $\sigma$
model for noninteracting disordered electron systems developed by Wegner, 
Efetov, and others \refs{\wegnert,\sw,\efetov}. 
A closely related model with internal 
supersymmetry, avoiding the replica trick used in \refs{\pru,\llp}, 
was subsequently derived by Weidenm\"uller \weidenm. In the present paper 
we shall study the latter model, extended by the addition of an extra pair
of indices to the $\sigma$ model matrix field. Such an extension is 
necessary for the calculation of disorder averages of products of $n$ 
retarded and $n$ advanced Green's functions $(n \ge 1)$, which is what 
is needed when one wants to go beyond the average of the conductance 
and calculate its variance and higher cumulants as well \alk. (Another 
benefit from making the extension is that it forces us to use the generic 
structures of the theory.) We will not give yet another derivation of the 
$\sigma$ model here but refer to the original papers for this 
\refs{\sw,\efetov,\pru,\weidenm}. 
The nonlinear $\sigma$ model is usually presented in 
terms of matrix fields $Q$ satisfying the nonlinear constraint $Q^2 = 1$. 
This presentation, while adequate for some purposes, is rather useless 
for most of what we intend to do in this paper. So, rather than wasting 
time by first writing down the $Q$-field formulation and then changing 
to something better, we shall present the ``good'' formulation and the 
proper structures of the model right away. 

We start from the associative algebra of $4n \times 4n$
supermatrices acting on the tensor product of three linear spaces:
(i) superspace or Boson-Fermion ($BF$) space, (ii) Advanced-Retarded
($AR$) space and (iii) replica or Extra ($E$) space. The first two
have dimension 2 and the last one dimension $n$. Invertible 
$4n\times 4n$ supermatrices $g$ satisfying the condition
\eqn\defG{
	g^{-1} = \eta g^\dagger \eta 
	\quad {\rm with} \quad
	\eta = \left( (E_{11})_{BF} \otimes (\sigma_3)_{AR}
	+ (E_{22})_{BF} \otimes 1_{AR} \right) \otimes 1_E}
form a noncompact Lie supergroup ${\bf G} = {\rm U}(n,n|2n)$. 
(Here as always $\{E_{ij}\}_{i,j=1,...,N}$ is the canonical basis 
of ${\rm End}(\CN^N)$ and $\sigma_1 = E_{12} + E_{21}$, 
$\sigma_2 = -i E_{12} + i E_{21}$,
$\sigma_3 = E_{11} - E_{22}$ for $N = 2$ are the Pauli matrices.)
The subgroup of elements in {\bf G} that commute with
	$$
	\Lambda = 1_{BF} \otimes (\sigma_3)_{AR} \otimes 1_E
	$$
is denoted by ${\bf K} = {\rm U}(n|n) \times {\rm U}(n|n)$. {\bf G} 
is projected onto {\bf G/K}, the space of left cosets $g{\bf K}$
$(g\in{\bf G})$, by $g \mapsto \pi(g) :=  g{\bf K}$. It is this 
coset space {\bf G/K} which is the field space of the nonlinear 
$\sigma$ model. The base manifold (or ``body'' in the language of 
deWitt \dewitt) of {\bf G/K} is the direct product of 
${\rm U}(n,n)/{\rm U}(n)\times{\rm U}(n)$ with
${\rm U}(2n)/{\rm U}(n)\times{\rm U}(n)$.
Both of these are Riemannian symmetric spaces \helgason, the former
of the noncompact and the latter of the compact type. We sometimes
refer to them as BB (Boson-Boson) and FF (Fermion-Fermion) space,
respectively. 

To write down the $\sigma$ model action -- and at the same time
prepare later analysis -- we identify various structures on
${\cal G} = {\rm Lie}({\bf G})$, the Lie algebra of {\bf G}. 
Defining orthogonality by the natural symmetric form 
\eqn\defB{
	{\rm B}(X,Y) = \str XY/2 \ ,}
where $\str$ is the supertrace,
we denote by ${\cal P}$ the orthogonal complement of ${\cal K} = 
{\rm Lie}({\bf K})$ in ${\cal G}$. The orthogonal decomposition
of an element $X\in{\cal G}$ by ${\cal G} = {\cal K} + {\cal P}$
is written $X = X_{\cal K} + X_{\cal P}$. (The explicit formulas 
are $X_{\cal K} = (X + \Lambda X \Lambda)/2$ and $X_{\cal P} = 
(X - \Lambda X \Lambda)/2$.) With these definitions the principal 
part of the $\sigma$ model Lagrangian can be written in the form
\eqn\defLnot{
	L_0(g,\partial g) =
	{\rm B} \left( (g^{-1}\partial_\mu g)_{\cal P} ,
	(g^{-1}\partial_\mu g)_{\cal P} \right) .}
(Repeated Greek indices are implicitly summed over the values $\mu = 0, 1$.)
Pruisken's Lagrangian \pru\ contains in addition to \defLnot\ another term, 
the so-called topological density $L_{\rm top}$, which is formulated as 
follows. Let a linear operator ${\rm J} : {\cal P} \to {\cal P}$ be 
defined by ${\rm J} X = i [ \Lambda , X ] / 2$. ${\rm J}$ squares to 
minus the identity and preserves ${\rm B}$. With its help we define
$\Omega(X,Y)$ for $X, Y \in {\cal P}$ by $\Omega(X,Y) = {\rm B}({\rm J} X,Y)$.
This two-linear form is skew symmetric:
	$$
	\Omega(Y,X) = {\rm B}(X,{\rm J} Y) = {\rm B}({\rm J} X,{\rm J}^2 Y)
	= - \Omega(X,Y) ,
	$$
owing to the symmetry of ${\rm B}$ and the properties of ${\rm J}$.
It induces the topological density by
\eqn\defLtop{
	L_{\rm top}(g,\partial g) = \epsilon^{\mu\nu}
	\Omega \left( (g^{-1}\partial_\mu g)_{\cal P} ,
	(g^{-1}\partial_\nu g)_{\cal P} \right)}
where $\epsilon^{\mu\nu}$ is the antisymmetric tensor. Both $L_0$ and 
$L_{\rm top}$ are invariant under a local change of representative 
$g(x) \mapsto g(x) k(x)$ $(k(x)\in{\bf K})$ and are therefore well-defined 
as Lagrangians on the coset field space {\bf G/K}. In addition, they are 
invariant under global transformations $g(x) \mapsto g_0 g(x)$ $(g_0 
\in{\bf G})$. Eqs.~\defLnot\ and \defLtop\ follow from the $Q$-field 
expressions \pru\ $L_0 = - \str (\partial Q)^2 / 8$ 
and $L_{\rm top} = \epsilon^{\mu\nu} \str Q \partial_\mu Q \partial_\nu 
Q / 8i$ on making the substitution $Q = g\Lambda g^{-1}$. 

After these preparations, we introduce the nonlinear $\sigma$ model for an 
integer quantum Hall system with longitudinal conductivity $\sigma_{xx}$
and Hall conductivity $\sigma_{xy}$ by the functional integral
\eqn\defZ{
	Z  = \int {\cal D}[g] \exp - \int_{\cal M} d^2 x
	\left( \sigma_{xx} L_0 + i \sigma_{xy}
	L_{\rm top} \right) .}
$\sigma_{xx}$ and $\sigma_{xy}$ are measured in their natural 
units $e^2 / h$. ${\cal M}$ is the configuration space of the 
two-dimensional electron gas, and the functional integration measure 
${\cal D}[g]$ equals $\prod_x dg_K(x)$ where $dg_K$ is the uniform 
super-integration measure for {\bf G/K}. (Note that the derivation of 
the nonlinear $\sigma$ model from a Gaussian white noise potential 
yields for $\sigma_{xx}$ and $\sigma_{xy}$ their bare (or SCBA) values, 
denoted by $\sigma_{xx}^0$ and $\sigma_{xy}^0$ in \refs{\pru,\llp}. 
We here imagine that these coupling constants have already been renormalized 
and simplify the notation by dropping the superscript 0.)
 
The functional integral \defZ\ calls for some additional explanation. 
First of all, the invariant symmetric form ${\rm B}$, Eq.~\defB, has 
the crucial property of being {\it positive definite} on the ordinary 
part (or body) of ${\cal P}$. This 
means that {\bf G/K} is {\it Riemann} in its natural geometry induced 
by ${\rm B}$. Second, an experimental quantum Hall system has a boundary 
$\partial {\cal M}$, which is composed of insulating and conducting parts,
giving rise to corresponding boundary conditions for the $\sigma$ model 
field in \defZ. What is particularly important is that the conducting 
boundary conditions cut off the spatially uniform fluctuations
of the noncompact $\sigma$ model field -- which would otherwise
cause a divergence -- by breaking the global {\bf G}-symmetry
of \defZ\ down to a global {\bf K}-symmetry. (The microscopic 
origin of the damping is the absorptive term $i\Gamma$ in the 
reduced Green's functions $G^\pm = (E_F - H\big|_{\cal M} \pm 
i\Gamma)^{-1}$.) This, in conjunction with the Riemannian nature of
{\bf G/K}, makes the functional integral \defZ\ well-defined and
convergent after UV regularization. Third, for the sequel we adopt the 
Corbino geometry ${\cal M} = [R_1,R_2] \times {\rm S}^1$, with two 
or more contacts placed along the boundary $\partial{\cal M}$.
Alternatively, for a particularly neat theoretical model, we might 
imagine ${\cal M}$ to be a two-torus ${\rm S}^1\times{\rm S}^1$, with 
several point contacts attached to it to cut off the divergent 
$\sigma$ model fluctuations. What is important for our derivation
of the quantum Hamiltonian in Sect.~3.5 is that one of the two
directions on ${\cal M}$ be {\it periodic}. This direction $\sim {\rm 
S}^1$ will be assigned the role of imaginary time. Finally, we remark that 
the functional integral $Z$, Eq.~\defZ, equals unity by supersymmetry 
[12] which serves to cancel all vacuum graphs. (This is true even in 
the presence of a boundary since the boundary conditions preserve 
{\bf K}-supersymmetry.) To obtain the Green's functions of the nonlinear 
$\sigma$ model one adds sources to the Lagrangian and takes derivatives 
as usual.

For $\sigma_{xy} = 0$, \defZ\ is the nonlinear $\sigma$ model 
describing the long wave length excitations of noninteracting 
disordered 2$d$ electron systems with time-reversal invariance 
broken by a weak magnetic field (unitary universality class). 
Perturbative renormalization group analysis \refs{\wegnert,\efetov} shows 
its coupling constant $1/\sigma_{xx}$ to be driven to strong coupling at 
large scales. This has led to the widely accepted conjecture that {\it 
all} stationary single-particle states of such systems are localized in 
$d = 2$. At the same time, the plateau-to-plateau transition in integer 
quantum Hall systems cannot be understood if all bulk electronic states 
are localized. There is ample numerical \refs{\cc,\hk} and experimental 
\expqp\ evidence that states do in fact become delocalized near the center 
of each Landau band. (The data points to delocalization occurring only 
at a {\it single} energy.) From accurate numerical studies for the
lowest Landau level, we know the correlation length exponent of this 
delocalization phase transition to be $\nu = 7/3$ (or close to that 
value). Hence, if \defZ\ is to describe the underlying physics, the 
addition of the topological density $L_{\rm top}$ must cause critical 
behavior, i.e. vanishing masses for some of the $\sigma$ model fields, 
at certain values of $\sigma_{xy}$. At first sight, though, $L_{\rm top}$ 
appears to be an unlikely candidate for the job it is supposed to do. 
$L_{\rm top}$ does not change the classical equations of motion, for 
it can be written locally as a total derivative. This is closely 
related to the fact that, for suitable boundary conditions, $\int 
d^2x \ L_{\rm top}$ is topologically quantized in integer multiples 
of $2\pi$ [10,12], making it invisible in weak-coupling perturbation 
theory. Nevertheless, Pruisken and collaborators [10] have argued
emphatically that the presence of the topological density does lead to 
decisive changes in the large-scale physics of the replica analog 
of the nonlinear $\sigma$ model \defZ. One of their arguments uses 
twisted boundary conditions and ideas stimulated by t'Hooft duality to 
show that some kind of phase transition must indeed occur somewhere in 
the range from $\sigma_{xy} = m$ to $\sigma_{xy} = m+1$. However, the 
precise nature of the phase transition is {\it not} specified by the 
duality argument. It could, in principle, be of first order with a 
finite correlation length. (This is in fact the conclusion reached by 
Affleck \afftwo: the ${\rm U}(2n)/{\rm U}(n)\times {\rm U}(n)$ nonlinear
$\sigma$ model for $n \ge 2$ has a {\it first-order} transition at 
$\theta = \pi$.) As was mentioned in the introduction, the current 
algebra of the model \defZ\ fails to meet the strict requirements set 
by conformal invariance in $d = 2$. By implication, the model \defZ\ 
cannot possess any nontrivial RG fixed points, contrary to what has 
been asserted by Pruisken et al. [10,11]. Rather, what must happen is 
that the model at $\sigma_{xy} = 1/2$ (mod 1) is driven under 
renormalization to the fixed point of some {\it other} field theory, 
which is not known at present but is likely to be of the Wess-Zumino 
type. To summarize the situation, a sceptic will say that fifteen years 
after the discovery of the integer quantum Hall effect, one of its 
crucial features, the occurrence of a localization-delocalization 
transition in the center of a Landau band, is still lacking an analytic 
description. This very unsatisfactory state of affairs has been the 
motivation for the efforts reported in the present paper. \bigskip

\noindent{\medbf 3.2 Gauge fixing}\medskip

\noindent The nonlinear $\sigma$ model for noninteracting disordered electron 
systems is usually presented as a theory of fields $Q$ taking values in some 
nonlinear space isomorphic to a coset space {\bf G/K}. In Sect.~3.1 we 
gave its alternative presentation [10] as a theory of {\bf G}-valued 
fields $g$ with an invariance under local gauge transformations $g(x) 
\mapsto g(x) k(x)$ for $k(x) \in {\bf K}$. In other words, we are here 
viewing the nonlinear $\sigma$ model as a kind of {\it gauge theory} 
with the gauge group {\bf K} acting on the field space {\bf G} by right 
multiplication. (However, as the gauge field has no dynamics, the nonlinear 
$\sigma$ model is not a gauge theory in the strict sense of the word.)
The practical implementation of the gauge theory point of view requires 
{\it gauge fixing}, i.e. some prescription for separating the unphysical 
(or gauge) degrees of freedom from the physical ones. Gauge fixing in the 
present context is best described as follows. We choose some (locally) 
smooth map $s : {\bf G/K} \to {\bf G}$. The only requirement on $s$ is 
that, with $\pi$ being the canonical projection ${\bf G} \to {\bf G/K}$, 
$s$ has to satisfy $\pi \circ s = {\rm identity}$. Equivalently, for every
$g\in{\bf G}$ the element $k(g)$ defined by the equation
\eqn\defk{
	g = s(\pi(g)) k(g) }
must be an element of {\bf K}. Technically speaking, an $s$ with
this property is called a {\it section} of the bundle $\pi : {\bf G}
\to {\bf G/K}$.
For the purpose of illustration, consider the example 
${\bf G} = {\rm SU(2)}$ and ${\bf K} = {\rm U(1)}$ (generated
by $i\sigma_3$). 
The coset space SU(2)/U(1) is isomorphic to the two-sphere
${\rm S}^2$ by the map $g{\rm U(1)} \mapsto g\sigma_3 g^{-1}
= \sum_{i=1}^3 n_i \sigma_i$, assigning to every coset $g{\rm U(1)}$
a unit vector ${\bf n} = (n_1,n_2,n_3)$. 
If we parametrize ${\rm S}^2$ by a polar angle $\theta$ and
an azimuthal angle $\phi$ in the usual way, two examples of a section
$s : {\rm S}^2 \to {\rm SU(2)}$ are
\eqn\examsec{
	s = e^{i\phi\sigma_3 /2} e^{i\theta\sigma_2 /2}
	\quad {\rm and} \quad
	s = e^{i\phi\sigma_3 /2} e^{i\theta\sigma_2 /2} 
	e^{-i\phi\sigma_3 /2} .}
The property $\pi \circ s = {\rm id}$ is easy to check using the
isomorphism ${\rm SU(2)/U(1)} \simeq {\rm S}^2$. 

After this example we continue our general considerations.
By definition, a section $s$ assigns to a coset $\pi(g) = g{\bf K}$
one representative $s(\pi(g)) \in {\bf G}$. Since this assignment
is (locally) one-to-one, $s$ distinguishes some submanifold of
{\bf G} which is (locally) isomorphic to {\bf G/K}. Thus,
making a choice of $s$ fixes the gauge: $s$ identifies the subspace
of physical degrees of freedom on the field space {\bf G}, which
are integrated over in the functional integral, while $k : {\bf G}
\to {\bf K}$ defined by Eq.~\defk\ accounts for the unphysical
gauge degrees of freedom (not integrated over).
There is of course much arbitrariness in the choice of $s$.
However, given any two sections $s$ and $s_1$ we can always find
a map $k_1 : {\bf G/K} \to {\bf K}$ such that $s$ and $s_1$ are 
related by a gauge transformation $s = s_1 k_1$.
Thus, the arbitrariness of $s$ precisely reflects the freedom we
have in choosing the gauge. 

We were careful to insert the qualification ``locally'' in various 
places of the above discussion. As a matter of fact, $s$ cannot be 
defined globally when {\bf G/K} is topologically nontrivial. To see 
this by way of example, let us focus again on ${\rm SU(2)/U(1)} 
\simeq {\rm S}^2$ and consider the first of the two sections given 
in \examsec. Since the azimuthal angle $\phi$ is ill-defined on the 
north and south pole, $s$ {\it is singular at these points}. (It is 
perhaps helpful to mention that a choice of $s$ specifies, by formula 
(27) below, a choice of vector potential {\bf A} for a magnetic 
monopole placed at the center of ${\rm S}^2$. The singularities of 
$s$ then translate into Dirac string singularities of {\bf A}.)
We can move the singularities of $s$ around at will by a gauge
transformation, but a basic result of fibre bundle theory \egh\ 
tells us that we can never make them disappear altogether. The 
same kind of topological obstruction occurs for our quantum Hall 
model space {\bf G/K}, due to the presence of the topologically 
nontrivial space ${\rm U}(2n)/{\rm U}(n)\times {\rm U}(n)$ in 
its base manifold. (In this case however the singularities are 
not points but are submanifolds of codimension two on 
${\rm U}(2n)/{\rm U}(n)\times {\rm U}(n)$.)

We need to make a small addendum concerning the most general choice of 
section $s : {\bf G/K} \to {\bf G}$ that will be permitted in this paper. 
It so turns out that one crucial step -- going from Eq.~(25) to Eq.~(28) 
below -- requires the values taken by $s$ to be restricted to ${\bf G}_0 = 
{\rm SU}(n,n|2n)$, the subgroup of elements of {\bf G} with superdeterminant 
equal to unity. There are good reasons for {\it not} making the restriction
to ${\bf G}_0$ throughout our development. (Working with ${\bf G}_0$ instead 
of ${\bf G}$ would cause a certain amount of inconvenience in Sect.~5.) Thus, 
we will base all our considerations on the pair {\bf G} and {\bf K}, except 
that a section $s$ of the bundle ${\bf G} \to {\bf G/K}$ will always have to 
be a map $s : {\bf G/K} \to {\bf G}_0$. \bigskip

\noindent{\medbf 3.3 Quantum Hamiltonian for $\sigma_{xy} = 0$}\medskip

\noindent To make a continuum field theory such as \defZ\ well-defined, 
one needs to specify some prescription for regularization of the
functional integral in the ultraviolet. Elementary particle physics 
has come up with quite a number of different regularization schemes
for us to use. The approach we adopt here is to reformulate the 
two-dimensional Euclidean field theory as a one-dimensional quantum 
theory defined on a lattice in space. This particular regularization 
scheme has been used successfully by Shankar and Read \sr\ in their 
frontal assault on the O(3) nonlinear $\sigma$ model with topological 
angle $\theta = \pi$. It is the optimal choice in the present context, 
too, as it will allow us to clearly identify the relevant low-energy 
degrees of freedom. 

The quantum Hamiltonian of a canonical 2$d$ Euclidean field theory can 
be constructed in the following standard way. First, one re-interprets 
the Euclidean theory as a Lorentz-invariant field theory in 1+1 
dimensional Minkowski space-time. Then, the Lagrangian of the latter 
is converted into a Hamiltonian function by carrying out a Legendre
transform. And finally, the classical Hamiltonian is turned into 
a quantum operator by the process of canonical quantization. 
What is attractive about this procedure is that it avoids
detailed manipulations of the functional integral.
It turns out, however, that the supersymmetric field theory \defZ\
does not have a Hilbert space for its space of quantum states,
so it is {\it not a quantum field theory of the canonical type}. 
Canonical quantization, though applicable as a purely formal
procedure, is therefore not really appropriate here. (The 
noncanonicity is probably one of the reasons why the model \defZ\ has 
not yet been embraced by the community at large but has been studied 
by a rather small group of people.) I emphasize that there is nothing 
ill-defined or suspect about our noncanonical field theory. After 
lattice regularization, we are faced with the bona fide statistical 
mechanics problem of computing a finite number of {\it well-defined 
and convergent} superintegrals.

The foregoing discussion motivates the no-nonsense approach we 
will adopt. Having UV-regularized the field theory by placing its
fields on a rectangular lattice, we will take the anisotropic 
(or Hamiltonian) limit and use no more -- and no less -- than 
straightforward manipulations of the integrals over field variables 
to derive the quantum Hamiltonian. More precisely, what we will do
is to carry out the familiar steps leading from Feynman's path
integral (in imaginary time) to the Schr\"odinger picture of quantum
mechanics. Aside from getting the proper structures into play, this 
approach has the added advantage of leading directly and without 
further ado to the expression for the quantum Hamiltonian optimally 
suited for later analysis. There exists one delicate aspect however: 
the topological term needs to be handled with care. To avoid 
dealing with too many issues at once, we here set $\sigma_{xy} = 0$ 
and postpone the inclusion of the topological term to Sect.~3.5.

As is usual in statistical mechanics, we define the quantum 
Hamiltonian $H$ as (minus) the logarithm of the transfer matrix
$T$ connecting two neighboring rows of lattice sites, both at 
constant time. (The first of the two directions on ${\cal M} =
I \times {\rm S}^1$ is decreed to be ``space'' $I = [R_1,R_2]$,
the other ``time'' or, rather, imaginary time.) In order for 
$H = - \ln T$ to come out to be a local and tractable operator,
we need to make the system anisotropic by lowering the coupling,
or ``temperature'', in time direction while strengthening it in 
space direction. Such an ad hoc modification of the system is 
permissible if we are concerned (as we are) with the study of 
critical or low-energy properties only, since spatial anisotropy 
is an irrelevant perturbation at a critical point. Imagining the 
anisotropy to be caused by our use of two different lattice 
constants -- a small one, $a_0$, in time direction and a larger 
one, $a_1$, in space direction -- we make the replacement
\eqn\discLnot{
	\int_{I\times{\rm S}^1}
	dx d\tau \ L_0(g,\partial g) \longrightarrow
	\sum_{\bf n} \left( {a_1 \over a_0} 
	\Delta(g_{\bf n},g_{{\bf n}+{\bf e_0}})+ {a_0 \over a_1}
	\Delta(g_{\bf n},g_{{\bf n}+{\bf e_1}})	\right) }
with the ``distance'' $\Delta(g,h)$ between $g$ and $h$ defined by
$\Delta(g,h) = \str g^{-1} h \Lambda h^{-1} g \Lambda / 4$. The sum 
runs over all lattice points ${\bf n}$, and ${\bf e_0}$ and ${\bf e_1}$ 
are the lattice vectors in time and space direction. Periodic boundary 
conditions $g_{{\bf n} + N_0 {\bf e_0}} = g_{\bf n}$ are assumed. Our 
discretization scheme has the right naive continuum limit, as is easily 
seen from $\Delta(g,g e^{aX}) = a^2 {\rm B}(X_{\cal P},X_{\cal P}) \ 
{\rm mod} \ a^3$. Note also that, since $\Delta(h g_1,h g_2) = 
\Delta(g_1,g_2) = \Delta(g_1 k_1,g_2 k_2)$ for $h \in {\bf G}$ and 
$k_1 , k_2 \in {\bf K}$, the replacement \discLnot\ preserves the global 
and local symmetries of the continuum theory. 

By its origin from ${\rm B} \left( (g^{-1}\partial g)_{\cal P} ,
(g^{-1}\partial g)_{\cal P} \right)$ with $\partial$ differentiating 
in time (space) direction, the summand with $a_0$ in the denominator 
(numerator) is naturally identified as a ``kinetic energy'' (``potential
energy''). We now wish to extract the logarithm of the transfer matrix 
in the Hamiltonian limit $a_0/a_1 \to 0$. To do so, we must first perform 
a standard transformation, turning kinetic energy into a second-order 
differential operator while inverting the coefficient $a_1/a_0$. A 
crucial simplification gained from making the system very anisotropic 
is that we can ignore the potential energy terms while doing this 
transformation. (Both potential energy and the differential operator 
that will eventually represent kinetic energy carry a factor of $a_0$, 
so their commutator is of negligible order $a_0^2$.) Thus, concentrating 
on a single site ${\bf n}$, setting $g_{\bf n} = g$ and $g_{{\bf n}+{\bf 
e_0}} = h$, and defining $\epsilon$ by $a_0 / a_1 = \sigma_{xx} \epsilon$,
what we need to do is to work out the integral
\eqn\tratau{
	(\tau f)(\pi(h)) := \int_{\bf G/K} dg_K \ \exp \left( - {1\over 
	\epsilon} \Delta(h,g) \right) f(\pi(g)) ,}
which transfers some function $f(\pi(g_{\bf n}))$ to the neighboring 
site ${\bf n} + {\bf e_0}$ in time direction. This is prepared by the
following little calculation:
\eqn\evaltau{
	(\tau f)(\pi(h)) = \int_{\bf G/K} dg_K \ 
	e^{- \Delta({1},g)/ \epsilon} f(\pi(hg))
	= \int_{\cal P} dX \ e^{-{\rm B}(X,X)/ \epsilon} f(\pi(he^X)).}
The first equality sign follows from the invariance of $dg_K$ 
under the substitution $g \mapsto h g$, and from the symmetry 
property $\Delta(h,hg) = \Delta({1},g)$ with ${1}$ being the unit 
element in {\bf G}. For the second equality sign we set $g = e^X$,
parametrizing {\bf G/K} by its tangent space ${\cal P}$ at the
origin $\pi({1})$. In the small-$\epsilon$ limit under consideration,
the integral receives contributions only from the immediate vicinity of 
$\pi({1})$. We may therefore replace $\Delta({1},e^X) = \str e^X 
\Lambda e^{-X} \Lambda / 4 = \str e^{2X} / 4 = \str \cosh(2X) / 4$ by
${\rm B}(X,X) = \str X^2 / 2$ and $dg_K$ by $dX$, the Euclidean
integration measure on ${\cal P}$. 

It remains to convert the last integral in \evaltau\ into a differential 
operator acting on $f$. To that end, we express $X$ in some basis 
$\{e_i\}_{i=1,...,{\rm dim}{\cal P}}$ as $X = \sum x^i e_i$, so that
\eqn\defmetr{
	{\rm B}(X,X) = \sum_{ij} x^i g_{ij} x^j}
with $g_{ij} = {\rm B}(e_i,e_j)$. By Taylor expanding $f$
and doing the resulting Gaussian integral, we easily get
\eqn\taures{
	(\tau f)(\pi(h)) = ( f + 4 \epsilon {\cal L} f )(\pi(h))
	\ {\rm mod} \ \epsilon^2}
where ${\cal L}$ is the second-order differential operator defined by
\eqn\deflapl{
	({\cal L} f)(\pi(g)) = \lambda(\partial/\partial x)
	f(\pi(g\exp {\scriptstyle\sum} x^i e_i))\Big|_{x=0} 
	\quad {\rm with} \quad
	\lambda(\partial/\partial x) = \sum_{ij} g^{ij}
	{\partial^2 \over\partial x^j \partial x^i} \ .}
(The indices of the metric tensor are raised by $\sum_j g^{ij} 
g_{jl} = {\delta^i}_l$ as usual.) Note that, in view of its 
derivation from a manifestly gauge-invariant expression,
${\cal L}$ is independent of the choice of gauge and thus makes 
sense as an operator acting on functions on {\bf G/K}. It is not 
difficult to see directly from \deflapl\ how this gauge invariance 
comes about. A change of representative $g \mapsto gk$ $(k \in 
{\bf K})$ amounts to the same as a coordinate transformation 
$\sum x^i e_i \mapsto \sum y^i e_i = k(\sum x^i e_i)k^{-1}$. Such 
a transformation leaves ${\rm B}(X,X)$, Eq.~\defmetr,  invariant 
and the process of raising the indices of the metric tensor confers 
this invariance upon $\lambda$.

We can now read off the quantum Hamiltonian $H$ from Eqs.~\defZ,
\discLnot, \tratau\ and \taures:
\eqn\defqHam{
	a_0 H := - \ln T = {a_0 \over a_1} \sum_{l=1}^{N_1}
	\left( - {4 \over \sigma_{xx}} {\cal L}_l + \sigma_{xx}
	\Delta(g_l,g_{l+1}) \right) \ {\rm mod} \ (a_0/a_1)^2 .}
The first term is site-diagonal, and for a single site is defined by the 
formula \deflapl. The second term couples neighboring sites and acts on 
``wave functions'' $f(\pi(g_1),...,\pi(g_{N_1}))$ by ordinary multiplication.

We should like to point out that the operator ${\cal L}$ is nothing 
but the Laplace operator $\nabla^2 = {\rm sdiv} \ {\rm grad}$ (gradient 
followed by superdivergence) for {\bf G/K}. To see this we note that, 
since a left translation $L_g : \pi(h) \mapsto \pi(gh)$ acts on the 
left while ${\cal L}$ acts on the right, $L_g$ and ${\cal L}$ {\rm 
commute}. Now, by a basic result of the theory of homogeneous spaces 
{\bf G/K} \helgason, any second-order differential operator with the 
property of commuting with left translations is a multiple of the 
Laplace operator. Hence ${\cal L} = c \nabla^2$ $(c\in \RN)$. 
The constant of proportionality $c$ equals unity since $\lambda$ 
is the correctly normalized differential operator associated with 
the quadratic form defining the metric. Choosing some system of 
super-coordinates $\xi^i$ in which the metric ${\rm B}\left( 
(g^{-1}{\rm d}g)_{\cal P} , (g^{-1}{\rm d}g)_{\cal P} \right)$ is 
expressed by $\sum {\rm d}\xi^i g_{ij}(\xi) {\rm d}\xi^j$,
we have for $\nabla^2$ the familiar coordinate expression
\eqn\nab{
	\nabla^2 = J(\xi)^{-1/2} \sum_{ij} (-)^{|i|} 
	{\partial\over\partial \xi^i} g^{ij}(\xi) J(\xi)^{1/2} 
	{\partial\over\partial \xi^j} }
where $J(\xi) = \sdet (g_{ij}(\xi))$ and $\sdet$ is the superdeterminant.
(The sign factor involving the superparity $|i|$ of $\xi^i$ is due to 
supersymmetry.) The coordinate expression \nab\ illustrates the meaning 
of the less familiar formula \deflapl. For practical calculations, it
is much less useful than \deflapl\ and we will actually never use it.

What can we say about the Hamiltonian $H$, Eq.~\defqHam? The normalization of
the functional integral $Z = 1$ translates into ${\rm trace}\exp(-\beta H) 
= 1$ for all $\beta$, $\sigma_{xx}$ and $N_1$. We therefore expect $H$ 
to have a unique singlet ground state with zero energy and excited 
states organized into supermultiplets consisting of an equal number of 
bosonic and fermionic states, whose contributions to the partition sum 
exactly cancel each other. Apart from these special features due to 
supersymmetry, we expect the physics of $H$ to be similar, at least 
qualitatively, to that of the quantum Hamiltonian of a conventional 
(i.e. nonsuper) nonlinear $\sigma$ model such as the O(3) one. The 
nature of the ground state will be determined by the battle between
kinetic energy and interaction energy (formerly called potential energy). 
The first of the two makes the field diffuse on {\bf G/K}, thereby
causing strong fluctuations in the ground state when $\sigma_{xx}$ is 
small. The interaction energy, in contrast, has the opposite tendency
of aligning neighboring field variables. For a space dimension larger 
than the lower critical dimension of 2, either one of the two tendencies 
may win. If $\sigma_{xx}$ is small enough, the quantum fluctuations
caused by $\sum {\cal L}_l$ disorder the ground state. As $\sigma_{xx}$ 
increases beyond some critical value $\sigma_{xx}^*$, the interaction 
energy takes over, driving a phase transition to an ordered ground state 
with spontaneously broken symmetry. This phase transition is the Anderson 
(metal-insulator) transition in noninteracting disordered electron systems 
with time reversal invariance broken by magnetic impurities or a weak 
magnetic field. In $d = 1 + 1$, the case at hand, renormalization group 
analysis of the continuum theory \defZ\ leads us to expect that quantum 
fluctuations {\it always disorder} the ground state on sufficiently large 
length scales, no matter how big is the value of $\sigma_{xx}$. In other 
words, the Hamiltonian \defqHam\ is expected to have a nondegenerate and 
disordered ground state with massive excitations for any $\sigma_{xx}$. 
This scenario must be changed drastically by the inclusion of the topological 
term. \bigskip

\noindent{\medbf 3.4 Topological density and Berry phase}\medskip

\noindent In Sect.~3.3, we put the Euclidean field theory 
\defZ\ for $\sigma_{xy} = 0$ on a rectangular lattice with 
lattice constants $a_0$ and $a_1$, and then took the extreme 
anisotropic limit $a_0 / a_1 \to 0$ to derive the expression 
\defqHam\ for  the quantum Hamiltonian $H$. Our present goal is 
to extend this procedure to the case of half-integral $\sigma_{xy}$.
We can imagine several ways of discretizing the topological density.
Recalling the $Q$-field expression $L_{\rm top} = \epsilon^{\mu\nu}
\str Q \partial_\mu Q \partial_\nu Q / 8i$
we might be led to the scheme
\eqn\disLtop{
	\int_{\cal M} d^2 x \
	\epsilon^{\mu\nu} \str Q \partial_\mu Q \partial_\nu Q
	\longrightarrow \sum_{\bf n} \str Q_{\bf n} 
	\left( Q_{{\bf n}+{\bf e_0}}-Q_{\bf n} \right)
	\left( Q_{{\bf n}+{\bf e_1}}-Q_{\bf n} \right) .}
Note the conspicuous absence of the lattice constants $a_0$ and $a_1$, 
indicating the topological (i.e. metric free) nature of the term.
Although the naive choice \disLtop\ pays no attention to the geometric 
meaning of the topological density, it should work just fine when 
field configurations are slowly varying. For our purposes, this would 
actually be good enough since we will be satisfied with a quantum 
Hamiltonian that gives the correct low-energy physics near criticality.
However, since triples of field variables $Q$ separated by one unit in 
both space and time direction are coupled together, the discretized 
topological density \disLtop\ is hard to make any analytic progress with. 
Fortunately, a discretization scheme much better than \disLtop\ exists. 
To formulate this scheme and understand 
its true meaning, we have to work through some differential geometry.
The crucial step will be to take a curve ${\rm S}^1 \to {\bf G/K}$,
$t \mapsto \gamma(t)$ and lift it to a curve $t \mapsto g(t)$ on
{\bf G}. In this way we will be able to relate the topological 
density to the geometric phase of Berry \berry. Discretization 
and the Hamiltonian limit will then be almost immediate. 

Before delving into our differential geometric project, let me inject
a few words of explanation concerning the use of terminology. 
What we are dealing with is a {\it supermanifold} (more precisely, 
a supermanifold of the Berezin-Kostant type \berezin). Such an object
is not made from points -- anticommuting c-numbers, unlike real numbers,
can't be considered as points on a line -- so the reader may wonder
just what is the meaning of the word ``curve'' in the present context?!
The answer is this. When constructing an ordinary (i.e. nonsuper) 
differential manifold one has two options: one may describe the manifold
either by its points (i.e. as a topological space with a differentiable
structure) or by its functions. Since a function $f$ can
be applied to a point $x$ to produce the number $f(x)$, functions 
are {\it dual} to the elements of the space they are functions on.
Because of this relationship, an ordinary manifold can 
be reconstructed from the knowledge of its functions. In the 
super case the point option is nonexistent, which is of course 
why Berezin \berezin\ defines a supermanifold by its (super-)functions.
This is not as horrible as it may sound, as the vast space of
functions -- on a supermanifold just as on an ordinary manifold --
can be generated from a basic set of {\it coordinate functions}.
The message following from this is that, on a pedantic level,
we ought to be using coordinate language throughout. For
example, when we talk about a ``curve'' on a supermanifold, what
we should really do is to describe the object ``curve'' in
terms of some set of superfunctions $\xi^i(t)$ depending on a parameter 
$t$ and taking values in the underlying parameter Grassmann algebra.
(In the ordinary case, the $\xi^i(t)$ would result from evaluation of 
a set of coordinate functions $x^i : {\bf G/K} \to \RN$ along 
the curve $t \mapsto \gamma(t)$.) Clearly, our notation will be more 
transparent, and our general train of thought much easier to follow, 
if we say ``curve'' rather than ``a set of functions depending on a 
parameter and taking values in ...'' This is why we shall stick to 
point manifold language for the most part, keeping however in mind 
that everything we do should be expressible (and ultimately must be 
expressed) in terms of supercoordinates. 

After this preamble, we consider for some smooth field configuration 
the integral of the topological density $L_{\rm top}$ over some 
connected and simply connected region $\Gamma$ of space-time ${\cal M}$:
\eqn\defvarG{
	\varphi_\Gamma = \int_{\Gamma} d^2x \ L_{\rm top} .}
We wish to elucidate the meaning of $\varphi_\Gamma$ as the geometric 
phase acquired by a ``{\bf G/K}-superparticle'' when transported once 
around the curve $\partial\Gamma \to \pi(g(\partial\Gamma))$. We begin 
by writing $L_{\rm top}$, Eq.~\defLtop, in the form
	$$
	L_{\rm top}(g,\partial g) = {i\over 2} \epsilon^{\mu\nu}
	\str \Lambda (g^{-1} \partial_\mu g)_{\cal P}
	(g^{-1} \partial_\nu g)_{\cal P} .
	$$
Our first step will be to convert the integral over $\Gamma$ 
into a line integral along the boundary $\partial\Gamma$. 
Since $L_{\rm top}$ is induced by the two-form
\eqn\defom{
	\omega = {i \over 2} \str \Lambda (g^{-1} {\rm d}g)_{\cal P}
	\wedge (g^{-1} {\rm d}g)_{\cal P} ,}
we will be able to apply Stokes' theorem if we can find a one-form
${\cal A}$ with exterior derivative ${\rm d}{\cal A} = \omega$. We 
claim that if $\pi_+ = (1+\Lambda)/2$ and 
$s$ is any section ${\bf G/K} \to {\bf G}_0$, the choice
\eqn\defcalA{
	{\cal A} = - i \str \pi_+ s^{-1} {\rm d}s}
does the job. The argument is as follows. Because $s^{-1}{\rm d}s$ takes 
values in ${\rm Lie}({\bf G}_0) = {\rm su}(n,n|2n)$, the supertrace 
$\str s^{-1}{\rm d}s$ vanishes and we may replace the expression for 
${\cal A}$ by ${\cal A} = \str \Lambda s^{-1}{\rm d}s/2i$. Differentiating 
the relation  $g = s(\pi(g)) k(g)$ we get $s^{-1} {\rm d}s = 
k(g^{-1}{\rm d}g - k^{-1}{\rm d}k)k^{-1}$, which we insert into 
${\rm d}{\cal A} = i \str \Lambda (s^{-1} {\rm d}s) \wedge (s^{-1} 
{\rm d}s) /2$. Evaluation of the two-form ${\rm d}{\cal A}$ on a
pair of tangent vectors $s X$ and $s Y$ gives $i \str \Lambda [X,Y]/2 
= i \str Y [\Lambda,X]/2$. This vanishes for $X \in {\cal K}$ and $Y 
\in {\cal K}+{\cal P}$, so only the ${\cal P}$-component of 
$s^{-1}{\rm d}s$ need be kept. Hence, ${\rm d}{\cal A} = i \str
\Lambda (s^{-1}{\rm d}s)_{\cal P} \wedge (s^{-1}{\rm d}s)_{\cal P}/2$, 
and since $(s^{-1}{\rm d}s)_{\cal P} = k (g^{-1}{\rm d}g)_{\cal P} 
k^{-1}$ and $k^{-1}\Lambda k = \Lambda$, comparison with \defom\ 
shows that ${\rm d}{\cal A} = \omega$, as claimed. 

Naive application of Stokes' theorem would now yield
$\varphi_\Gamma =  -i \int_{0}^1 dt \str \pi_+ s^{-1} \dot s$,
where  $\dot s$ is the total derivative of $s$ along the curve 
$\partial\Gamma$, which we parametrize by $t \in [0,1)$. This, 
however, is not quite correct. There exists a difficulty
of topological origin. For a {\it fixed} choice of gauge $s$, the 
gauge potential $-i \str \pi_+ s^{-1} \dot s$ induced from 
${\cal A}$ by the field configuration along the loop $\partial\Gamma$, 
may not have a smooth extension to the interior of the loop. This 
is because in the presence of topological excitations the field
may wind once -- or several times -- around a Dirac-string type
singularity of $s$ as we move along $\partial\Gamma$. Correct application
of Stokes' theorem would then require subtraction of contributions from
the topological singularity. Fortunately, there exists a way to get around 
this problem, which is to {\it use a gauge transformation to move the string 
singularities somewhere else} on the manifold, far from where the fields 
are. This works locally, i.e. for a small loop $\Gamma$. In general
we are forced to patch together several coordinate charts in each
of which $s$ is well-defined. The transition functions for going from 
one chart to another are given by gauge transformations. Thus, 
we are led to consider what happens if we make a $t$-dependent gauge 
transformation $s(\gamma(t)) \mapsto s(\gamma(t)) u(t)$. Such a
transformation changes the expression for $\varphi_\Gamma$ by the 
addition of the integral $-i \int_0^1 dt \str \pi_+ u(t)^{-1} \dot u(t) 
= -i \int_0^1 dt {d\over dt} \str \pi_+ \ln u(t)$. This integral 
is {\it quantized in integer multiples of $2\pi$}, by the following
argument. In a matrix notation where $\Lambda = 1_{BF} \otimes
(\sigma_3)_{AR} \otimes 1_E$ is represented by 
$\left( \mymatrix{1_{2n} &0\cr 0 &-1_{2n}\cr} \right)$ and $u(t)$
by $\left(  \mymatrix{A(t) &0\cr 0 &D(t)\cr} \right)$, we have
$\str \pi_+ \ln u(t) = \ln \sdet A(t)$. The curve $t \mapsto \gamma(t)$
traced out on {\bf G/K} by the $\sigma$ model field along $\partial
\Gamma$ is closed, so $u(1) = u(0)$ and $A(1) = A(0)$. Integrating
the total derivative we get
	$$
	-i \int_0^1 dt {d\over dt} \str \pi_+ \ln u(t) = 
	i \ln \sdet A(0) - i \ln \sdet A(1) ,
	$$
which proves the statement by the single-valuedness of the function
${\rm U}(n|n) \to {\rm U}(1)$, $A \mapsto \sdet A$. We conclude that 
the correct formula for $\varphi_\Gamma$ is
\eqn\varpGres{
	\varphi_\Gamma =  - i \int_{0}^1 dt
	\str \pi_+ s^{-1} \dot s + 2\pi q ,}
with $q$ some integer. The gauge-ambiguous integral $\oint dt
\str \pi_+ s^{-1} \dot s$ is a one-dimensional version of what
is called a {\it Wess-Zumino term} in field theory.

In the next step we perform one further manipulation, which will 
eventually allow us to completely integrate the expression for 
$\varphi_\Gamma$. (This is not yet possible in the present 
form since $s$ and $\Lambda$ do not commute.) Given the 
curve $t \mapsto \gamma(t)$ on the base space {\bf G/K}, 
we define the lifted curve $t \mapsto g(t)$ on the total space 
{\bf G} by the equations
	$$\eqalign{
	&({\rm A}) \quad \pi(g(t)) = \gamma(t) ,			\cr
	&({\rm B}) \quad (g(t)^{-1} \dot g(t))_{\cal K} = 0 .		\cr}
	$$
Eq.~(B) has the following geometric interpretation. The time derivative
$\dot g(t)$ is tangent to {\bf G} at $g(t)$. The left translated vector
$g(t)^{-1}\dot g(t)$ lies in the tangent space of {\bf G} at the group
unit. It can therefore be regarded as an element of ${\cal G} = {\rm Lie}
({\bf G})$. Now recall the orthogonal decomposition ${\cal G} = {\cal K}
+ {\cal P}$ where ${\cal K} = {\rm Lie}({\bf K})$. $\dot g(t)$ is called
{\it vertical} if $g(t)^{-1}\dot g(t)$ lies in ${\cal K}$. ($\dot g(t)$
is then tangent to a fibre of the {\bf K}-bundle ${\bf G}\to{\bf G/K}$.)
In contrast, Eq.~(B) requires $g(t)^{-1}\dot g(t)$ to be orthogonal to 
${\cal K}$ for all $t$. This is expressed by saying that a solution of
Eq.~(B) is a curve on {\bf G} which is everywhere {\it horizontal} (Figure 2).

To solve Eqs.~(A) and (B), we use the ansatz $g(t) = s(\gamma(t)) u(t)$
with unknown $u(t) = k(g(t)) \in {\bf K}$. (A) is then satisfied 
automatically while (B) turns into
	$$
	({\rm B'}) \quad \dot u(t) u(t)^{-1} = -
	(s^{-1} \dot s)(\gamma(t))_{\cal K} \ ,
	$$
which is a first-order differential equation determining $u(t)$
uniquely once $u(0)$ has been specified. Using $({\rm B}')$ in 
Eq.~\varpGres\ we can do the integral for $\varphi_\Gamma$:
\eqn\varfin{
	\varphi_\Gamma = i \int_0^1 dt
	\str \pi_+ \dot u(t) u(t)^{-1} + 2\pi q = 
	i \str \pi_+ \ln k(g(0))^{-1} k(g(1)) + 2\pi q}
since $\pi_+$ and $u(t) \in {\bf K}$ commute. 

We conclude this subsection with a more ``global'' view of what we have 
done. Given a smooth field configuration and some value for $\sigma_{xy}$, 
the topological density assigns to a region $\Gamma \subset {\cal M}$ 
the topological phase 
$p_\Gamma = \exp (-i\sigma_{xy} \int_\Gamma d^2 x \ L_{\rm top})$. 
The field configuration maps the boundary of $\Gamma$ onto a closed
curve on {\bf G/K} which we parametrize by $t \in [0,1) \mapsto \gamma(t)$.
This curve is lifted to a curve $\partial\Gamma \to {\bf G}$ (not closed
in general) by solving $(g^{-1}\dot g)_{\cal K} = 0$. Making some choice 
of gauge $k(g) = s(\pi(g))^{-1} g$ (more precisely, patching together 
local charts) we get $p_\Gamma = \exp \left( \sigma_{xy} \str \pi_+ 
\ln k(g(0))^{-1} k(g(1)) \right)$. This way of writing $p_\Gamma$
identifies it as a Berry phase. The expression for 
$p_\Gamma$ is gauge-invariant if $\sigma_{xy} = m \in \IN \ $, 
in which case we can write 
\eqn\exppG{
	p_\Gamma = \mu \bigl( k(g(0))^{-1} k(g(1)) \bigr)}
where
\eqn\defmuk{
	\mu(k) = \exp \left( m \str \pi_+ \ln k \right)}
is {\it globally well-defined} as a representation $\mu : {\bf K} \to 
{\rm U}(1)$. By the analogy with the Dirac monopole problem that
results on setting ${\bf G/K} = {\rm SU}(2)/{\rm U}(1)$, we call
the integer $m$ the {\it monopole charge}. The calculation we have 
done in this subsection is an application of the beautiful theory \kn\ 
of connexions on the hermitean line bundles that are associated with a 
principal fibre bundle. We refrain from elaborating on this wider context 
here. \bigskip

\noindent{\medbf 3.5 Quantum Hamiltonian for $\sigma_{xy}\not=0$}\medskip

\noindent The functional integral \defZ\ carries the topological phase 
factor $p_{\cal M} = \exp(-i\sigma_{xy} \int_{\cal M} d^2x$ $L_{\rm top})$.
As was explained in the preceding subsection, we can use Stokes' theorem 
to convert $\int_{\cal M} d^2x L_{\rm top}$ into a gauge-ambiguous line 
integral along the boundary of ${\cal M}$. The gauge ambiguity is
unobservable if and only if $\sigma_{xy}$ is an integer. We will derive 
the quantum Hamiltonian of the $\sigma$ model for this special case first. 
The generalization to the more interesting case of half-integral 
$\sigma_{xy}$ will be presented afterwards.

Let $\sigma_{xy} = m \in \IN \ $. For a configuration space ${\cal M}$
without boundary, Eq.~\varpGres\ leads to $p_{\cal M} = \exp( -im \varphi_
{\cal M}) = 1$, which is just another way of stating the quantization 
of the topological term. Hence, the quantum Hamiltonian $H$ of the 
$\sigma$ model is given by Eq.~\defqHam, just as for the case of vanishing 
$\sigma_{xy}$. As was said earlier, $H$ is expected
to have a disordered ground state with an excitation gap, corresponding 
to localization of all states. A less dull situation arises when ${\cal 
M}$ does have a boundary $\partial{\cal M} \not= 0$. For the Corbino
disk geometry, $\partial{\cal M}$ consists of two pieces, an inner and 
an outer edge, both of which are circles $\sim {\rm S}^1$ with radii 
$R_1$ and $R_2$ respectively. The orientations they inherit from 
${\cal M}$ are opposite to each other. Using Eq.~\varpGres\ we get a sum 
of two Wess-Zumino terms, one for each of the two edges:
\eqn\edges{
	\int_{\cal M} d^2 x \ L_{\rm top} = - i \oint d\tau 
	\str \pi_+ \left( s_N^{-1} \partial_\tau s_N - s_1^{-1} 
	\partial_\tau s_1 \right) \ {\rm mod} \ 2\pi\IN}
where $s_1$ ($s_N$) is $s$ evaluated at $x = R_1$ ($x = R_2$). The
appearance of these terms changes the $\sigma$ model dynamics at
the edges only. More precisely, the time evolution of the edge 
degrees of freedom is governed no longer by the free-particle 
Hamiltonian $-{\cal L}$ but by another
operator which we are now going to construct. As before, we 
discretize the field theory \defZ\ on a rectangular lattice with 
lattice vectors ${\bf e_0}$, ${\bf e_1}$ and lattice constants $a_0$, 
$a_1$ in time and space direction, respectively. With all the
preparations that were made in Sect.~3.3, we can focus attention 
on the kinetic energy in combination with the topological phase. To 
be concrete, we consider the outer edge and take its orientation to 
follow the positive direction of time. Visiting a set of lattice sites 
${\bf n} \to {\bf n}+{\bf e_0} \to {\bf n}+2{\bf e_0} \to ...$ we 
encounter a sequence of field variables which we parametrize by 
$\pi(g_{\bf n}) \to \pi(g_{\bf n}e^{X_1}) \to \pi(g_{\bf n} e^{X_1} 
e^{X_2}) \to ...$ $(X_i \in {\cal P})$. In the time-continuum limit 
$a_0 \to 0$ this discrete sequence condenses into a continuous curve 
${\rm S}^1 \to {\bf G/K}$ which is closed by periodic boundary 
conditions for time. To get a discrete approximation to the line 
integral \varfin, we need to lift the discrete sequence of field 
variables to the total space {\bf G}. Since $t \mapsto g(t) = g(0) 
e^{tX}$ satisfies $(g^{-1} \dot g)_{\cal K} = 0$ for $X \in {\cal P}$, 
the lifted sequence is $g_{\bf n} \to g_{\bf n}e^{X_1} \to g_{\bf n} 
e^{X_1} e^{X_2}$ etc. By formula \exppG, the total topological phase 
factor accumulated along the outer edge is $\mu( k(g_{\bf n})^{-1} 
k(g_{\bf n}\prod_i e^{X_i}) )$ (time-ordered product). Using the 
representation property $\mu({k_0}^{-1}k_2) = \mu({k_0}^{-1}k_1) 
\mu({k_1}^{-1}k_2)$ we see that the kinetic plus topological part of 
the transfer matrix $\tau$ for a single site and a single step in negative 
time direction is given by
	$$
	(\tau f)(\pi(g)) = \int_{\cal P} dX \ e^{-{\rm B}(X,X)/\epsilon}
	\mu( k(g)^{-1} k(g e^X) ) f(\pi(g e^X)) ,
	$$
compare Eq.~\evaltau. Reversal of the time direction is equivalent to 
the substitution $\mu \to \mu^{-1}$. Taking the time-continuum limit 
$\epsilon \sim a_0 \to 0$ we obtain for the $\sigma$ model degrees of 
freedom at the outer edge the single-site Hamiltonian $- 4 {\cal L}^m 
/ a_1 \sigma_{xx}$ where ${\cal L}^m$ is the second-order differential 
operator defined by
\eqn\defmonH{
	({\cal L}^m f)(\pi(g)) = \lambda(\partial/\partial x)
	\mu(k(g) k(g\exp {\scriptstyle\sum} x^i e_i)^{-1})
	f(\pi(g\exp{\scriptstyle\sum} x^i e_i)) \Big|_{x=0} .}
By setting $\mu = \exp m \str \pi_+ \ln = 1$ we retrieve Eq.~\deflapl,
defining the Laplace operator ${\cal L}$. In view of the relation 
between $\mu$ and the gauge connection ${\cal A}$, Eq.~\defcalA, it 
should not be surprising that ${\cal L}^m$ can be shown to be identical 
to the operator $(\nabla-im{\cal A})^2$ (symbolic notation). (More 
precisely, if ${\cal A} = \sum A_j(\xi) {\rm d}\xi^j$, ${\cal L}^m$ 
is the operator that results from \nab\ by making the replacement
$\partial/\partial\xi^j \to \partial/\partial\xi^j - imA_j(\xi)$. We 
will never use this equivalence.) Eq.~\defmonH\ expresses ${\cal L}^m$ 
through derivatives acting on the {\it right}, but for some purposes 
it is better to switch to derivatives acting on the {\it left}. This 
is done in Appendix A, where we derive the alternative formula
\eqn\defmonHp{
	({\cal L}^m f)(\pi(g)) = \lambda_{\cal G}(\partial/\partial x)
	\mu(k(g)k(e^{\Sigma x^i e_i}g)^{-1}) f(\pi(e^{\Sigma x^i e_i}g))
	\Big|_{x=0} .}
Here $\lambda_{\cal G}(\partial/\partial x) = {\scriptstyle\sum}_{i,j=1}
^{{\rm dim}{\cal G}} g^{ij} \partial^2/\partial x^j \partial x^i$ is
defined by inverting the quadratic form ${\rm B}$ on the Lie algebra
${\cal G}$ with basis $\{ e_i \}$. We will see in Sect.~4.6 that the 
single-site Hamiltonian $- 4 {\cal L}^m / a_1 \sigma_{xx}$ describes 
uni-directional transport through a number $|m|$ of edge channels.
 
To get the site-diagonal part of the Hamiltonian for the degrees of 
freedom at the inner edge of the Corbino disk, we simply reverse the 
sign of the monopole charge $m$. (Recall that there is a relative minus 
sign in Eq.~\edges\ from opposite orientation of the two edges.) We will 
construct the ground states of $-{\cal L}^m$ for all $m$ in Sect.~4. 

Let us consider half-integral values of $\sigma_{xy}$ next. In this 
case, the gauge ambiguity of the Wess-Zumino integral \varpGres\ prevents 
us from wiping the interior of ${\cal M}$ clean of topological phase.
Decomposing the topological density $\sigma_{xy} L_{\rm top}$ into its 
integer and fractional parts, we can convert the former part into a
boundary term in the same fashion as before, but the latter part cannot 
be removed. We must treat it differently. This we do by modifying the 
continuum field theory \defZ\ as follows. We organize the space-time 
manifold ${\cal M}$ into a set of equidistant strips. The strips have a 
width of one lattice spacing $a_1$ and are parallel to the time direction. 
Every other strip is shaded, see Figure 3. We set the fractional part $(=
1/2)$ of the topological coupling $\sigma_{xy}$ to zero in the empty regions 
while raising it to twice its original value in the shaded ones. We are 
allowed to make such a drastic modification since, once again, all we 
are concerned with are the universal low-energy properties. These are 
determined by slowly varying fields, for which the staggered value of the 
modified topological coupling balances out. Put in formulas, we replace 
$0.5 \int_{\cal M} d^2x \ L_{\rm top}$ in \defZ\ by $\int_{{\cal M}_1} d^2x
\ L_{\rm top}$ where ${\cal M}_1$ is the union of the shaded regions. By 
using Stokes' theorem for ${\cal M}_1$ we get a {\it sum of line integrals}:
\eqn\zebra{
	{1 \over 2} \int_{\cal M} d^2 x \ L_{\rm top} \longrightarrow
	-i\sum_l (-)^l \oint d\tau \str \pi_+ s_l^{-1} \partial_\tau s_l \ .}
The sign of the line integrals alternates because the orientation
of the loops forming the boundary of ${\cal M}_1$ does.

The sceptical reader might have the impression that by modifying
the topological coupling we are making a novel and unnecessary
approximation. {\it This impression is wrong.} A close look at the
literature (see \fradkin\ for a very readable account) reveals that the 
same approximation is {\it always} made in this context and, as a 
matter of fact, there is no way of avoiding it. Our presentation is 
simply inverted compared to most published ones, which start from the 
sum of line integrals \zebra\ (derived by using coherent states on an 
antiferromagnetic quantum spin chain) and end up with the topological 
term of the $\sigma$ model. The argument usually given is the following. 
One takes the alternating sum \zebra\ and breaks translational invariance 
by organizing the line integrals into groups of two:
	$$
	i \sum_l \oint d\tau \str \pi_+
	(s_{2l+1}^{-1} \partial_\tau s_{2l+1} - 
	s_{2l}^{-1} \partial_\tau s_{2l} ) .
	$$
By Taylor expanding up to linear order in the lattice constant
$a_1$, one gets $i \sum_l a_1 \oint d\tau \str$ $\pi_+
(\partial_x s^{-1} \partial_\tau s)_{2l}$. Because periodic boundary
conditions in time direction are used, this expression is not changed
by subtraction of a term with $\partial_x \leftrightarrow \partial_\tau$,
whence it becomes
	$$
	{i\over 2} \sum_l a_1 \oint d\tau \str \Lambda 
	[ s^{-1}\partial_\tau s , s^{-1} \partial_x s ]_{2l} \ .
	$$
The expression $i \str\Lambda [s^{-1}\partial_\tau s,s^{-1} \partial_x s]$
is recognized as twice the topological density $L_{\rm top}$. Note
in particular that the sum runs over {\it even} lattice sites only
(i.e. the topological coupling is staggered). By replacing $\sum_{{\rm
even}\ l} \to \sum_{{\rm all}\ l} / 2$ and taking the naive continuum
limit, one recovers the topological term $\int_{\cal M} dx d\tau \ 
L_{\rm top} / 2$.

We are finally ready to write down the quantum Hamiltonian $H$ for 
$\sigma_{xy} \in \IN + 1/2$. Putting everything together, we get the formula
\eqn\dfquaH{
	a_1 H = - {4 \over \sigma_{xx}} \left( \sum_{{\rm even} \ l} 
	{\cal L}_l^+ + \sum_{{\rm odd} \ l} {\cal L}_l^{-} \right)
	+ \sigma_{xx} \sum_{{\rm all} \ l} \Delta(g_l,g_{l+1}) .}
where ${\cal L}^\pm := {\cal L}^{\pm 1}$ is defined by Eq.~\defmonH. 
This expression for $H$ replaces the formula \defqHam\ and is correct 
modulo terms of higher order in the small parameter $a_0 / a_1$. The 
sign of the monopole charge $m = \pm 1$ alternates on even and odd 
sites because the sign multiplying the line integrals \zebra\ does. 
Eq.~\dfquaH\ applies to case of a two-torus ${\cal M} = {\rm S}^1
\times {\rm S}^1$. For the disk geometry it must be modified at the 
edges to account for the integer part of $\sigma_{xy}$ in the manner
described above. 

So far, we have explained how to pass to the Hamiltonian formulation 
of the $\sigma$ model when $\sigma_{xy} \in \IN + 1/2$. This is not 
yet entirely satisfactory since, in order to compute critical indices 
such as the correlation length exponent $\nu$, we may wish to study 
the $\sigma$ model slightly off these points. We can do so as follows 
\afftwo. Suppose $\sigma_{xy} = (1-\epsilon)/2$ with $|\epsilon| \ll 1$. 
For definiteness, let $\epsilon > 0$. If we choose ${\cal M}_1$, the 
union of the shaded regions in Figure 3, to satisfy the condition
${\rm area}({\cal M}_1) = {\rm area}({\cal M})/2$ as before, the
alternating sum of line integrals \zebra\ will produce an {\it excess} 
topological phase. We can eliminate the excess by making ${\rm area}
({\cal M}_1)$ smaller by a factor of $1-\epsilon$. This modification 
amounts to choosing an alternating sequence of lattice constants 
$(1-\epsilon)a_1$ (width of the shaded regions) and $(1+\epsilon)a_1$ 
(width of the empty regions). It is clear that the alternating spatial 
distance between sites causes a staggering of the site-site interaction. 
Thus, to perturb the $\sigma$ model off the point $\sigma_{xy}^* = 1/2$, 
we may simply add to the quantum Hamiltonian \dfquaH\ an alternating
term $\tau \sum_l (-)^l \Delta(g_l,g_{l+1})$ where $\tau \sim
\sigma_{xy} - \sigma_{xy}^*$. \bigskip

\noindent{\bigbf 4 \ Strong-Coupling Ground States} \medskip

\noindent Low-energy properties of quantum chromodynamics in $3 +
1$ dimensions, such as confinement of quarks and gluons, are
beyond the reach of weak-coupling perturbation theory. In the 
late seventies, it was hoped that the incorporation of topological
excitations such as instantons would lead to progress. These
hopes, however, were frustrated by severe infrared divergences
which make it difficult if not impossible to formulate a
sensible expansion around the dilute instanton gas limit. On
the other hand, confinement occurs very naturally in lattice 
QCD, by the area law for Wilson loops in the strong-coupling
limit. The situation is similar for the field theory \defZ\ of the 
integer quantum Hall effect. Although weak-coupling perturbation 
theory augmented by instanton calculus \refs{\prutwo,\wz} has 
yielded some qualitative insights, it has not led to any conclusive or 
quantitative statements about low-energy (or large-scale) 
phenomena such as quantization of the Hall conductance and the 
delocalization transition between Hall plateaus. In this case, too, 
an approach starting from the opposite limit of strong coupling
looks more promising. Diagrammatic perturbation theory predicts
that the renormalization group flow takes $\sigma_{xx}$ to small 
values. Therefore, an expansion around the small-$\sigma_{xx}$
(or strong-coupling) limit is more likely to capture the essential 
long wave length physics.

The quantum Hamiltonian $H$, Eq.~\dfquaH, is a sum of two terms: $a_1 H = 
4 H_0 / \sigma_{xx} + \sigma_{xx} H_1$. For $\sigma_{xx} \to 0$, the 
term $H_0 = - \sum_l ( {\cal L}_{2l}^+ + {\cal L}_{2l+1}^-)$ dominates 
and $H_1 = \sum_l \Delta(g_l,g_{l+1})$ can be treated as a perturbation.
The first step towards making a strong-coupling expansion is to
diagonalize $H_0$. Since $H_0$ does not couple fields on neighboring
lattice sites, diagonalization of $H_0$ reduces to a one-site problem.
We shall denote the one-site Hamiltonian by ${\cal H}_s = -{\cal L}^m
= -(\nabla-{\cal A})^2$, concealing the dependence on the monopole
charge $m = \pm 1$ but making the dependence on the choice of gauge
$s$ by ${\cal A} = m \str \pi_+ s^{-1} {\rm d}s$ explicit. The purpose
of the current section is to construct the ground state of ${\cal H}_s$
for all $m \in \IN \ $. (Doing the calculation for a general value of $m$ 
will enable us to discuss edge state transport in the Hall plateau 
regions in Sect.~4.6.) We will see that this ground state is infinitely 
degenerate unless $m = 0$ in which case $\mu = \exp m \str \pi_+ \ln$ 
is trivial. States in the degenerate ground state (as in any eigenspace) 
of ${\cal H}_s$ transform according to some representation of the 
symmetry group {\bf G}. It turns out that for $m > 0$ $(m < 0)$, this 
representation possesses a lowest (highest) weight. Our strategy will 
make use of this fact: we will first identify the lowest (highest) weight 
and then generate the entire space of ground states by symmetry 
transformations. Part of what we do takes place at the edge of current 
mathematical understanding. This is why we need to begin with some 
preparation in Sect.~4.1. \bigskip

\noindent{\medbf 4.1 Principal remarks}\medskip

\noindent We begin by recalling some of the obvious structures 
at hand. The basis for our considerations is the super coset space 
${\bf G/K} = {\rm U}(n,n|2n)/{\rm U}(n|n)\times {\rm U}(n|n)$.
{\bf G} acts on {\bf G/K} by left translation $x = \pi(h) \mapsto 
g\cdot x = \pi(gh)$. This induces an action of {\bf G} on functions 
on {\bf G/K} by
\eqn\defLg{
	(L_g f)(x) = f(g^{-1}\cdot x).}
(Since the construction of $L_g f$ from $f$ involves Grassmann-analytic
continuation \berezin\ of $f$ w.r.t. the Grassmann parameters of $g$, 
$f$ must have sufficiently good analyticity properties in order for 
$L_g f$ to be well-defined.) The action \defLg\ has the representation 
property $L_{gh} = L_g L_h$.

Integration of functions on {\bf G/K} is defined in the sense of
Berezin \berezin. (Again, since the definition of the Berezin integral
involves differentiation w.r.t. the fermionic coordinates, functions
must be analytic.) {\bf G/K} carries with it a Berezin integration 
measure $dx$ which is invariant under left translations. 
$dx$ is unique within multiplication by a constant and is
alternatively denoted by $dg_K$. By the invariance of $dx$,
	$$
	\int_{\bf G/K} f(x)dx = \int_{\bf G/K} (L_g f)(x)dx 
	\quad (g\in{\bf G}) .
	$$
Given two analytic functions $f_1$ and $f_2$, we define their
scalar product $(f_1,f_2)$ by
\eqn\scalprod{
	(f_1,f_2) = \int_{\bf G/K} \overline{f_1(x)} f_2(x) dx}
whenever the right-hand side exists. The bar denotes complex
conjugation. The scalar product \scalprod\ is preserved by left
translation,
	$$
	(L_g f_1 , L_g f_2) = (f_1 , f_2) \quad (g\in{\bf G}).
	$$
We will see later that the operator ${\cal H}_s = - (\nabla - 
{\cal A})^2$ is hermitean with respect to it. In this respect, 
the scalar product \scalprod\ is similar to the scalar product 
on the space of square-integrable functions in standard quantum 
mechanics. There is, however, one important difference: the scalar 
product \scalprod\ {\it is not positive definite and does not define 
a Hilbert space}. (See Eq.~(69) below.) From a rigorous mathematical 
point of view, this raises a number of interesting questions: 
convergence of sequences of functions, completeness of the function 
space etc. These are the standard questions of functional analysis 
that physicists never need to worry about because they always deal 
with Hilbert spaces where these questions have been answered by 
mathematicians long ago. It seems to me that the proper definition 
of norm and topology for function spaces on supermanifolds is not well 
understood at present. (In any case, {\it I} don't understand it very well.) 

The loss of positivity is worrysome not just on a rigorous level
but may sometimes cause severe problems for the practical user.
As an example we mention the difficulties that one encounters when 
trying to establish a workable theory of Fourier analysis on compact 
supergroups \huffmann. (By Fourier analysis we mean expansion in 
eigenfunctions of the Laplace operators.) In this case the lack of 
a positive definite invariant hermitean form causes Laplace-Casimir 
operators to be generically {\it non-diagonalizable}! Fortunately, 
the situation for Riemannian supersymmetric spaces {\bf G/K} -- as 
opposed to supergroups {\bf G} -- appears to be much better, in spite 
of the lack of positivity. The first proof of a Fourier expansion 
theorem, for the simple case of a rank-one supersymmetric space, was 
published in \mrzcmp. The method of \mrzcmp\ was subsequently extended 
\mmz\ to all of Efetov's spaces \efetov, i.e. the supersymmetric spaces 
{\bf G/K} that appear in the theory of noninteracting disordered 
electron systems. It was found that the problem of non-diagonalizability
of Laplace operators is absent for these spaces. It turns out 
that the method invented in \mrzcmp\ can be further generalized to
prove a spectral expansion theorem for the operator ${\cal H}_s$.
(Note that the hermitecity of ${\cal H}_s$ implies orthogonality
of its eigenspaces w.r.t. the scalar product \scalprod.) In the present 
paper we are concerned with the strong-coupling limit of the Hamiltonian 
$H$, Eq.~\dfquaH. For that, all we need are the ground states of ${\cal 
H}_s$ and its conjugate. We will therefore concentrate on these ground 
states and relegate the full spectral expansion of ${\cal H}_s$ to a 
future complete account of my work on supersymmetric Fourier analysis. 

With superanalytic theory being incompletely developed at present, even 
the modest goal of constructing the ground state of ${\cal H}_s$ is not 
entirely straightforward. Suppose we have found some eigenfunction of 
${\cal H}_s$. How do we know it occurs in the spectral expansion? (Note 
that invariant differential operators on noncompact spaces are bounded 
neither from above nor from below. They have a large number of 
eigenfunctions that do {\it not} occur in their spectral expansion. 
Examples of such eigenfunctions are the functions that correspond 
to the scaling fields of Sect.~2.1.) On a Hilbert space the answer is 
easy: we simply look to see whether the function is (properly or 
improperly) normalizable. If it is not, we reject it. For reasons 
that are not entirely understood, the same principle works in the 
present case, in spite of the lacking Hilbert space property.
Although the proof of the Fourier expansion theorem for 
Efetov's spaces makes no reference to the hermitean form \scalprod, 
it turns out that the {\it eigenfunctions occurring in the Fourier
expansion are precisely those with ``proper'' behavior with respect 
to \scalprod.} (By this I mean that eigenfunctions belonging to a discrete 
series have finite norm in a certain sense, while those belonging to a 
continuous series are normalized to a $\delta$-function.) This then 
is the heuristic principle we adopt: given some eigenfunction of 
${\cal H}_s$ we shall declare it to be a ``physical state'' (occurring 
in the spectral expansion) if it is normalizable w.r.t. \scalprod.
\vfill\eject

\noindent{\medbf 4.2 Induced Representation}\medskip

\noindent We again begin with some motivation. Eq.~\defLg\ defines an
action of the symmetry group {\bf G} on functions on  {\bf G/K}. This 
action is not yet the ``good'' one for our purposes, since it does not 
commute with the Hamiltonian ${\cal H}_s = - (\nabla - {\cal A})^2$.  
If this is not obvious, consider the simple example of the motion of a 
charged particle in the field of a magnetic monopole. Such a system is 
invariant under rotations that fix the magnetic monopole. However, 
rotational invariance is not manifest in the Hamiltonian formulation 
since writing $(\nabla - {\cal A})^2$ requires making some choice of 
${\cal A}$, which inevitably breaks spherical symmetry. Consequently, 
the symmetry transformations of ${\cal H}_s$ are not just rotations but 
are {\it rotations followed by a suitable gauge transformation}. The 
same happens in the present case. Thus, the good action of {\bf G} will 
combine left translations with gauge transformations. 

To construct the good action we proceed as follows. For a fixed value
of the monopole charge $m$, let us denote the linear space of functions
which ${\cal H}_s$ acts on by $W_s$. The subscript $s$ does not mean
inequivalence of $W_s$ and $W_{s'}$ but is simply a reminder that the 
elements of $W_s$ are tied to some choice of gauge ${\cal A} = m
\str \pi_+ s^{-1}{\rm d}s$. When $s$ is changed, the functions 
change. Thus, $W_s$ is a space of {\it gauge-dependent} functions on 
{\bf G/K}. Since a change of section $s \mapsto s k$ takes ${\cal A}$ into
${\cal A} + {\rm d}( m \str \pi_+ \ln k)$, we expect the
behavior under gauge transformations to be governed by the one-dimensional 
representation $\mu$ of the gauge group {\bf K} which is defined by 
$\mu(k) = \exp\left( m \str \pi_+ \ln k \right)$. 
To confirm this expectation we use an idea that
has been described to physicists in Refs. \refs{\egh,\stone}. 
It is to get rid of the gauge dependence by ``working with all 
gauges at once'' and regarding an element $f \in W_s$ as a function 
$F$ on the total space ${\bf G}$. In detail this works as follows. 
Recall that a choice of section $s$ gives rise to a map $k : {\bf G} \to
{\bf K}$ by the equation $g = s(\pi(g)) k(g)$. Given $k(g)$, we 
associate with $f \in W_s$ a function $F$ on ${\bf G}$ by
\eqn\defsecF{
	F(g) = \mu(k(g))^{-1} f(\pi(g)) .}
Since $k(gk_1) = k(g) k_1$ for $k_1 \in {\bf K}$ and since $\mu$ is a 
representation, $F$ satisfies
\eqn\symcon{
	F(gk_1) = \mu(k_1)^{-1} F(g) .}
In this way, $W_s$ corresponds to the linear space, $W$, of functions $F$ 
on {\bf G} subject to the symmetry condition \symcon. The correspondence 
is one-to-one, with the inverse of relation \defsecF\ being
\eqn\defsef{
	f(x) = \mu \big( k (s(x)) \big) F( s(x) ) \qquad (x \in {\bf G/K}).}
In other words, Eqs.~\defsef\ and \defsecF\ define a bijection 
$\psi_s : W \to W_s$. (An element $f \in W_s$ is technically called 
a section of an associated bundle \kn.) 

Let us illustrate the above correpondence between $f$ and $F$ 
at the simple example of SU(2)/U(1). With the choices
	$$
	g = e^{i\phi\sigma_3/2} e^{i\theta\sigma_2/2} e^{i\psi\sigma_3/2},
	\quad s = e^{i\phi\sigma_3/2} e^{i\theta\sigma_2/2}, \quad 
	\mu(e^{i \alpha \sigma_3 /2}) = e^{iK\alpha},
	$$
the eigenfunctions of the Dirac monopole Hamiltonian are
${\cal D}_{K M}^J (s^{-1}) = \tilde{\cal D}_{K M}^J (0,-\theta,-\phi)$ $= f$
where $\tilde {\cal D}$ are Wigner's D-functions expressed in Euler 
angles $\phi, \theta, \psi$. They extend to functions 
${\cal D}_{K M}^J (g^{-1}) = \tilde{\cal D}_{K M}^J (-\psi,-\theta,-\phi) 
= F$, which are functions on SU(2) obeying the symmetry condition
	$$
	{\cal D}_{K M}^J \left( (g e^{i\alpha\sigma_3 /2})^{-1}
	\right) = e^{-iK\alpha} {\cal D}_{K M}^J (g^{-1}) .
	$$ 

How does the monopole Hamiltonian operate on functions
$F \in W$? Making the similarity transformation ${\cal H} := 
\psi_s^{-1} {\cal H}_s \psi_s$ we get from Eq.~\defmonH\ for
${\cal H}_s = - {\cal L}^m$:
\eqn\calH{
	({\cal H} F)(g) = - \lambda(\partial/\partial x)
	F(g \exp {\scriptstyle\sum} x^i e_i)\Big|_{x=0} .}
Note that ${\cal H}$ preserves the symmetry condition \symcon.
In fact, assuming $F$ to satisfy \symcon\ and using the invariance of 
$\lambda(\partial/\partial x)$ under $\sum x^i e_i \mapsto 
k(\sum x^i e_i)k^{-1}$, we get
	$$\eqalign{
	({\cal H} F)(gk) &= 
	- \lambda(\partial/\partial x) F(gk \exp
	{\scriptstyle\sum} x^i e_i)\Big|_{x=0}	\cr
	&= - \mu(k)^{-1} \lambda(\partial/\partial x) 
	F \left( g \exp(k{\scriptstyle\sum} 
	x^i e_i k^{-1}) \right)\Big|_{x=0} = 
	\mu(k)^{-1} ({\cal H}F)(g) . 		\cr}
	$$
Therefore, ${\cal H}$ is well-defined as an operator $W \to W$. 

Consider now the natural representation of the symmetry group 
{\bf G} on $W$ by left translation:
\eqn\LgFh{
	(L_g F)(h) = F(g^{-1} h) .}
(Right translations are unnatural because they interfere with the 
constraint \symcon.) The representation \LgFh\ commutes with the
operator ${\cal H}$, Eq.~\calH. This is obvious since left translations 
act on the left while ${\cal H}$ acts on the right. At this point we 
recognize the advantage of making the similarity transformation from 
${\cal H}_s$ to ${\cal H}$: the symmetry transformations of ${\cal H}$
are simply the left translations! The good action, $T_g$, of the 
symmetry group {\bf G} on gauge-dependent functions
$f \in W_s$ now follows by transforming from $W$ back to
$W_s$. We have $T_g = \psi_s L_g {\psi_s}^{-1}$ or, in explicit terms,
\eqn\symtrans{
	(T_g f)(\pi(h)) = \mu(k(h) k(g^{-1}h)^{-1})
	f(\pi(g^{-1}h)) .}
These are the symmetry transformations of ${\cal H}_s$. They form a 
representation of {\bf G}, by virtue of their construction from \LgFh\ 
via similarity transformation. The action \symtrans\ is called an {\it 
induced representation}. It depends on $\mu$ and on the choice of 
gauge $s$. This fact is not made explicit in our notation.

We claim that the action \symtrans\ preserves the hermitean form \scalprod:
	$$
	(T_g f_1 , T_g f_2) = (f_1 , f_2) \quad (g\in{\bf G}).
	$$
(If $(\cdot,\cdot)$ were positive definite this would amount to saying that 
the representation is unitary.) In view of Eq.~\symtrans\ and the invariance 
of $dx$ under left translations $x = \pi(h) \mapsto g^{-1}\cdot x = 
\pi(g^{-1}h)$, our statement is obvious if we can show that
\eqn\barmu{
	\overline{\mu(k(g))} = \mu(k(g))^{-1}}
holds for $g\in{\bf G}$. The relation \barmu\ is established as 
follows. From Eq.~\defmuk\ and $\str A = \str A^{\rm T}$ (superscript 
T denoting supertranspose) and $\bar A^{\rm T} = A^\dagger$, we get 
$\overline{\mu(k(g))} = \mu(\overline{k(g)}) = \mu(k(g)^\dagger)$. 
Now, by Eq.~\defG\ $k(g)\in{\bf K}
\subset{\bf G}$ satisfies $k(g)^\dagger = \eta k(g)^{-1} \eta$ and, 
since $\mu$ is a representation and $\mu(\eta)^2 = 1$ (note $\eta\in
{\bf K}$), $\mu(k(g)^\dagger) = \mu(k(g))^{-1}$, so \barmu\ is proved. 

It follows from \barmu\ that the monopole Hamiltonian ${\cal H}_s$ is 
hermitean w.r.t. the scalar product \scalprod. The proof is simply a matter 
of writing down
	$$
	(f_2,{\cal H}_s f_1) = \int_{\bf G/K} dg_K \ 
	\overline{f_2(\pi(g))} ({\cal H}_s f_1)(\pi(g)) ,
	$$
using the formula \defmonHp\ for ${\cal H}_s = - {\cal L}^m$, making the 
change of integration variable $g \mapsto e^{-\Sigma x^i e_i} g$ (which
leaves $dg_K$ invariant), and then using Eq.~\barmu. \bigskip

\noindent{\medbf 4.3 Choice of gauge, and coordinate presentation}\medskip

\noindent To proceed further, we pick suitable coordinates for {\bf G/K}, 
and we fix the gauge by choosing some section $s : {\bf G/K} \to {\bf G}$.
Our choice of coordinates is motivated by the requirement that the
projection $\pi : {\bf G} \to {\bf G/K}$ be expressed in a particularly
simple form.

An element $g \in {\bf G}$ is written
\eqn\gABCD{
	g = \pmatrix{	A	&B	\cr
			C	&D	\cr}.}
Here, and in all that follows, the displayed matrix structure refers 
to advanced-retarded space. Each of the blocks $A$, $B$, $C$, $D$ is 
a $2n\times 2n$ supermatrix acting in the tensor product of 
Boson-Fermion and Extra space. They are subject to a certain set
of constraints that follow from the defining equation for {\bf G},
Eq.~\defG. In particular, these constraints imply that both $A$ 
and $D$ have an inverse in 
some neighborhood of the unit element of {\bf G}. Since $\Lambda 
= {\rm diag}(1_{BF\otimes E},-1_{BF\otimes E})$ in the presentation 
adopted, an element $k \in {\bf K}$ is of the block-diagonal form 
$k = \left( \mymatrix{ k_+ &0\cr 0 &k_-\cr} \right)$. Invariants 
for the action $g \mapsto gk$ $(k \in {\bf K})$ are the combinations
$CA^{-1}$ and $BD^{-1}$. Counting shows that the number of independent
variables in $CA^{-1}$ and $BD^{-1}$ equals the number of degrees
of freedom on {\bf G/K}. We put $Z := CA^{-1}$, $\tilde Z := BD^{-1}$
and take $Z$, $\tilde Z$ as local coordinates for the coset space
{\bf G/K}. Thus, $f(\pi(g))$ will from now on be written $f(Z,\tilde Z)$. 
(For clean mathematics we should work with several coordinate charts. 
We will however get away with using just one.) Recalling the defining 
equation for {\bf G} one can show with a little calculation that $Z$ 
and $\tilde Z$ are related by $\tilde Z = Z^\dagger \tau_3$ where 
$\tau_3 = (\sigma_3)_{BF} \otimes 1_E$. 

The meaning of the coordinates $Z$, $\tilde Z$ is best understood by 
looking at some examples. Let us specialize to $n = 1$, in which
case ${\bf G/K}$ has a two-sphere ${\rm S}^2 \simeq 
{\rm SU(2)}/{\rm U(1)} \simeq {\rm U(2)}/{\rm U(1)}\times{\rm U(1)}$
for its FF-space. By projecting the pair of supermatrices $(Z,\tilde Z)$ 
on the FF-sector, we get a pair of complex numbers $(-z,\bar z)$. 
(The minus sign is dictated by the unitarity condition $\tilde Z
= Z^\dagger \tau_3$.) The special element $\Lambda$ is replaced by 
the Pauli matrix $\sigma_3$. Parametrizing ${\rm SU(2)}$ by Euler angles 
\eqn\Eulang{
	g = \exp(i\phi\sigma_3/2) \exp(i\theta\sigma_2/2)
	\exp(i\psi\sigma_3/2) = \pmatrix{\alpha &\bar\beta\cr
	-\beta &\bar\alpha} ,}
we get $z = \beta/\alpha = \tan(\theta/2)\exp(-i\phi)$. It follows that, 
if ${\bf n} = (n_1,n_2,n_3) = (\sin\theta\cos\phi,\sin\theta$ $\sin\phi,
\cos\theta)$ is a unit vector parametrizing the two-sphere ${\rm S}^2$,
$z$ has the expression $z = (n_1-in_2)/(1+n_3)$. In other words, $z$ is 
the usual complex coordinate associated with the stereographic mapping 
of ${\rm S}^2 \simeq {\rm SU(2)/U(1)}$ onto the complex plane $\CN \ $. 
Consider next the noncompact analog space ${\rm U(1,1)}/{\rm U(1)}\times
{\rm U(1)} \simeq {\rm SU(1,1)/U(1)}$, which is the BB-space of {\bf G/K} 
for $n = 1$ and is isomorphic to the two-hyperboloid (or hyperbolic plane) 
${\rm H}^2$. For this we replace $(Z,\tilde Z) \to (z,\bar z)$ (without 
the minus sign). Changing to Euler angles for ${\rm SU(1,1)}$ by making 
in the exponent of the middle factor in Eq.~\Eulang\ the substitution 
$i\sigma_2 \to \sigma_1$, we get $z = \tanh(\theta/2) \exp(-i\phi)$. 
${\rm H}^2$ can be parametrized by a hyperbolic unit vector ${\bf n} = 
(n_1,n_2,n_3) = (\sinh\theta\cos\phi, \sinh\theta\sin\phi, \cosh\theta)$. 
As before, $z = (n_1-in_2)/(1+n_3)$. What this means is that $z$ is the 
complex coordinate of the Poincar\'e disk model $(|z|^2 < 1)$ of ${\rm 
H}^2$ \bv. Finally, we mention that $Z$ and $\tilde Z$ coincide with 
the matrices $w_{21}$ and $-w_{12}$ used for the parametrization of $Q$ 
in Ref.~[12] for $n = 1$. To verify this, one expresses $Q = g\Lambda 
g^{-1}$ in terms of $Z, \tilde Z$ and compares the result with Eq.~(3.2) 
of that reference.

Let us now fix $s : {\bf G/K} \to {\bf G}_0 = {\rm SU}(n,n|2n)$. For 
various reasons that will become clear as we go along, the optimal choice 
is $s(\pi(g)) = (g \Lambda g^{-1} \Lambda)^{1/2}$ with coordinate expression
\eqn\sectis{
	s(Z,\tilde Z) = \pmatrix{ 	1	&\tilde Z	\cr
					Z	&1		\cr}
	\pmatrix{	(1-{\tilde Z}Z)^{-1/2}	&0		\cr
			0		&(1-Z\tilde Z)^{-1/2}	\cr}.}
It is easy to verify that (i) $s(Z,Z^\dagger \tau_3) \in {\bf G}_0$ and
(ii) $\pi \circ s = {\rm id}$. (These properties are sufficient in order 
for $s$ to be a section of the bundle ${\bf G} \to {\bf G/K}$.) Specializing 
to the FF-sector for $n = 1$ and using the Euler angle parametrization 
\Eulang, we see that the choice $s(\pi(g)) = (g\sigma_3 g^{-1} 
\sigma_3)^{1/2}$ corresponds to the second of the two sections in \examsec.

We now wish to re-express the central formulas \defmonH\ and 
\symtrans\ in terms of the complex coordinates $Z$ and $\tilde Z$, 
starting with the latter. For that we need to figure out 
how left translations $\pi(h) \mapsto \pi(gh)$ act on $Z$ and 
$\tilde Z$. We put $h = \left( \mymatrix{A' &B'\cr C' &D'\cr} \right)$
and characterize the coset $\pi(h)$ by $Z = C'{A'}^{-1}$ and $\tilde Z 
= B'{D'}^{-1}$. For the product $gh$ with $g$ given by Eq.~\gABCD\ we 
write $gh = \left( \mymatrix{A'' &B''\cr C'' &D''\cr} \right)$.
The complex coordinates of the left translated coset $\pi(gh)$
are denoted by $g \cdot Z = C''{A''}^{-1}$ and $g\cdot\tilde Z = 
B''{D''}^{-1}$. Carrying out the matrix multiplication we find
\eqn\gactsZ{
	\eqalign{
	g \cdot Z &= (DZ + C)(A + BZ)^{-1} ,				\cr
	g \cdot \tilde Z &= (A \tilde Z + B)(D + C\tilde Z)^{-1}.	\cr}}
We mention in passing that these relations generalize the rational action 
of ${\rm SL}(2,\CN)$ on $\CN$ by $z \mapsto {az+b\over cz+d}$. By 
solving Eqs.~\gactsZ\ for $Z$, $\tilde Z$ we get the inverse transformations
\eqn\ginZ{
	\eqalign{
	g^{-1} \cdot Z &= (D - ZB)^{-1} (ZA - C), 	\cr
	g^{-1} \cdot\tilde Z 
	&= (A - \tilde Z C)^{-1} (\tilde Z D - B).	\cr}}
The other ingredient on the right-hand side of Eq.~\symtrans\ is the
multiplier built from $\mu$ and $k(\cdot)$. Unfortunately, the derivation
of its explicit expression in terms of $Z$, $\tilde Z$ requires a certain 
amount of algebra. To avoid cluttering the main text with a lot of 
formulas, I have relegated these details to Appendix B. Using the
information given there, we arrive at the following coordinate expression 
for the representation \symtrans:
\eqn\symtraco{
	(T_g f)(Z,\tilde Z) = (\sdet g)^{m/2} 
	\sdet \left( A-\tilde ZC \over D-ZB \right)^{m/2}
	f ( g^{-1}\cdot Z , g^{-1}\cdot \tilde Z ) .}
(Eq.~\gABCD\ is understood.) Next, we write down the coordinate expression 
of the second-order differential operator ${\cal H}_s = - {\cal L}^m$ defined
in \defmonH. Again, the calculations are somewhat involved which is why they 
are done in Appendix B. The result however is quite transparent and can be 
stated as follows. Let $X$, $\tilde X$ be a pair of supermatrices with the 
same dimensions and symmetries as $Z$, $\tilde Z$. If we define a
differential operator $\partial_{X,\tilde X}$ analogous to 
$\lambda(\partial/\partial x)$ in \defmonH\ by
\eqn\invdiff{
	\partial_{X,\tilde X} = \sum_{ij} (-)^{|i|} {\partial^2 
	\over \partial X_{ij} \partial\tilde X_{ji}} \ ,}
the operator ${\cal H}_s$ has the coordinate expression
\eqn\calHcos{
	({\cal H}_s f)(Z,\tilde Z) = - \partial_{X,\tilde X} 
	\sdet \left( {1+Z\tilde X \over 1+\tilde Z X} \right)^{m/2} 
	f \bigl((Z+X)(1+\tilde Z X)^{-1},...\bigr) \Big|_{X=\tilde X=0} .}
The second argument of $f$ on the right-hand side is obtained from the 
first one by $Z \leftrightarrow \tilde Z$ and $X \leftrightarrow \tilde X$. 

Since the formula \calHcos\ for the monopole Hamiltonian ${\cal H}_s$ 
may look unfamiliar we turn, once again, to the prototypical example 
${\rm SU(2)/U(1)}$ for illustration. In this case it is well-known \sr\
that, for $m = 1$, ${\cal H}_s$ has a spin 1/2 ground state spanned by 
the functions ${\cal D}_{-1/2,-1/2}^{1/2}(s^{-1}) = \cos(\theta/2)$ and
${\cal D}_{-1/2,1/2}^{1/2}(s^{-1}) = \sin(\theta/2)\exp(-i\phi)$. To 
verify this from the present formalism, we use
\eqn\calHSt{
	({\cal H}_s f)(z,\bar z) = - {\partial^2 \over \partial\zeta
	\partial\bar\zeta} \left( {1-\bar z\zeta \over 1-z\bar\zeta}
	\right)^{m/2} f\left({z+\zeta \over 1 - \bar z \zeta} \ , \
	{\rm c.c.}\right)\Big|_{\zeta=\bar\zeta=0} ,}
as results from \calHcos\ by specialization to the FF-sector. (The 
multiplier got inverted because $\sdet$ puts the determinant 
of the FF-block in the {\it denominator}.) The expressions for the 
functions $\cos(\theta/2)$ and $\sin(\theta/2) \exp(-i\phi)$ in 
terms of the complex coordinates $z$, $\bar z$ are $(1+\bar z z)^{
-1/2}$ and $z(1+\bar z z)^{-1/2}$. From Eq.~\calHSt\ these are easily
seen to be eigenfunctions, for $m = 1$, of ${\cal H}_s$ with eigenvalue
1/2. More generally, for any positive monopole charge $m$ the functions 
$z^k (1+\bar z z)^{-m/2}$ $(0 \le k \le m)$ are seen to be eigenfunctions 
with eigenvalue $m/2$. They form an irreducible spin $m/2$ multiplet of 
${\rm SU(2)}$. Indeed, by specializing Eq.~\symtraco\ and making the 
similarity transformation $f(z,\bar z) = (1+\bar z z)^{-m/2} \varphi(z)$ 
we get the transformation law
\eqn\tprime{
	(T'_{ \left( \mymatrix{\alpha &\bar\beta\cr -\beta &\bar\alpha\cr}
	\right) } \varphi)(z) = (\bar\alpha + \bar\beta z)^m \varphi
	\left( {\alpha z-\beta \over \bar\alpha+\bar\beta z} \right) .}
{}From this we read off the representation of the generators $\sigma_+ 
= (\sigma_1+i\sigma_2)/2$, $\sigma_- = (\sigma_1-i\sigma_2)/2$, and 
$\sigma_3$ by the differential operators $-z^2 \partial_z + mz$, 
$\partial_z$, and $2z\partial_z - m$, respectively. It can be shown 
that the $(m+1)$-dimensional space of functions $f(z,\bar z) = (1+\bar 
z z)^{-m/2} \varphi(z)$ with holomorphic $\varphi(z) = \sum_{k=0}^m 
c_k z^k$ is the ground state of ${\cal H}_s$. (Note that in the zero 
curvature limit $m\to\infty$ we recover the lowest Landau level wave 
functions $f(z,\bar z) = \exp(-\bar z z) \varphi(z)$ of the Hamiltonian 
$-(\nabla-{\cal A})^2$ on the complex plane in the symmetric gauge ${\cal 
A} = (\bar z {\rm d}z - z {\rm d}\bar z)/2$.) The representation of ${\rm 
SU(2)}$ on holomorphic functions has been used \hr\ in numerical studies 
of the fractional quantum Hall effect. ($z = v/u$ in the notation of 
Ref. \hr.) Eq.~\tprime\ applies to the case of positive $m$. When $m$ is 
negative, we get a ground state spanned by antiholomorphic functions.

All of the above can be transcribed without difficulty to the noncompact 
analog space ${\rm SU(1,1)/U(1)}$. To cut things short, we just mention 
that the ground states $z^k (1+\bar z z)^{-m/2}$ $(0 \le k \le m)$ 
translate into the ground states $z^k (1-\bar z z)^{m/2}$ 
$(0 \le k < \infty)$. The latter transform according to a unitary
infinite-dimensional lowest-weight representation of the noncompact
group ${\rm SU(1,1)}$. \bigskip

\noindent{\medbf 4.4 Holomorphic sections}\medskip

\noindent At the end of the preceding subsection we recalled the fact 
that the degenerate ground state of the monopole Hamiltonian ${\cal H}_s$ 
for ${\rm S}^2$ and monopole charge $m > 0$ is spanned by the 
functions $(1+\bar z z)^{-m/2} \varphi(z)$ with holomorphic $\varphi(z) = 
\sum_{k=0}^m c_k z^k$. This observation holds the clue how to make 
progress with ${\cal H}_s$ in the general case, Eq.~\calHcos, and
nonnegative $m$. The functions $(1+\bar z z)^{-m/2}$ for ${\rm S}^2$ 
and $(1-\bar z z)^{m/2}$ for ${\rm H}^2$, generalize to $\sdet
(1-\tilde Z Z)^{m/2}$ for our quantum Hall model space {\bf G/K}. 
Let us therefore set $v_m(Z,\tilde Z) = \sdet (1-\tilde Z Z)^{m/2}$ 
and make a similarity transformation from ${\cal H}_s$ to ${\cal H}_+ = 
v_m^{-1} {\cal H}_s v_m$. The transformation is carried out with the 
help of the identity
	$$
	1 - (Z+X)(1+\tilde ZX)^{-1}(\tilde Z+\tilde X)(1+Z\tilde X)^{-1}
	= (1-Z\tilde Z)(1+X\tilde Z)^{-1}(1-X\tilde X)(1+Z\tilde X)^{-1}.
	$$
Substituting $f = v_m w$ into Eq.~\calHcos\ we obtain
\eqn\calHplus{
	({\cal H}_+ w)(Z,\tilde Z) = - \partial_{X,\tilde X}
	\sdet \left( { (1-X\tilde X)^{1/2} \over 1+\tilde ZX} 
	\right)^m w \bigl( (Z+X)(1+\tilde ZX)^{-1} , ... \bigr)	
	\Big|_{X=\tilde X=0} .}
What happens for holomorphic $w(Z,\tilde Z) = \varphi(Z)$? The crucial 
feature of Eq.~\calHplus\ is the {\it absence of any dependence on 
$\tilde X$} in the denominator of the multiplier. With this observation, 
and recalling the definition of $\partial_{X,\tilde X}$ in \invdiff, we 
immediately find $\varphi(Z)$ to be an eigenfunction of ${\cal H}_+$ 
with eigenvalue
\eqn\zeroze{
	- \partial_{X,\tilde X} \sdet (1-X\tilde X)^{m/2} 
	\Big|_{X=\tilde X=0} .}
By the property of perfect grading, i.e. the equal number of bosonic and 
fermionic degrees of freedom, the eigenvalue \zeroze\ {\it vanishes}. We 
thus arrive at the important conclusion that functions $f$ of the form
	$$
	f(Z,\tilde Z) = \sdet (1-Z\tilde Z)^{m/2} \varphi(Z)
	$$
with holomorphic $\varphi$ are {\it zero modes} of the Hamiltonian ${\cal 
H}_s$ for monopole charge $m > 0$. For the general reason of stability of 
the functional integral \defZ, we expect only eigenfunctions with nonnegative 
eigenvalues to contribute to the spectral expansion of ${\cal H}_s$. 
(The operator ${\cal H}_s$ does have a vast number of eigenfunctions 
with negative eigenvalues. These however are not normalizable and do 
not occur in its spectral expansion.) We thus perceive the {\it central
role played by holomorphic functions}: they span the
degenerate ground state of ${\cal H}_s$ not only for ${\rm SU(2)/U(1)}$ 
but for our quantum Hall model space {\bf G/K} (and other spaces 
{\bf G/K}) as well. The story does not end here, since the requirement 
of decent behavior under transition to another coordinate chart imposes 
a constraint on $\varphi(Z)$. (Recall that for ${\rm SU(2)/U(1)}$ the 
degree of the polynomial $\varphi(z)$ must not exceed $m$.) 
We can avoid getting into the boring technicalities of switching
charts by first identifying some special zero mode 
and then generating the entire space of zero modes by the action 
\symtraco\ of the symmetry group {\bf G}. This is what we will do in
Sect.~4.5. To prepare the second step, we now specialize the 
representation \symtraco\ to the case of holomorphic sections.

As is shown in Appendix B, by making the similarity transformation
$T_g^+ = v_m^{-1} T_g v_m$ and keeping only the dependence on the
holomorphic coordinate $Z$, we obtain from \symtraco:
\eqn\Tgplus{
	( T_g^+ \varphi)(Z) = (\sdet g)^m \sdet(D-ZB)^{-m} 
	\varphi(g^{-1}\cdot Z) .}
(Again, Eq.~\gABCD\ is understood.) Eq.~\Tgplus\ gives rise to a 
representation of ${\cal G} = {\rm Lie}({\bf G})$ by the usual
procedure of linearization around the group unit, and this 
representation extends to the complexified Lie algebra 
${\cal G}_{\Bbb C} = {\rm gl}(2n,2n)$ by linearity. The motivation 
for going to the complexification is that we desire the convenience 
of working with the usual raising and lowering operators. (These do 
not exist inside ${\cal G}$.) Making the extension and then returning 
to group level, we get a representation of the complexified group 
${\bf G}_{\Bbb C} = {\rm GL}(2n,2n)$. It is given by the formula 
\Tgplus\ without any change. (In fact, the condition $g^{-1} = \eta 
g^\dagger \eta$ defining the pseudo-unitary group {\bf G} is never 
used in the derivation of \Tgplus. All we need is the existence of the 
inverse $g^{-1}$.) Having made the extension to ${\bf G}_{\Bbb C}$ we 
can now write down the action of raising operators $\left( \mymatrix{1 
&B\cr 0 &1\cr} \right)$ and lowering operators $\left( \mymatrix{1 &0\cr 
C &1\cr} \right)$ at the group level. These are seen from \Tgplus\ and
\ginZ\ to be represented by
\eqna\railow
	$$\eqalignno{
	(T_{ \left( \mymatrix{1 &B\cr 0 &1\cr}\right) }^+ \varphi) (Z)
	&= \sdet (1 - ZB)^{-m} \varphi ((1-ZB)^{-1}Z),&\railow{\rm a} \cr
	(T_{ \left( \mymatrix{1 &0\cr C &1\cr} \right)}^+ \varphi) (Z)
	&= \varphi(Z-C).			      &\railow{\rm b} \cr}
	$$
Another special case of importance is the action of elements of 
${\bf K}_{\Bbb C}$:
\eqn\Tkp{
	( T_{ \left( \mymatrix{A &0\cr 0 &D\cr} \right) }^+ \varphi) (Z)
	= \sdet A^m \varphi( D^{-1} Z A ).}
All of the above applies to the case of positive monopole charge $m > 0$ 
but is easily transcribed to $m < 0$. In the latter case the zero modes of 
${\cal H}_s$ are functions $f = v_{-m} \varphi$ with {\it antiholomorphic} 
$\varphi(\tilde Z)$. Carrying out the similarity transformation 
$T_g^- = v_{-m}^{-1} T_g v_{-m}$, we obtain
\eqn\Tgminus{
	(T_g^- \varphi) (\tilde Z) = \sdet (A-\tilde ZC)^m 
	\varphi(g^{-1}\cdot \tilde Z) .}\bigskip

\noindent{\medbf 4.5 Zero modes and coherent states}\medskip

\noindent Our results of Sect.~4.4 reveal the ground state of 
${\cal H}_s$ for $m\not=0$ to be vastly degenerate: any function
$f$ of the form $f = v_m\varphi$ ($f=v_{-m}\varphi$) with holomorphic 
(antiholomorphic) $\varphi$ is a zero mode of ${\cal H}_s$ for 
$m>0$ $(m<0)$. However, not all of these are physical states and 
contribute to the spectral expansion of ${\cal H}_s$. It is the 
purpose of the current subsection to describe which of them do. 

We put $m \ge 1$ for definiteness and consider the special vector 
$v_m(Z,\tilde Z) = \sdet (1-Z\tilde Z)^{m/2}$ obtained by
setting $\varphi = 1$. Our first task will be to show that $v_m$
has finite length. To do this with a minimum amount of effort, 
we put $v_m$ in coordinate-free form. Using $g = s(\pi(g)) k(g)$ and
the expression \sectis\ for $s$, we get
\eqn\vmcfree{
	v_m(\pi(g)) = \sdet \left( 2 + g \Lambda g^{-1} 
	\Lambda + \Lambda g \Lambda g^{-1} \right)^{-m/2} .}
Note that the coordinate-free formula \vmcfree\ guarantees good
behavior of $v_m$ under transition to another coordinate chart.
Aside from decaying sufficiently fast at infinity -- so that $(v_m,v_m)$ 
in fact exists -- $v_m$ has the special property $v_m(k\cdot x) = v_m(x)$
for $k \in {\bf K}$. Functions with this property are called
{\bf K}-{\it radial}. For $f$ any such function (with the additional
property of being integrable), the so-called Parisi-Sourlas-Efetov-Wegner 
supersymmetric integral formula states that
\eqn\PSEW{
	\int_{\bf G/K} f(x) dx = f(o)}
where $o = \pi(1)$ is the unique element of {\bf G/K} which is left 
invariant by {\bf K}. (Incidentally, Eq.~\PSEW\ fixes a natural 
normalization of $dx$.) A pedestrian proof of \PSEW\ for $n = 1$ can 
be found in \mrznpb. The general proof has been worked out in \bundschuh. 
Application of theorem \PSEW\ to $f = \bar v_m v_m$ yields $(v_m,v_m) 
= 1$ since $v(o) = 1$ from Eq.~\vmcfree. This says that $v_m$ is a physical 
state.

Given $v_m$, we can apply a symmetry transformation $T_g$ to form the 
vector $v_g = T_g v_m$. Since $T_g$ preserves the scalar product \scalprod\ 
for $g \in {\bf G}$ we have $(v_g,v_g) = (v_m,v_m) = 1$ and, since $T_g$
commutes with ${\cal H}_s$, $v_g$ is an eigenstate of ${\cal H}_s$ 
with zero energy. Under the assumption of irreducibility of 
the representations entering the spectral expansion of ${\cal H}_s$, 
we can generate the entire zero-energy sector in this way. 

For the explicit construction of all zero-energy states it is 
convenient to go to the complexification ${\bf G}_{\Bbb C}$ and 
work with raising and lowering operators, which are norm-changing. 
A unique feature of $v_m$ is that it satisfies the equations
\eqn\lowway{
	T_{ \left( \mymatrix{1 &0\cr C &1\cr} \right) } v_m = v_m 
	\qquad {\rm and} \qquad T_k v_m = \mu(k) v_m ,}
which are easily verified using \railow{\rm b}, \Tkp\ and \defmuk. 
Thus, $v_m$ is 
a {\it lowest-weight vector carrying the one-dimensional representation
$\mu$ of} {\bf K}. To see how the raising operators act on $v_m$ we
set $\varphi=1$ in Eq.~\railow{\rm a}\ and find
\eqn\raisop{
	(T_{ \left( \mymatrix{1 &B\cr 0 &1\cr} \right) } v_m)(Z,\tilde Z)
	= \sdet (1 - BZ)^{-m} v_m(Z,\tilde Z) .}
The space of states that can be reached by repeated application of
a raising operator to $v_m$ will be denoted by $V_m$. It is called a
lowest-weight module of {\bf G}.

A vector of the form \raisop, obtained by applying a raising operator
$\exp \left( \mymatrix{0 &X\cr 0 &0\cr} \right)$ parametrized by $X$
to a lowest-weight vector, is commonly called a coherent state. To 
enhance the readability of our notation, we temporarily promote the 
subscript of $T$ to argument and write
	$$
	v_X := T \left( \mymatrix{1 &X\cr 0 &1} \right) v_m \ .
	$$
We will now calculate the coherent state overlap function 
${\cal N}(X,Y) = (v_X , v_Y)$. Knowing it we will be able to
deduce the norms and overlaps of vectors by Taylor expansion. 
Since $\left( \mymatrix{0 &X\cr \tau_3 X^\dagger &0\cr} \right)$ 
is an element of Lie({\bf G}) and $T_g$ for $g \in {\bf G}$
preserves the scalar product,
	$$
	{\cal N}(X,Y) = (v_m ,
	T \left( \mymatrix{1 &0\cr -\tau_3 X^\dagger &1\cr} \right)
	T \left( \mymatrix{1 &Y\cr 0 &1\cr} \right) v_m ) .
	$$
Using the representation property and carrying out the matrix
multiplication, we get
	$$
	{\cal N}(X,Y) = (v_m ,
	T \left( \mymatrix{1 &Y\cr -\tau_3 X^\dagger 	
	&1-\tau_3 X^\dagger Y\cr} \right) v_m) .
	$$
To calculate this, we use the following trick. There exist matrices 
$B$, $C$, $k_+$ and $k_-$ such that
\eqn\tricky{
	\pmatrix{1 &Y\cr -\tau_3 X^\dagger 
	&1-\tau_3 X^\dagger Y \cr} = 
	\pmatrix{1 &B\cr 0 &1\cr}
	\pmatrix{k_+ &0\cr 0 &k_-\cr}
	\pmatrix{1 &0\cr C &1\cr} .}
{}From \lowway\ we have 
	$$
	(v_m, T \left( \mymatrix{1 &B\cr 0 &1\cr} \right) 
	T \left( \mymatrix{k_+ &0\cr 0 &k_-\cr} \right) 
	T \left( \mymatrix{1 &0\cr C &1\cr} \right) v_m) = 
	\mu( \left( \mymatrix{k_+ &0\cr 0 &k_- \cr} \right) ) (v_m,v_m) .
	$$ 
Hence, ${\cal N} = \exp\left( m \str\ln k_+ \right) = (\sdet k_+)^m$, 
and solving Eq.~\tricky\ for $k_+$ we obtain
\eqn\overlap{
	{\cal N}(X,Y) = \sdet (1-\tau_3 X^\dagger Y)^{-m} .} 
{}From this result in combination with Eqs.~\railow{}\ and \Tkp, all 
desired information about the state vector content and norms and overlaps 
of vectors in $V_m$ can be deduced by Taylor expansion.

Let us now take a close-up look at $V_m$ for the minimal model $n = 1$. 
The base manifold of the super coset space 
{\bf G/K} for $n = 1$ is the direct product of a two-sphere ${\rm S}^2 
\simeq {\rm SU(2)}/{\rm U(1)}$ and a two-hyperboloid ${\rm H}^2 \simeq 
{\rm SU(1,1)}/{\rm U(1)}$. The Cartan algebra of ${\rm Lie}({\bf G})$ 
is four-dimensional in this case and can be taken to be generated by
	$$
	H_i = (E_{ii})_{BF} \otimes (\sigma_3)_{AR} \quad (i=1,2), 
	\quad H_3 = (\sigma_3)_{BF} \otimes 1_{AR} , \quad
	H_4 = 1_{BF} \otimes 1_{AR} .
	$$
The eigenvalues (also called weights) of the generators $H_1$ and $H_2$
will be used to label vectors in $V_m$. $H_1$ ($H_2$) is the generator 
of the ${\rm U(1)}$ subgroup of rotations fixing the north pole on
${\rm H}^2$ (${\rm S}^2$). Introducing $B_{ij} = (E_{ij})_{BF} \otimes 
(E_{12})_{AR}$ $(i,j=1,2)$, we define differential operators
${\cal B}_{ij} : V_m \to V_m$ by
\eqn\calBij{
	{\cal B}_{ij} f = {\rm d}T(B_{ij}) f = {d \over dt} 
	T_{\exp t B_{ij}} f \Big|_{t=0}}
where $t$ is a commuting or anticommuting element of the parameter 
Grassmann algebra according to whether $B_{ij}$ is even or odd, 
respectively. Their coordinate expressions are easily found from 
Eq.~\railow{\rm a}. Acting repeatedly with elements of the set $\{{\cal 
B}_{ij}\}$ on $v_m$, we get $v_m$ times a polynomial in the complex 
coordinates $Z$. By applying ${\cal B}_{ij}$ once, we raise 
the degree of the polynomial by one unit. (We say that $v_m$ is 
the {\it vacuum} and one action of an operator ${\cal B}_{ij}$
creates one {\it quantum of excitation}.) It is quite straightforward
to construct all polynomials that are generated in this way. The 
degree in the anticommuting variables $Z_{ij}$ $(i\not= j)$ cannot 
exceed one, and the degree in the variable $Z_{22}$ (coordinatizing 
the compact FF-space) turns out to be $\le m$. Omitting the
details of this straightforward calculation, we now give a summary
of the results found. For every pair of integers $p,q$ in the range
$p\ge m$, $-m \le q \le m$ $(a_{pq})$, $p\ge m+1$, $-m+1 \le q \le m-1$ 
($\beta_{pq}$ and $\gamma_{pq}$), and $p\ge m+2$, $-m+2 \le q \le m-2$ 
($d_{pq})$, we define the state vectors
	$$\eqalign{
	a_{pq} &= {\cal B}_{11}^{(p-m)/2} {\cal B}_{22}^{(q+m)/2} v_m, 	\cr
	\beta_{pq} &= {\cal B}_{11}^{(p-m-1)/2} 
	{\cal B}_{22}^{(q+m-1)/2} {\cal B}_{12} v_m, 			\cr
	\gamma_{pq} &= {\cal B}_{11}^{(p-m-1)/2} 
	{\cal B}_{22}^{(q+m-1)/2} {\cal B}_{21} v_m,			\cr
	d_{pq} &= {\cal B}_{11}^{(p-m-2)/2} {\cal B}_{22}^{(q+m-2)/2} 
	({\cal B}_{12} {\cal B}_{21} - m^{-1} 
	{\cal B}_{11} {\cal B}_{22})v_m	.				\cr}
	$$
These vectors form an orthogonal basis of $V_m$. (The allowed 
values of $p$ and $q$ differ by two units in each case since 
exponents must be integer-valued.) All of them are eigenvectors 
of ${\rm d}T(H_1)$ and ${\rm d}T(H_2)$ with eigenvalue $p$ and $q$, 
respectively. Furthermore, ${\rm d}T(H_3) a_{pq} = {\rm d}T(H_3) 
d_{pq} = 0$ and ${\rm d}T(H_3) \beta_{pq} = +2 \beta_{pq}$, ${\rm 
d}T(H_3) \gamma_{pq} = -2 \gamma_{pq}$. All vectors are annihilated 
by ${\rm d}T(H_4)$. The vectors $a_{pq}$, $\beta_{pq}$, $\gamma_{pq}$, 
$d_{pq}$ for fixed $p$ and variable $q$ form an ${\rm SU}(2)$-multiplet 
with spin $S = m/2$, $(m-1)/2$, $(m-1)/2$, $(m-2)/2$, respectively. 
When $q$ is held fixed and $p$ is varied, they form a unitary 
lowest-weight representation of ${\rm SU(1,1)}$, again  with 
spin $S = m/2$, $(m-1)/2$, $(m-1)/2$, $(m-2)/2$, respectively.
The norms $|f|^2 = (f,f)$ are
\eqn\normform{\eqalign{
	|a_{pq}|^2 &= N \left( {p-m\over 2} \ , \ {q+m\over 2} \right) ,\cr
	|\beta_{pq}|^2 &= - |\gamma_{pq}|^2 = m \
	N \left( {p-m-1\over 2} \ , \ {q+m-1\over 2} \right) ,		\cr
	|d_{pq}|^2 &= (m^2-1) \ N \left( {p-m-2\over 2} \ , \ 
	{q+m-2\over 2} \right)	,					\cr}}
where $N$ is defined by
	$$
	N(p,q) = p!^2 \pmatrix{m+p-1\cr p\cr} 
	q!^2 \pmatrix{m\cr q\cr} .
	$$
The $d_{pq}$ are seen to be null vectors for $m = 1$. The weight
diagram for $m = 2$ is shown in Figure 4.

We thus have a complete and detailed description of the 
lowest-weight module $V_m$ for $m > 0$ and $n = 1$. The 
transcription to $m < 0$ can be made very simply by complex 
conjugation, taking the lowest-weight vector $v_m$ into the 
highest-weight vector $v_{-m}$ and holomorphic functions into 
antiholomorphic ones. There remain two important questions that 
are left open by our analysis. i) Does the module $V_m$ really 
exhaust the space of zero-energy states? (It is conceivable in 
principle that $V_m$ might be an invariant subspace of a bigger 
indecomposable representation. If this were the case, there would 
exist other zero-energy states that cannot be reached by applying 
raising operators to the lowest-weight vector $v_m$.) ii) Are the 
excited states of ${\cal H}_s$ separated from the ground state by 
a gap? Unfortunately, giving a convincing answer to these questions 
is beyond the scope of this paper. Without proof I claim that the 
answer to both questions is {\it yes}. As was explained in 
Sect.~4.1, I hope to publish the proof in a future complete
account of my work on supersymmetric Fourier analysis. \bigskip

\noindent{\medbf 4.6 Edge dynamics and Hall conductance}\medskip

\noindent Our derivation of the quantum Hamiltonian of the nonlinear 
$\sigma$ model goes through even when the configuration space ${\cal 
M}$ of the $2d$ electron gas has a boundary. The only 
condition on ${\cal M}$ is that one of its two directions be periodic. 
We will now use this flexibility for a semi-realistic calculation of 
the quantized Hall conductance in the plateau regions. The experimental 
setup we imagine is shown in Figure 5. We take for ${\cal M}$ a disk 
$[R_1,R_2] \times {\rm S}^1$ and attach various voltage and current 
probes to its outer edge. 

Let us begin by summarizing what phenomenological theory \butt\ tells 
us about this kind of system. When the magnetic field is located in 
a Hall plateau region, all stationary electron states in the interior 
of the disk are localized. Transport proceeds via edge states and is 
effectively one-dimensional. The number $M$ of edge channels equals 
the number of Landau levels below the Fermi energy. Edge channels are 
{\it directed}, i.e. electrons can propagate in one direction only. 
The direction of propagation is determined by the sign of the magnetic 
field ($E \times B$ drift). For ideal probes, the conductance
coefficients $g_{pq}$ are given by
\eqn\concoe{\eqalign{
	g_{21} &= g_{32} = ... = g_{1N} = M ,		\cr
	g_{12} &= g_{23} = ... = g_{N1} = 0 ,		\cr
	g_{11} &= g_{22} = ... = g_{NN} = -M ,		\cr}}
if the ordering of the contacts follows the orientation induced by the
magnetic field. All other conductance coefficients vanish. We will now 
show how the relations \concoe\ follow from the effective field theory \defZ. 

The disorder averages of the $g_{pq}$ are given by certain correlation 
functions of the field theory whose form is specified below. To take the 
Hamiltonian limit of the field theory, we assign to the radial and angular 
directions of ${\cal M}$ the roles of space and time (or, rather, imaginary 
time), respectively. By proceeding as in Sects.~3.3-5, we get a quantum 
Hamiltonian $H$ governing the evolution in time (i.e. around the disk) of a 
discrete set of degrees of freedom. Each of these takes values in {\bf G/K} 
and is located on one site of a lattice in radial direction. The degrees of 
freedom will be referred to as {\bf G/K}-superparticles (SPs). We are going 
to consider values of the magnetic field at the center of a Hall plateau 
region, where the Hall conductivity $\sigma_{xy}$ is an integer $m\in\IN\ $. 
The dynamics of the SPs in the interior of the disk is then governed by the 
Hamiltonian specified in Eq.~\defqHam. In contrast, the two SPs at the inner 
and the outer edge sense the additional field of a fictitious ``magnetic 
monopole'' -- the remnant of the topological density $L_{\rm top}$ -- with 
charge $\pm m$. The sign of the monopole charge is determined by the sign 
of $\sigma_{xy}$ and the orientation of the integration contour for the 
corresponding Wess-Zumino term $\oint dt \str \pi_+ s^{-1} \dot s$, see 
Figure 5 and Sect.~3.4. For definiteness, we take $m < 0$ and $m > 0$ for 
the inner and outer edge, respectively. 

Because of the influence of the probes, the SP at the outer 
edge gets perturbed in some way every time it passes a contact. 
(Qualitatively, passage past the contact collapses the wave function 
and puts the SP close to the origin of {\bf G/K}; see Eq.~(71) below.) 
When $\sigma_{xx}$ is large (high Landau level), the edge SP passes 
the perturbation on to its neighbors by the action of the coupling 
term of the Hamiltonian \defqHam, thereby producing a complicated 
many-superparticle problem which we are unable to solve. Nevertheless, 
we can deal with such a situation by the following argument. We expect
that there exists a {\it mass gap}, that is to say, a finite distance 
over which correlations decay. (This expectation comes from weak-coupling 
perturbation theory for the field theory \defZ, as was mentioned earlier. 
Ultimately, we would like to establish the existence of a mass gap by 
direct calculation for the quantum Hamiltonian $H$. This however remains 
as a project for the future. Here we will simply {\it assume} the 
existence of a gap and see what we can deduce from it.) When the length 
over which correlations decay is much larger than the lattice spacing, 
we are facing the many-SP problem we are unable to solve. However, we can 
use the renormalization group to improve the situation. Let us imagine 
changing the short-distance cutoff of the field theory \defZ. (We may 
think of this cutoff scale as being the lattice constant $a_1$ for $H$.) 
As we increase the cutoff, the correlation length (being a length) becomes 
shorter when measured in units of the cutoff. Continuing the process, we 
eventually get a correlation length of order unity or less. In this limit, 
the SPs propagate {\it independently} of each other, and the perturbation 
caused by the probes is not transferred to the interior of the disk but 
affects only the SP at the outer edge. With the assistance of the 
renormalization group, we thus get a one-SP problem. This problem 
we can solve. (Of course, our use of the renormalization group 
is naive and overly simplistic. No argument from the renormalization 
group should be trusted without further work when the correlation length 
is of the order of the lattice spacing or less. Nevertheless, the
independent edge superparticle approximation is not without justification, 
for this approximation corresponds to transport occurring only via edge 
states localized near the sample boundary, see below.)

In the independent edge-SP approximation, the average conductance 
coefficients reduce to the dynamic one-SP correlation functions ${\rm Tr} 
\prod_i O_i \exp \left( -(t_i-t_{i-1}) {\cal H}_s \right)$ with each operator 
insertion $O_i$ representing one contact. ${\cal H}_s = - {\cal L}^m$ 
is the one-SP Hamiltonian, Eq.~\defmonH. The imaginary time intervals are 
given by $t_i - t_{i-1} = 4 L_{i,i-1} / a_1 \sigma_{xx}$ where $\sigma_{xx}$ 
is the renormalized longitudinal conductivity, and $L_{i,i-1} / a_1$ is 
the distance between the $(i-1)^{\rm th}$ and the $i^{\rm th}$ contact, 
measured in units of the lattice constant $a_1$.

We now need the expressions for the $\sigma$ model operators $O_i$ 
when the conductance coefficient is $g_{pq}$. It will be sufficient to 
consider the case $p \not= q$ since the diagonal coefficients $g_{qq}$ 
are determined through the off-diagonal ones by current conservation 
($\sum_p g_{pq} = 0$). To avoid getting into lengthy calculations that 
would detract from our main purpose, we will give a heuristic derivation. 
Let us first consider a diffusive quantum dot in a weak magnetic field 
(unitary universality class). In this case the formulas for the operators 
$O_i$ can be extracted from Appendix B of Ref.~\altland. Expressing them in 
terms of the complex coordinates $Z, \tilde Z$ we find
	$$\eqalign{
	O_p &=M_p\bar Z_{11}\sdet(1-\tilde Z Z)^{M_p} \quad({\rm current}),\cr
	O_q &=M_q Z_{11}\sdet(1-\tilde Z Z)^{M_q} \quad ({\rm voltage}),\cr
	O_r &= \sdet(1-\tilde Z Z)^{M_r} \quad ({\rm none}) ,\cr}
	$$
where $M_i$ is the number of scattering channels in lead $i$. (We are 
assuming the ideal limit of unit transmission into every open channel.) From 
these equations we can guess the analogous expressions for quantum 
Hall systems. The main novelty that occurs for a strong magnetic field as 
compared to a weak field is {\it spatial separation} of incoming and outgoing
channels in the leads. Figure 6 depicts the situation when the incoming 
and outgoing channels are located on the left and right edges of the 
lead, respectively. For an ideal contact there are {\it no direct 
transitions} between the incoming and outgoing channels of the same
lead. As a result, every contact operator must separate into two factors,
one each for the left $(L)$ and the right $(R)$ side of the contact. For
example, the expression for a voltage contact factors as
	$$
	O_q = M_q (Z_L)_{11} \sdet(1-\tilde Z_L Z_L)^{M_q/2}
	\times \sdet(1-\tilde Z_R Z_R)^{M_q/2} .
	$$
The number of edge channels in the leads equals the number of occupied
Landau levels $m = \sigma_{xy}$. Making the identification $M_p = M_q = 
... = m$ and changing to coordinate-free notation we obtain
\eqn\opers{\eqalign{
	O_p &= v_m ({\cal B}_{11}v_m,\cdot )  \quad ({\rm current}),\cr
	O_q &= {\cal B}_{11} v_m (v_m,\cdot ) \quad ({\rm voltage}),\cr
	O_r &= v_m (v_m,\cdot ) \quad ({\rm none}) ,		\cr}}
where $v_m(Z,\tilde Z) = \sdet(1-\tilde Z Z)^{m/2}$, $({\cal B}_{11} v_m)
(Z,\tilde Z) = m Z_{11} \sdet(1-\tilde Z Z)^{m/2}$, and $(\cdot,\cdot)$ 
is the inner product \scalprod.

After these preparations, we are ready to proceed to the essential point 
of this subsection. We are going to calculate the correlation function 
${\rm Tr} \prod_i O_i \exp \left( -(t_i-t_{i-1}) {\cal H}_s \right)$,
using the expressions \opers\ for the operators $O_i$. The following two
observations enable us to do the calculation with ease. First, since
$m > 0$, the ground state of ${\cal H}_s$ is spanned by functions $f = 
v_m\varphi$ with holomorphic (as opposed to antiholomorphic) $\varphi$. 
The space of such functions is denoted by $V_m$ as before, and we set
$v = v_m$ for brevity. The state ${\cal B}_{11} v$ lies in $V_m$. Second, 
when inserting a complete set of intermediate states in the expression for 
the correlation function, we may restrict the sum over states to $V_m$, 
since any contribution from the excited states of ${\cal H}_s$ will be 
exponentially small of order $\exp ( - {\rm const} \times {\scriptstyle 
\Delta} L / \sigma_{xx})$, where ${\scriptstyle\Delta}L$ is the distance 
between neighboring contacts. This allows us to replace $\exp(-(t_i-t_{i-1}) 
{\cal H}_s)$ by the projection operator onto $V_m$, denoted by $P$.

We must now distinguish cases. Let us first consider $g_{pq}$ for
$p = q+1$. As the time evolution outside the interval from $t_q$ to 
$t_{q+1}$ does no more than project onto $v$, we obtain
	$$
	\langle g_{q+1,q} \rangle = ({\cal B}_{11} v,
	e^{-(t_{q+1}-t_q){\cal H}_s} {\cal B}_{11} v) .
	$$
Making an exponentially small error by replacing $\exp\left( -(t_{q+1}-
t_q){\cal H}_s \right)$ by the projector $P$, and using $P {\cal B}_{11}
v = {\cal B}_{11} v$, we get $\langle g_{q+1,q} \rangle = ({\cal B}_{11} 
v, {\cal B}_{11} v)$. Since $({\cal B}_{11}v, {\cal B}_{11}v) = 
|a_{m+2,-m}|^2 = m$ by the first of Eqs.~\normform, we arrive at 
$\langle g_{q+1,q}\rangle = m$, which is precisely the result 
expected from the phenomenological picture of transmission through 
edge states. What happens when we interchange $p \leftrightarrow q$? 
As before, the time evolution outside the interval $[t_{q-1},t_{q+2}]$ 
projects on $v$, but now $\langle g_{q,q+1}\rangle = (v,P{\cal B}_{11}v) 
(v,Pv) ({\cal B}_{11}v,Pv)$ vanishes by the orthogonality relation 
$(v,{\cal B}_{11}v) = 0$. Similarly, all $\langle g_{pq} \rangle$ with 
$p$ and $q$ separated by more than one unit vanish by orthogonality. 

In the same way, we can calculate the conductance fluctuations. To get 
the second moment, what we do \mmz\ is to replace $Z_{11}$ ($\bar Z_{11}$) 
in the expression for a voltage (current) contact by $Z_{11} Z_{22}$ 
($\bar Z_{11} \bar Z_{22}$). This amounts to substituting ${\cal B}_{11}
{\cal B}_{22}v$ for ${\cal B}_{11}v$ in \opers. The factorization property 
$({\cal B}_{11} {\cal B}_{22} v , {\cal B}_{11} {\cal B}_{22} v) = 
({\cal B}_{11} v , {\cal B}_{11} v) ({\cal B}_{22} v , {\cal B}_{22} v )$ 
leads to ${\rm var}(g_{pq}) = 0$, so conductance coefficients do not 
fluctuate. All of the above results hold modulo exponentially small 
corrections and for ideal contacts and leads.

As it stands, our argument applies only to integral values of $\sigma_
{xy}$. It extends to other values by Khmel'nitskii's flow diagram \khmel\ 
or, more precisely, by assuming the nonlinear $\sigma$ model at 
$(\sigma_{xx},\sigma_{xy}) = (0,m\in\IN)$ to be an attractive fixed 
point of the renormalization group -- with the basin of attraction 
being the strip $m - 1/2 < \sigma_{xy} < m + 1/2$. (In the 
introduction to this paper we reviewed the argument why the 
critical nonlinear $\sigma$ model at $\sigma_{xy} = 1/2$ (mod 1) 
cannot be renormalizable but must flow under renormalization to
another, conformal invariant field theory. There is no doubt,
however, that the noncritical system will be attracted to the 
$\sigma$ model with $\sigma_{xx} \to 0$ and $\sigma_{xy}$ an integer.)

In summary, for $\sigma_{xy}$ close to $m \in \NN$
we have argued that the conductance coefficients
	$$
	g_{21} = g_{32} = ... = m
	$$
are integers, while all others (except for the diagonal ones, which 
are not easily accessible by direct calculation but are determined 
by current conservation) vanish. By the reasoning of Ref.~\butt, this 
explains the experimental observation of the integer quantum Hall 
effect for systems with ideal probes. \bigskip

\noindent{\bigbf 5 \ Quantum Spin Chain}\medskip

\noindent Apart from demonstrating how quantization of the Hall 
conductance arises in the $\sigma$ model framework, the calculation 
of Sect.~4.6 serves another useful purpose by throwing much light 
on the physical meaning of the holomorphic zero-energy module $V_m$ 
($m>0$) and its antiholomorphic dual $V_{-m}$. The lesson we have learnt is 
this. The voltage contact $q$ injects probability flux into the $m$ edge 
states emanating from the lead $q$. In our field-theoretic formalism 
the process of injection corresponds to preparing the edge superparticle 
(SP) in the quantum state ${\cal B}_{11} v_m$. This state, being a {\it 
zero-energy state} of the one-SP Hamiltonian $-{\cal L}^m$, evolves in 
imaginary time without attenuation. A rather different behavior obtains 
when we replace ${\cal L}^m$ by its dual ${\cal L}^{-m}$, which 
corresponds to reversing the sign of the magnetic field. In this 
case ${\cal B}_{11} v_m$ {\it lies outside} the zero-energy space of 
$-{\cal L}^{-m}$, and therefore $\exp(t{\cal L}^{-m}) {\cal B}_{11} 
v_m$ decays exponentially as $t$ increases, with the decay time being 
determined by the smallest excitation energy of $-{\cal L}^{-m}$. (Of 
course, here as always ``time'' is one of the two directions of the 
2$d$ quantum Hall system.) We thus see that evolution of edge SP states 
over long times (i.e. distances) is {\it directed} by the magnetic field. 
This directedness is the field-theoretic equivalent of $E \times B$ 
drift along a boundary.

Having understood this, we now turn to the critical value $\sigma_{xy}
= 1/2 \ ({\rm mod} \ 1)$ corresponding to a point of transition
between Hall plateaus. The quantum Hamiltonian, $H$, of the nonlinear
$\sigma$ model for such a point has been given in Eq.~\dfquaH. $H$ 
differs from the Hamiltonian for a plateau region by the alternating 
sequence of monopole Laplacians ${\cal L}^\pm$ replacing the plain 
Laplacian ${\cal L}$. This difference is crucial: while $-{\cal L}$ 
has a nondegenerate ground state with a gap, the operators ${\cal H} 
:= -{\cal L}^+$ and ${\cal H}^* := -{\cal L}^-$ possess infinitely 
many zero-energy states, see Sect.~4.5. These are given by holomorphic 
and antiholomorphic functions for ${\cal H}$ and ${\cal H}^*$, 
respectively. In the absence of the coupling $H_1 = \sum_l \Delta(g_l,
g_{l+1})$, a superparticle prepared in such a zero-energy state evolves
in time without attenuation. Note that the unattenuated propagation of 
superparticles in holomorphic/antiholomorphic states in the bulk is 
strongly reminiscent of the semiclassical picture of electron guiding 
center drift along percolating equipotential lines. Our interest is 
in the critical long-range correlations. These will not be affected 
by the high-energy modes of ${\cal H}$ and ${\cal H}^*$ -- they do 
no more than give rise to complicated and irrelevant short-distance 
physics -- but are determined by the zero-energy modes coupled together 
by the interaction Hamiltonian $H_1$. Degenerate perturbation theory
tells us how to construct the critical low-energy theory: we are to 
project the full space of quantum states of the $\sigma$ model on the 
alternating product of zero-energy modules $V_{+1} \otimes V_{-1} 
\otimes V_{+1} \otimes V_{-1} \otimes ...$ . Calculating the matrix 
elements of $H_1$ between the states of this multi-``spin'' space, we
will get the Hamiltonian of a {\it quantum spin chain}. 

To provide further motivation and background to Sect.~5.1, let us
anticipate how the spin chain Hamiltonian is related to established 
theory. Starting from a quantum Hall system in the semiclassical 
high-field limit, where electrons drift slowly along the equipotential 
lines of a smooth random potential while performing rapid cyclotron 
motion, Chalker and Coddington \cc\ formulated a network model that 
mimics the relevant effects of quantum tunnelling near saddle points. 
The links of the network are directed, see Figure 7a. Disorder is 
modelled by random $2 \times 2$ scattering matrices $S$ connecting the 
incoming and outgoing channels at each node of the network. To bring 
the Chalker-Coddington model closer to the model that will emerge in 
Sect.~5.1, we take its Hamiltonian (or anisotropic) limit. This is done 
by setting $S = \exp(-i\epsilon h)$, with $h$ some random hermitean 
$2 \times 2$ matrix, and sending $\epsilon \to 0$. The resulting 
Hamiltonian dynamics is illustrated graphically in Figure 7b. Electrons 
run up or down on straight lines of alternating direction, and with 
a small probability amplitude ($\sim \epsilon$) they tunnel from one 
line to a neighboring one, thereby reversing their direction of motion 
and acquiring a random change of phase. One can carry out the disorder 
average for this Hamiltonian version of the Chalker-Coddington model by 
using the supersymmetry method. Doing so, one arrives rather directly 
at the spin chain Hamiltonian of Sect.~5.1 \nread, with electrons 
running up and down straight lines translating into degrees of freedom 
whose quantum states are specified by holomorphic and antiholomorphic 
functions, respectively. 

The nonlinear $\sigma$ model is derived from the quantum limit of a 
random potential with correlation length $l_c$ short compared to the
magnetic length $l_B$. We are discovering that its low-energy limit 
converges to a model associated with the {\it opposite} limit $l_c 
\gg l_B$. This coincidence is a beautiful vindication of the 
hypothesis of low-energy universality at a critical point.\bigskip

\noindent{\medbf 5.1 Mapping on a Spin Chain}\medskip

\noindent We start with some basic definitions underlying the notion of 
what we are going to call a generalized ``spin'' chain. (We really ought
to say ``superspin'' instead of just ``spin'', but we prefer to use the 
shorter and less pretentious word when there is no risk of confusion.) As 
always, let $E_{ij}$ $(i,j = 1, ...,p+q)$ be a set of canonical generators 
of ${\rm End} (\CN^{p+q})$ obeying the multiplication law $E_{ij} E_{kl} = 
\delta_{jk} E_{il}$. ${\rm End}(\CN^{p+q})$ is ${\rm Z}_2$-graded
by $|E_{ij}| = |i| + |j| \ {\rm mod} \ 2$, where $|i| = 0$ for 
$1 \le i \le p$ and $|i| = 1$ for $p+1 \le i \le p+q$. The introduction
of a bracket operation (``supercommutator'')
\eqn\comrel{
	[ E_{ij} , E_{kl} ] = \delta_{jk} E_{il} -  
	(-)^{(|i|+|j|)(|k|+|l|)} \delta_{li} E_{kj}}
turns ${\rm End}(\CN^{p+q})$ into the Lie superalgebra ${\rm gl}(p,q;\CN)$, 
with the quadratic Laplace-Casimir element being $\sum_{ij} (-)^{|j|} 
E_{ij} E_{ji}$. ${\rm gl}(2n,2n;\CN)$ can be identified with the complex Lie
superalgebra of the symmetry group {\bf G} of the quantum Hall model space 
{\bf G/K} by the canonical isomorphism
\eqn\caniso{
	{\rm gl}(2n,2n;\CN) \simeq {\rm gl}(1,1;\CN)
	\otimes {\rm gl}(2) \otimes {\rm gl}(n) .}
(Recall that the three factors in the tensor product relate to superspace, 
advanced-retarded space and extra space, in this order.) 

With $V := V_{+1}$ and the isomorphism \caniso\ being understood, we 
introduce ${\cal E}_{ij} : V \to V$ by ${\cal E}_{ij} = {\rm d} T(E_{ij})$, 
where ${\rm d}T$ 
is the differential taken at the unit element of the representation 
$g \mapsto T_g$ defined by Eq.~\symtraco. The ${\cal E}_{ij}$ are linear 
differential operators on the space of functions $f\in V$. We shall refer 
to them as ``spin operators''. They obey the supercommutation relations
\eqn\commrel{
	[ {\cal E}_{ij} , {\cal E}_{kl} ] = (-)^{(|i|+|j|)(|k|+|l|)} 
	\delta_{jk} {\cal E}_{il} -  \delta_{li} {\cal E}_{kj} .}
It is seen that these do not properly represent the superalgebra \comrel,
because of the appearance of overall extra minus signs. (Their origin
is explained in Appendix C.) Nevertheless, the bracket \commrel\ does define
a Lie superalgebra and, as a matter of fact, it is a perfectly sensible
object to work with. Operators ${\cal E}^*_{ij}$ on the dual space $V^* := 
V_{-1}$ are defined in identical fashion. They, too, obey the superalgebra 
\commrel. By Eq.~\gABCD, the set $\{ {\cal E}_{ij} \}_{i,j=1,...,4n}$ 
naturally decomposes into four sets of operators ${\cal A}_{kl}$, 
${\cal B}_{kl}$, ${\cal C}_{kl}$, and ${\cal D}_{kl}$ $(k,l=1,...,2n)$. 
The same decomposition is made for the spin operators ${\cal E}^*_{ij}$.
For future reference we note the action of the spin operators on
the special vectors $v = v_1 \in V$ and $v^* = v_1 \in V^*$:
\eqn\spinactv{\eqalign{
	{\cal C}_{ij} v &= {\cal B}^*_{ij} v^* = 0 ,	\cr
	{\cal A}_{ij} v &= {\cal D}_{ij} v = {\cal A}^*_{ij} v^*
	= {\cal D}^*_{ij} v^* = 0 \quad (i \not= j) ,	\cr
	{\cal A}_{ii} v &= (-)^{|i|} v / 2,
	\quad {\cal D}_{ii} v = 0 ,			\cr
	{\cal A}^*_{ii} v^* &= (-)^{|i|+1} v^* / 2 ,
	\quad {\cal D}^*_{ii} v^* = 0 .			\cr}}
These express the fact that $v$ and $v^*$ are lowest- and highest-weight
vectors carrying the one-dimensional {\bf K}-representations $\mu = \exp
\str \pi_+ \ln$ and $\mu^{-1} = \exp - \str \pi_+ \ln$, respectively.

We turn to the formulation of the two-spin Hamiltonian. For two sites, the 
$\sigma$ model Hamiltonian $H$, Eq.~\dfquaH, has $V \otimes V^*$ for its 
degenerate space of strong-coupling ground states. The degeneracy is split 
by the two-site interaction $H_1=\sigma_{xx}\Delta(g,h)/a_1$ orthogonally 
projected onto $V \otimes V^*$. Orthogonality is defined w.r.t. to the 
invariant scalar product \scalprod, extended to tensor product space by
\eqn\extscap{
	(a \otimes b^* , c \otimes d^*) = 
	(-)^{|b^*| |c|} (a,c) (b^*,d^*) .}
We wish to express the projected Hamiltonian $H_{\rm eff} := P H_1 P$ 
in terms of spin operators. Since $\Delta(g,h)$ is {\bf G}-invariant and 
invariance is preserved by orthogonal projection, $H_{\rm eff}$ must be 
some Laplace-Casimir element of {\bf G} formed from the spin operators 
on $V$ and $V^*$. The question is now, which Laplace-Casimir? Introducing 
$Q(g) = g \Lambda g^{-1}$ we can write $\Delta(g,h) = \str Q(g) Q(h) / 4$. 
A symmetry transformation $g \mapsto x g$ $(x\in{\bf G})$ takes $Q(g)$ 
into $x Q(g) x^{-1}$. The unique set of operators $V \to V$ ($V^* \to V^*$) 
with this transformation behavior, which is identical to that of the adjoint 
representation, are the spin operators ${\cal E}_{ij}$ $({\cal E}^*_{ij})$. 
It follows that $H_{\rm eff}$ must be proportional to the {\it quadratic}
Laplace-Casimir element:
	$$
	H_{\rm eff} = J \sum_{ij} (-)^{|i|+1}
	{\cal E}_{ij} \otimes {\cal E}^*_{ji} \ .
	$$
This operator commutes with the two-spin generators ${\cal E}_{kl} 
\otimes 1 + 1 \otimes {\cal E}^*_{kl}$, as is easily verified by using 
the supercommutation relations \commrel. By construction, $H_{\rm eff}$ is 
hermitean w.r.t. to the invariant scalar product \extscap. (The interaction 
Hamiltonian $\Delta(g,h)$ obviously is, and hermitecity is preserved by 
orthogonal projection.) The constant of proportionality $J = 4\sigma_{xx} 
/ a_1 > 0$ is finite and is found by calculating some special nonzero 
matrix element of $H_1$. This is done in Appendix D. From now on we will 
often omit the tensor-product symbol in multi-spin operators, keeping it 
for clarity however when writing multi-spin states.

It is clear how to generalize all this to the case of $2N$ spins.
The degenerate space of strong-coupling ground states becomes $V 
\otimes V^* \otimes V \otimes V^* \otimes ...$ ($2N$ factors), and
the scalar product \extscap\ is extended in the natural way. (We just 
keep track of the minus signs that arise from changing the ordering of 
the fermions.) Orthogonal projection yields the multi-spin Hamiltonian
\eqn\spinHam{
	H_S = \sum_{{\rm even} \ l} \sum_{ij} (-)^{|i|+1} \left( 	
	J_+ {\cal E}_{l,ij} {\cal E}^*_{l+1,ji} + J_- {\cal E}_{l,ij} 
	{\cal E}^*_{l-1,ji} \right), \qquad J_+ = J_- = J .}
We have thus arrived at a spin chain reformulation of the nonlinear $\sigma$ 

model with topological coupling $\sigma_{xy}^* = 1/2 \ ({\rm mod} \ 1)$. 
The spin Hamiltonian \spinHam\ is completely specified by the commutation 
relations \commrel\ for the spin operators ${\cal E}_{ij}$ and ${\cal 
E}^*_{ij}$ and their action \spinactv\ on the lowest- and highest-weight 
vectors $v$ and $v^*$. The spatial boundary conditions are chosen to be 
periodic. (In the case of open boundary conditions, we need to modify 
the modules $V, V^*$ for the edge spins in general; see Sect.~4.6.)
According to what was said at the end of Sect.~3.5, we can incorporate 
deviations of $\sigma_{xy}$ from $\sigma_{xy}^*$ by staggering the 
spin-spin interaction, i.e. by putting $J_{\pm} = (1\pm\epsilon)J$ with 
$\epsilon \sim \sigma_{xy}-\sigma_{xy}^*$.

Because the spin chain is built from alternating modules $V$ and 
$V^*$, translation by one site cannot ever be a symmetry of the 
Hamiltonian $H_S$. However, for $J_+ = J_-$ and periodic boundary
conditions, $H_S$ possesses another symmetry which is just as good.
There exists a canonical isomorphism ${\cal T} : V \to V^*$, $f \mapsto
\bar f$ (complex conjugation). ${\cal T}$ squares to superparity by our 
using an adjoint of the second kind, $\overline{\bar f} = (-)^{|f|} f$ 
\berezin. It extends to a map ${\cal T} : V \otimes V^* \otimes ... \to 
V^* \otimes V \otimes ...$ in the natural way. From Eqs.~\symtrans\ and
\barmu\ one can show that the two-spin Hamiltonian behaves under complex 
conjugation as follows:
	$$
	{\cal T}^{-1} \left( \sum_{ij} (-)^{|i|} {\cal E}_{ij}
	\otimes {\cal E}^*_{ji} \right) {\cal T} = 
	\sum_{ij} (-)^{|j|} {\cal E}^*_{ji} \otimes {\cal E}_{ij} .
	$$
This equation implies that the operator composed of translation by one site 
followed by complex conjugation, {\it commutes} with the Hamiltonian $H_S$ 
of a periodic chain with $J_+ = J_-$.

The Hamiltonian $H_S$ has been defined over the spin spaces $V = 
V_{+1}$ and $V^* = V_{-1}$. Actually, we can formulate a whole family 
of Hamiltonians $H_S^{(m)}$ by taking $V = V_m$ and $V^* = V_{-m}$ 
with $m$ any positive integer. Since we know $m$ to be the number
 of ``edge states'', it is clear which role to assign to $H_S^{(m)}$:
it applies to a situation where $m$ Landau levels are strongly
mixed by the disorder. This is of experimental relevance for $m = 2$
since, for a weak magnetic field, the broadening caused by the disorder
may exceed the Zeeman splitting and render Landau levels spin-degenerate.
A plausible global phase diagram for this case has been suggested by
D.K.K.~Lee and Chalker \lc. (See also a recent preprint by D.H.~Lee 
\dhlee\ in this context.) Here we simply mention that, by using the
coherent states of Sect.~4.5 to derive a functional integral 
representation of the $m = 2$ chain, we retrieve the nonlinear 
$\sigma$ model \defZ\ with $\sigma_{xy} = 0$ (mod 1). This implies
a finite localization length in the center of the Landau band, in
accordance with the conclusions reached in Ref.~\lc.

We have argued that projection of the $\sigma$ model on the superspin 
chain should retain the essential long wave length quantum Hall physics. 
Therefore, we expect the chain for $m = 1$ and $J_+ = J_-$ where a 
delocalization transition is known to occur, to be massless. Clearly, 
a direct proof of masslessness would be very welcome. A useful pool 
of ideas we can draw from is provided by the Lieb-Schultz-Mattis 
theorem \lsm\ for isotropic antiferromagnetic (ordinary) spin $S 
= 1/2$ chains, which has been generalized to arbitrary half-integer 
$S$ by Affleck and Lieb \al. What these authors do is to construct 
a variational state (orthogonal to the ground state) with excitation 
energy of order $1/L$ for a chain of length $L$. By the variational 
principle, the existence of such a state implies that the Hamiltonian 
either has zero gap, or else a doubly degenerate ground state with 
a finite gap, in the thermodynamic limit $L\to\infty$. 
Because of the resemblance between the $S = 1/2$ spin chain and the 
superspin chain with Hamiltonian \spinHam, it seems that it should be 
possible to transcribe the theorem and its proof. Unfortunately, the 
transcription has not been done successfully up to now, for but one 
reason which however is vital: in order for the variational principle 
to apply, we need the Hamiltonian to be hermitean with respect to a 
quadratic form which is positive definite. As we have seen, positivity 
is not a property enjoyed by the invariant scalar product \scalprod. 
Nevertheless, as we shall see, the two-superspin Hamiltonian is 
completely diagonalizable and has positive excitation energies. We are 
thus encouraged to think that the difficulty is not of a fundamental 
kind but is a technical complication that can be overcome.

To compute from the (super-)spin chain such physical observables as the 
conductance coefficients $\langle g_{pq} \rangle$, we need the spin chain 
expressions for the ``contact operators'' $O_i$ of Sect.~2.2. For an ideal
contact connected to a single spin, we know these to be given by $O_p = 
v({\cal B}_{11}v,\cdot)$ for current, $O_q = {\cal B}_{11} v (v,\cdot)$ 
for voltage, and $P_v = v(v,\cdot)$ or $P_v^* = v^*(v^*,\cdot)$ for a 
contact which is neither voltage nor current; see Sect.~4.6. In that 
subsection, we calculated the Hall conductance for a disk geometry using 
the independent edge spin approximation. Having constructed the Hamiltonian 
$H_S$ governing the evolution of the interacting spin chain, we can now 
study the conductance coefficients in principle in the following, more 
interesting setting. We assume periodic boundary conditions in both 
time and space direction with lengths $L_0$ and $L_1$, respectively. 
(Periodicity in space eliminates the boundary current.) $L_0$ 
is chosen to be very large. We select two times $t_i < t_f$ where all 
spins are connected to ``particle reservoirs''. Recall that the effect 
of the reservoirs is to hit every degree of freedom in $V$ or $V^*$ with 
a projector $P_v$ or $P_v^*$, at the corresponding time. For $L_0 \to 
\infty$, the time evolution outside the fixed interval $[t_i,t_f]$ 
projects on the ground state of $H_S$. In Sect.~5.2 we will show that 
this ground state has zero energy and contains the state $\psi_0 = v 
\otimes v^* \otimes v \otimes v^* \otimes ...$ with unit amplitude. To 
measure a conductance $\langle g \rangle$, we connect the $k^{\rm th}$ 
$V$-spin, say, to a voltage probe at $t_i$ and the $l^{\rm th}$ $V$-spin 
to a current probe at $t_f$. From the above expressions for $O_p$ and 
$O_q$, we then get
\eqn\twospig{
	\langle g \rangle = (\psi_f , \exp\left( - |t_f-t_i| H_S 
	\right) \psi_i)}
where $\psi_i$ ($\psi_f$) is the multi-spin state obtained from $\psi_0$ 
by creating one quantum of excitation $v \to {\cal B}_{11}v$ in the $k^{\rm 
th}$ $(l^{\rm th})$ $V$-spin. When $L_1$ is held fixed, $\langle g \rangle$ 
must decay exponentially with distance $|t_f-t_i|$ over some length $\xi$, 
the localization length. For the critical chain $\xi$ will be proportional 
to $L_1$ for $L_1$ large. $\langle g \rangle$ is expected to be governed by 
the laws of conformal invariance. However, since the definition of $\langle 
g \rangle$ involves two {\it strings} of operator insertions at the times 
$t_i$ and $t_f$, its behavior under conformal transformations is not 
elementary. (The two strings impose boundary conditions.) Things simplify 
if we close off all the contacts which are neither voltage nor current. 
Then $\langle g \rangle$ turns into a spin-spin correlation function which 
measures conductance in a two-terminal (small contact) geometry of the kind 
discussed in Sect.~2.2. In this case, the hypothesis of conformal invariance 
relates the constant of proportionality between $L_1$ and the localization 
length $\xi$ to the critical index governing the algebraic decay of $\langle 
g \rangle$ in the infinite system \cft.

To conclude this subsection, we mention that the $\sigma$ model 
expressions for the scaling fields of Sect.~2.1 are polynomials in 
the elements of the matrix-valued function $Q_{ij}(g) = (g\Lambda 
g^{-1})_{ij}$; see e.g. Ref. \wegone. Calculating the matrix element 
of such a polynomial $Q_{i_1 j_1} Q_{i_2 j_2} ... Q_{i_r j_r}$ between 
two states in $V$ or $V^*$, we get an integral which is {\it divergent} 
for $r \ge 2$ because of the noncompactness of {\bf G/K}. (The matrix 
elements between excited states of the monopole Hamiltonian are divergent 
for {\it all} $r \ge 1$.) In other words, these $\sigma$ model operators 
do not have a sensible restriction to the spin chain. (They are 
well-defined only when the system is perturbed by a finite frequency 
$\omega + i\varepsilon$.) This is a technical manifestation of the 
distinction between the different observables discussed in more direct 
physical terms in Sects.~2.1-2. \bigskip

\noindent{\medbf 5.2 Solution of the two-spin problem}\medskip

\noindent In Sect.~5.1 we constructed a spin Hamiltonian $H_S$ by 
projection of the quantum Hamiltonian of the nonlinear $\sigma$
model. Solving for the ground state and the low-energy excitations 
of $H_S$ we will get an analytic description of the critical
properties of integer quantum Hall systems. However, such a solution 
is not easy to find and has not been found yet. Here we will have to be 
satisfied with doing some preparatory work toward that ambitious goal.

The two-spin Hamiltonian $H := \sum_{ij} (-)^{|i|+1} {\cal E}_{ij} 
{\cal E}^*_{ji}$ is an invariant second-order differential operator 
acting on wave functions
	$$
	(f \otimes f^*)(Z,\tilde Z,Z',\tilde Z') = v(Z,\tilde Z) 
	v(Z',\tilde Z') \varphi(Z) \varphi^*(\tilde Z') 
	$$
with holomorphic $\varphi$ and antiholomorphic $\varphi^*$. The 
following trick allows us to reduce the problem of diagonalizing $H$ 
to a problem that has been solved before.
Because $f$ and $f^*$ are completely specified by the dependence
of $\varphi$ and $\varphi^*$ on $Z$ and $\tilde Z'$, respectively,
we may identify $Z \equiv Z'$ and $\tilde Z \equiv \tilde Z'$ without
incurring any loss of information. On making this identification,
$H$ turns into an invariant second-order differential operator 
acting on functions
\eqn\idZbarZ{
	F(Z,\tilde Z) = v(Z,\tilde Z)^2 \varphi(Z) \varphi^*(\tilde Z) .}
There exists only one such invariant differential operator. This is
the Laplacian ${\cal L}$ for {\bf G/K} or, to be precise, ${\cal L}$
restricted to the truncated space of functions on {\bf G/K} of the
form \idZbarZ. Therefore, $H = c {\cal L}$ with some constant $c$ and in 
the restricted sense just specified. In Appendix E we show $c = -1/2$. 

The Laplacian ${\cal L}$ for Riemannian supersymmetric spaces {\bf G/K} 
has been the object of some amount of study. A general procedure for the
construction of its eigenfunctions, using such concepts as the Iwasawa
decomposition of a semisimple complex Lie group and Harish Chandra's
formula, has been described in \mmz. We are not going to review this
procedure here but will specialize to $n = 1$ for brevity. In this
case, an accidental simplification occurs which enables us to short cut
the general theory. 

The key to making progress -- for $n = 1$ as in general -- is to
consider the action of ${\cal L}$ on {\bf K}-radial functions
$F(\pi(g)) = F(\pi(kg))$ $(k\in{\bf K})$. By the Cartan decomposition
${\bf G} = {\bf K} {\bf A}^+ {\bf K}$ \helgason\ such functions can be
regarded as functions on the positive part ${\bf A}^+$ of a maximal
abelian subgroup ${\bf A} \subset {\bf G}$ with Lie algebra contained
in ${\cal P}$, the tangent space of {\bf G/K} at $\pi(1)$. In the
present case, ${\bf A} \simeq \RN \times {\rm S}^1$, and ${\bf A}^+$
is conveniently parametrized by two coordinates $x,y$ in the range
$1 \le x < \infty$ and $-1 \le y \le 1$. (These are the polar
coordinates of Efetov, denoted by $\lambda_1, \lambda$ in \efetov.)
The origin $\pi(1)$ has the coordinates $x = y = 1$.
The invariant Berezin measure $dg_K$ on {\bf G/K} induces a scalar
product on the space of {\bf K}-radial functions defined by
\eqn\auxscap{
	(F_1,F_2)^\# = \int_1^\infty dx \int_{-1}^1 dy \ (x-y)^{-2}
	\ \overline{F_1(x,y)} F_2(x,y) .}
With $S$ being the function $S(x,y) = x-y$, the {\bf K}-radial part
${\cal L}^\#$ of ${\cal L}$ is the operator \refs{\efetov,\mmz}
	$$
	{\cal L}^\# = S \left( {\cal L}_{{\rm H}^2}^\# + 
	{\cal L}_{{\rm S}^2}^\# \right)  S^{-1}
	$$
where ${\cal L}_{{\rm H}^2}^\# = 4\partial_x (x^2-1) \partial_x$ and 
${\cal L}_{{\rm S}^2}^\# = 4\partial_y (1-y^2) \partial_y$ are the 
radial parts of the Laplacians on the two-hyperboloid and the 
two-sphere, respectively. Note that ${\cal L}^\#$ is hermitean 
w.r.t. the scalar product \auxscap\ on the space of {\bf K}-radial 
functions that vanish as $x-y$ at the point $x = y = 1$.

The separable form of ${\cal L}^\#$ makes it possible to write down
its eigenfunctions $\phi_{\lambda,l}$ immediately:
	$$
	\phi_{\lambda,l}(x,y) = (x-y) {\cal P}_{-(i\lambda+1)/2}(x)
	P_{(l-1)/2}(y) .
	$$
Here, ${\cal P}_\nu(x)$ is a Legendre function and $P_n(y)$ a 
Legendre polynomial \abram. The eigenvalue is $-(\lambda^2 + l^2)$,
and the quantum numbers have the range $\lambda \in \RN^+$ and 
$l \in 2\NN - 1$. Because of the special form of the functions 
\idZbarZ, not all of these quantum numbers occur in the spectral 
expansion of the two-spin Hamiltonian $H$. Since $V = V_{+1}$ and 
$V^* = V_{-1}$ both carry angular momentum (or SU(2)-spin) $\le 
1/2$, the tensor product $V \otimes V^*$ carries at most angular 
momentum 1. The largest angular momentum present in $\phi_
{\lambda,l}$ is seen to be $(l+1)/2$. Therefore, the only $l$-quantum 
number that occurs for $m = 1$ is $l = 1$. (There is no restriction on 
$\lambda \in \RN^+$.) 

The functions $\phi_{\lambda,l}$ all vanish at the point $x = y = 1$
corresponding to $Z = \tilde Z = 0$. The two-spin quantum space
$V \otimes V^*$ contains the state $v \otimes v^*$ whose wave function
does {\it not} vanish at $Z = \tilde Z = 0$ so, clearly, something
is still missing. What is missing and must be added for completeness
is the constant function $\phi_0 = 1$, which trivially is an 
eigenfunction of
	$$
	{\cal L}^\# = 4 (x-y)^2 \left( {\partial \over \partial x} 
	{x^2-1 \over (x-y)^2} {\partial \over \partial x} + 
	{\partial \over \partial y} {1-y^2 \over (x-y)^2} {\partial
	\over \partial y} \right)
	$$
with eigenvalue zero. Using $\phi_0 = 1$ together with the functions
$\phi_{\lambda,l}$ we can expand any {\bf K}-radial function on
{\bf G/K} with ``good'' behavior at infinity. (See \mrzcmp\ for more
insight into the precise role played by $\phi_0$ in a prototypical
example.)

Given a {\bf K}-invariant stationary two-spin state $\psi$ we can 
generate {\bf K}-noninvariant stationary states by applying symmetry 
transformations to the equation $H\psi = E\psi$. Doing this in a 
systematic fashion by using the methods of supersymmetric Fourier 
analysis \refs{\mrzcmp,\mmz}, we obtain a continuous set of {\bf G}-multiplets
labelled by $\lambda, l$ (with $l=1$ being the only quantum number occurring
for $m=1$) that are eigenspaces of $H$ and completely exhaust $V \otimes 
V^*$. 

To put this in perspective, recall the indefiniteness of the invariant 
scalar product \scalprod, which jeopardizes the diagonalizability of $H : 
V \otimes V^* \to V \otimes V^*$. (In fact, it is precisely because of 
this indefiniteness that Laplace-Casimir elements of Lie superalgebras 
{\it generically cannot} be brought to diagonal form.) What saves the 
day is the existence of an auxiliary scalar product $(\cdot,\cdot)^\#$ 
on a {\it restricted} space $(V \otimes V^*)^\#$ of {\bf K}-invariant 
states. This auxiliary scalar product has all the attributes of a
(Hilbert space) scalar product, and $H : (V\otimes V^*)^\# \to (V\otimes 
V^*)^\#$ is hermitean with respect to it. This in conjunction with
{\bf G}-symmetry guarantees the diagonalizability of $H$.

Note the pronounced ``antiferromagnetic'' character of $H$: it has a 
({\bf G}-)singlet ground state separated by a gap from a continuum of 
non-singlet excited states. The existence of a gap implies exponential 
decay of correlations in the two-spin system, which is related to 
exponential localization of all states in a Chalker-Coddington network
model with only two links in the short, spatial direction. For the
spin chain, exponential decay comes about because the repeated
action of $H$ on the state $\psi_i = ({\cal B}_{11}v)\otimes v^*$
creates on average an ever increasing number of quanta of excitation,
eventually producing a time-evolved state which has exponentially
small overlap with $\psi_f = \psi_i$. In the network model, 
exponential localization sets in when the elements of the transfer
matrix become exponentially large. Thus, growth in the number of
${\cal B}$- (and ${\cal C}^*$-) quanta is the spin chain equivalent
of growth in the transfer matrix amplitudes of the network model.
Using the Fourier decomposition of the Dirac delta distribution 
$\delta_o$ for {\bf G/K} \mmz,
\eqn\Diracd{
	\delta_o = \phi_0 + 2^2 \sum_{l \in 2\NN-1} \int_0^\infty
	d\lambda \ {\lambda\tanh(\pi\lambda/2) l \over 
	(\lambda^2 + l^2)^2 } \ \phi_{\lambda,l} ,}
and doing the same calculations as in \mmz, one can derive an exact 
expression for the (two-spin) conductance $\langle g \rangle$ of 
Eq.~\twospig. $\langle g \rangle$ decays as $|t_f-t_i|^{-3/2} \exp\left(
-J|t_f-t_i|/2\right)$ for large distance $|t_f-t_i|$.

Let us now present an independent argument showing why the ground state 
of $H$ has zero energy. This argument has the merit of permitting easy 
generalization to many spins. We write the two-spin Hamiltonian in the 
form
	$$
	H = \sum_{ij} (-)^{|i|+1} \left(
	{\cal A}_{ij} {\cal A}^*_{ji} + {\cal B}_{ij} {\cal C}^*_{ji} +
	{\cal C}_{ij} {\cal B}^*_{ji} + {\cal D}_{ij} {\cal D}^*_{ji}
	\right) 
	$$
and consider the state $v \otimes v^*$. It is a {\bf K}-singlet since 
$T_k v = \mu(k) v$ and $T_k v^* = \mu(k)^{-1} v^*$. Acting with $H$ on 
it we get
\eqn\Hactvvs{
	H (v\otimes v^*) = \sum_{ij} (-)^{|i|+1} \left( {\cal B}_{ij}v
	\right) \otimes \left( {\cal C}^*_{ji} v^* \right) ,}
as follows from Eqs.~\spinactv. Let $L_M$ $(M \ge 0)$ denote the 
$(M+1)$-dimensional space of states generated by at most $M$ 
actions of $H$ on $v \otimes v^*$. Since $H$ commutes with the 
generators of {\bf K}, all elements in $L_M$ are {\bf K}-invariant.
Taking $M\to\infty$ we obtain a space, $L_{\infty}$, which is closed 
under the action of $H$. (We here pay no attention to mathematical 
subtleties that might arise from the infinite-dimensionality of 
$L_\infty$.) This space is decomposed by $L_\infty = L_{M=0} \oplus
L_c$ (direct and orthogonal sum). We claim that $H L_\infty \subset 
L_c$. To prove this statement it suffices to prove $H \psi \in L_c$ 
for $\psi \in L_c$, since $H (v\otimes v^*) \in L_c$ from Eq.~\Hactvvs. 
Let ${\cal N} = \sum_i \left( {\cal A}_{ii} - {\cal D}_{ii} \right) 
/ 2$ be the operator counting the number, $N$, of quanta of excitation
in the first factor of $V \otimes V^*$. We have ${\cal N}(v\otimes 
v^*) = 0$. The commutation relations $[{\cal N},{\cal B}_{ij}] = 
{\cal B}_{ij}$, $[{\cal N},{\cal C}_{ij}] = - {\cal C}_{ij}$,
$[{\cal N},{\cal A}_{ij}] = [{\cal N},{\cal D}_{ij}] = 0$ imply that 
$H$ changes $N$ by at most one unit. $L_c$ is a direct sum of spaces 
with $N \ge 1$. Let $\psi$ now be a state with $N = 1$. There 
exists only one such state which has the additional property of being 
{\bf K}-invariant. This is the state on the right-hand side of 
Eq.~\Hactvvs. Using the relations \commrel\ and \spinactv, we easily see that 
$\sum_{ij} (-)^{|i|+1} {\cal C}_{ij} {\cal B}^*_{ji}$ annihilates $\psi = 
\sum_{ij} (-)^{|i|+1} ({\cal B}_{ij}v) \otimes ({\cal C}^*_{ji} v^*)$,
so $H\psi$ has zero intersection with $v \otimes v^*$. But the image
$H\psi$ of any state $\psi \in L_c$ with more than one quantum of excitation
in $V$ carries at least one such quantum, which proves $H L_c \subset 
L_c$. We conclude that the cokernel of $H : L_{\infty} \to L_\infty$ 
is (at least) one-dimensional. Therefore, if $H$ can be brought to 
diagonal form -- and we know it can be from the previous discussion 
-- it must have an eigenstate with zero energy. On physical grounds, 
this eigenstate is expected to be the ground state. (Because the
partition sum equals unity, the existence of a negative-energy
state would give rise to unphysical spin correlations that increase
exponentially with distance.)

The result $H L_\infty \subset L_c$ has the following consequence
which is worth mentioning. $L_0$ and $L_c$ are orthogonal spaces 
w.r.t. the scalar product \extscap. Hence, by the hermitecity of $H$,
	$$
	( H^N (v\otimes v^*) , H^N (v\otimes v^*) ) = 
	( v\otimes v^* , H^{2N} (v\otimes v^*) ) = 0 ,
	$$
so the space $L_c$ is {\it null}, i.e. all of its elements have
vanishing norm.

The above argument readily generalizes to an arbitrary even number of 
spins. Let $L_0$ denote the linear space spanned by $\psi_0 = v \otimes 
v^* \otimes ... \otimes v \otimes v^*$. If $\psi$ is some state that 
differs from $\psi_0$ on more than two neighboring sites, then clearly 
$H_S \psi \cap L_0 = \{ 0 \}$. If $\psi \in H_S^{M-1} L_0$ ($M\ge 1)$ 
differs from $\psi_0$ on no more than two neighboring sites, we are back 
to the two-spin problem analyzed before and again we have $H_S \psi \cap 
L_0 = \{ 0 \}$. This shows $H_S^M L_0$ to have zero intersection with 
$L_0$. Hence, the rank of $H_S$ is reduced by one and we expect a 
zero-energy (ground) state as before. 

A neat application of the results of this subsection is the following.
When the process of quantum tunnelling of electrons across saddle
points is switched off, the Chalker-Coddington model (or, rather, a
slight generalization thereof \lwk) reduces to a network model which 
undergoes a classical percolation transition. This phase transition is 
in a different universality class and has correlation length exponent 
$\nu = 4/3$ \trugman. With a complete solution of the two-spin problem in 
hand, we can now write down a spin chain Hamiltonian for the classical 
percolation transition, too. To do so, it is easiest to make a 
checkerboard decomposition of the Hamiltonian and pass to a vertex model 
\baxter\ by exponentiation. The vertex model is specified by its $R$-matrix 
$R : V \otimes V^*\to V^* \otimes V$ (Figure 8). Because of the connection 
of the supersymmetric vertex model with an underlying disordered network
model, it is clear that the $R$-matrix at the critical point of the
percolation transition must be
\eqn\classR{
	R_{ab^*,c^*d} = {1\over 2} \left( \delta_{ad} \delta_{b^*c^*}
	+ \delta_{\bar ab^*} \delta_{c^*\bar d} \right).}
(In the network model, electrons turn {\it either} right {\it or} left at 
a node, with equal probabilities. Correspondingly, the spin degrees of 
freedom of the vertex model must be transferred by the $R$-matrix to the 
right or left, with equal weights.) On general grounds, the $R$-matrix 
\classR\ must meet the requirement of {\bf G}-invariance, and it does. 
With ${\rm P} : a \otimes b^* \mapsto (-)^{|b^*||a|} b^* \otimes a$ the 
graded permutation operator, $R$ can be written in the form $R = {\rm P} 
(1 + \Pi_0)/2$ where $\Pi_0$ projects on the unique {\bf G}-singlet state 
contained in the tensor product $V \otimes V^*$. (Of course, this singlet 
state is nothing but the ground state of the two-spin Hamiltonian 
discussed above.) Restricting the resolution of unity \Diracd\ to $V 
\otimes V^*$, we obtain
	$$
	R = {\rm P} \tilde R , \quad
	\tilde R = 1 - {1 \over 2} \int_{\RN^+} 
	d\mu(\lambda) \Pi_\lambda .
	$$
Here $d\mu(\lambda) = 2^2 \lambda\tanh(\pi\lambda/2)(\lambda^2+1)^{-2}
d\lambda$, and $\Pi_\lambda$ is the spectral projector onto total 
SU(1,1)-spin with quantum number $\lambda$, normalized by the condition
	$$
	\left( \int_{\RN^+} d\mu(\lambda) h(\lambda) \Pi_\lambda
	\right)^2 = \int_{\RN^+} d\mu(\lambda) h(\lambda)^2 \Pi_\lambda .
	$$
To return to a Hamiltonian formulation, we have to compute the logarithm
of the transfer matrix. We put $\tilde R^\epsilon = \exp(-\epsilon 
H_{\rm cp})$ and get the two-spin Hamiltonian
\eqn\Hcp{
	H_{\rm cp} = \int_{\RN^+} d\mu(\lambda) h(\lambda) 
	\Pi_\lambda, \qquad h(\lambda) = 1/2 ,}
in the limit $\epsilon \to 0$. The Hamiltonian of the percolating chain 
is a translationally invariant sum of two-spin Hamiltonians. To move the 
system off the critical point, we stagger the interaction strength.

The constancy of the excitation energy $h(\lambda)$ for $H_{\rm cp}$ 
reflects the additional symmetries that distinguish classical 
percolation from the generic case of quantum percolation, where 
$h(\lambda)$ is nonconstant. We note that $H_{\rm qp} = \sum_{ij} 
(-)^{|i|+1} {\cal E}_{ij} {\cal E}_{ji}^*$ can be written in the 
form \Hcp\ with $h(\lambda) = \lambda^2 + 1$.\bigskip

\noindent{\bigbf 6 \ Summary and Outlook}\medskip

\noindent In this paper the study of the supersymmetric version of 
Pruisken's nonlinear $\sigma$ model for the integer quantum Hall 
effect has been resumed. By generalizing the work of Shankar and 
Read, we derived a $1d$ Hamiltonian lattice formulation of the 
$\sigma$ model which has the right (naive) continuum limit. The 
lattice Hamiltonian consists of a site-diagonal part and a site-site 
interaction, the former being a supersymmetric generalization of the 
Hamiltonian for a charged particle moving in the field of a magnetic 
monopole. The monopole charge $m$ is determined by the value of the 
Hall conductivity $\sigma_{xy}$. In the center of a Hall plateau
region, where $\sigma_{xy}$ is an integer, $m$ vanishes for all 
sites in the bulk of the system and equals $\pm \sigma_{xy}$ at 
the edges. In this special case we used a renormalization group 
assisted argument to calculate the conductance coefficient $g$ 
pertaining to neighboring probes in a Corbino disk geometry. As is 
expected from the microscopic picture of directed transmission through 
channels localized near the sample boundary, we obtain either $g = 0$ 
or $g = |\sigma_{xy}|$ depending on the sign of $\sigma_{xy}$. This 
result suggests an interpretation of the monopole charge as the number 
of ``edge states''.

For half-integral values of the Hall conductivity $\sigma_{xy}$
corresponding to the points of transition between neighboring plateaus, the 
monopole charge in the bulk of the system alternates on sites $(m = \pm 1)$.
Projection onto the degenerate strong-coupling ground states of the lattice 
Hamiltonian yields a superspin chain Hamiltonian effectively describing the 
critical point of the $\sigma$ model. The space of quantum states for even 
and odd sites of the chain is a space of holomorphic and antiholomorphic 
functions $V$ and $V^*$, respectively. This alternating structure corresponds 
to the alternating orientation of links in the Chalker-Coddington network 
model, reflecting the directedness of guiding center motion in a strong 
magnetic field and a smooth random potential. A direct derivation of the 
superspin chain from the network model is possible \nread. While Pruisken's 
nonlinear $\sigma$ model is derived from the ``quantum'' limit of a random 
potential with correlation length $l_c$ short compared to the magnetic 
length $l_B$, the network model applies to the opposite semiclassical
limit $l_c \gg l_B$. Since both models map on the very same superspin chain, 
we conclude that the ratio $l_c / l_B$ is an irrelevant parameter for the 
IQHE transition. By the hypothesis of universality at a critical point, 
such irrelevance is not unexpected.

We recalled the fact that the dynamic structure factor and other
frequency-dependent correlation functions as well, have critical amplitudes
that diverge at zero frequency. Within the field-theoretic framework
this divergence is related to the fact that such correlators are not 
expressible as superspin correlation functions: the matrix elements of the 
corresponding $\sigma$ model operators between states in the superspin 
quantum spaces $V$ and $V^*$ just do not exist. A quantity that does
restrict to the superspin chain in a natural manner is d.c. conductance.

While the nature of the subject is such as to force extensive use
of mathematical argument, we tried to keep the mathematics to the 
minimum needed for clarity and completeness. For example, we did 
not mention the role of the topological density as pullback of the 
K\"ahler form of the K\"ahler supermanifold {\bf G/K}. The theory 
of connexions on hermitian line bundles associated with a principal 
fibre bundle (central to Sects.~3.4-5) and of holomorphic sections 
of such bundles (central to Sects.~4.2-5), and the theory of Fourier 
analysis on supersymmetric spaces {\bf G/K} (central to Sects.~4 and 
5.2) were only mentioned parenthetically. We made these omissions in 
an effort to keep the paper within reasonable size and accessible to 
readers without a background in mathematical physics.

Although the main body of this paper was concerned with the $\sigma$
model for 2$d$ electrons in a strong magnetic field, most of what we 
did can be transcribed with minor modifications to the symplectic 
$\sigma$ model describing time-reversal invariant systems with spin-orbit
scattering. The role played by the state vectors $v$ and $v^*$ is taken 
by the zero mode of the Laplacian discovered in Ref.~\mmz. This zero mode 
is one state in an infinite-dimensional space of strong-coupling ground 
states of the quantum Hamiltonian of the symplectic $\sigma$ model. The 
strong-coupling ground states transform according to a representation of 
the symmetry group which is neither lowest-weight nor highest-weight nor 
irreducible. It is not clear at present whether the resulting superspin 
chain can serve as the universal critical theory of the Anderson 
metal-insulator transition for symplectic symmetry in $d = 2$. (The 
transition caused by spin-orbit scattering differs from that for 
quantum Hall systems in that a continuous symmetry is spontaneously 
broken.)

An issue we briefly addressed is how the classical limit $l_B \to 0$
fits into the superspin chain picture. This limit is not immediately
accessible from our approach since the derivation of Pruisken's model, 
as it stands, applies only to the limit $l_B \gg l_c$. Nevertheless,
by using the relation to the Chalker-Coddington model we were able to 
write down a superspin chain Hamiltonian for the classical percolation 
transition, too. This Hamiltonian has a higher degree of symmetry than 
the generic critical Hamiltonian describing quantum percolation. In this 
way, the superspin chain provides a unified framework in which to discuss 
both quantum and classical percolation and the crossover from the latter 
to the former.

Since the mapping on a superspin chain retains the essential long wave 
length quantum Hall physics, the chain for $\sigma_{xy}^* = 1/2$ (mod 1), 
where a delocalization transition occurs, should be critical. More 
precisely, we expect a unique {\bf G}-singlet ground state with local 
``antiferromagnetic'' order and massless (non-Goldstone) excitations. 
The staggering of the spin-spin interaction for $\sigma_{xy} \not= 
\sigma_{xy}^*$ favors singlets on strong bonds and opens a mass gap.
Gaplessness of the critical chain should be guaranteed by the same 
mechanism that underlies the Lieb-Schultz-Mattis theorem for 
isotropic antiferromagnetic spin chains with half-integer spin. In 
spite of the tantalizing analogy, I have not yet succeeded in
transcribing the theorem and its proof. The problem is not lack of
translational invariance but the indefiniteness of the invariant 
scalar product \scalprod, which prevents the application of the 
variational principle in its usual form.

The ground-state energy of the superspin Hamiltonian vanishes 
identically for a chain of any length and for arbitrary boundary 
conditions, as is guaranteed by unbroken {\bf K}-supersymmetry. 
The Virasoro algebra underlying the massless chain for $m = 1$ 
must therefore have central charge $c = 0$. Note that this does 
not imply triviality of a theory which is nonunitary. (The Kac-Moody 
algebra of conserved currents {\it will} have a nontrivial central 
extension.)

In addition to the Hamiltonian \spinHam\ for $m = 1$, we can formulate 
quantum superspin chain Hamiltonians for all values of the monopole
charge $m > 0$. By the interpretation of $m$ as the number of
transmitting channels, we know that such a generalized Hamiltonian 
describes a situation where $m$ Landau levels are strongly mixed
by the disorder. The chain for monopole charge $m$ resembles a 1$d$ 
isotropic antiferromagnet with spin $S = m/2$. Large values of $m$ 
offer the possibility of making a $1/m$ expansion. A natural 
starting point for such an expansion is the N\'eel state $v \otimes 
v^* \otimes v \otimes v^* \otimes ...$ . Approximating the spin 
operators by their large-$m$ limits one gets a quadratic Hamiltonian 
which can be diagonalized by a Boboliubov transformation in the usual 
manner. The low-energy excitations of this Hamiltonian, the superspin 
waves, are found to have a linear dispersion relation, as one would 
expect from the close analogy to antiferromagnetic spin chains. These 
low-energy modes are the Goldstone modes restoring the continuous {\bf 
G}-symmetry broken by the N\'eel state. Unfortunately, this type of
analysis is not too useful for understanding the critical properties, for 
the $1/m$ expansion is ignorant of the expected profound difference between 
massive superspin chains for even $m$ and massless ones for odd $m$.

What are the prospects for obtaining some sort of exact solution?
Although not too much work in this direction has been done as yet, 
a few statements can be made. The superspin chain Hamiltonian \spinHam\
is unlikely to be integrable. (The corresponding ordinary spin chains 
for spin $S > 1/2$ are not integrable.) It is therefore very fortunate 
that the universality we expect for $m = 1$ gives us much freedom to 
modify the superspin chain: any Hamiltonian which has the ``right'' 
symmetries and is not too remote in parameter space, should give the 
same low-energy physics --- the physics of the unique RG-fixed point 
that describes the IQHE transition. Making use of this freedom, one 
may be able to construct a one-parameter family of commuting transfer 
matrices (with the superspin Hamiltonian being the logarithmic
derivative at a special point) by solving the quantum Yang-Baxter 
equation (QYBE). This, unfortunately, is not immediately possible but 
requires an extension of existing theory, for the following reason. 
We are looking for a {\bf G}-invariant solution of QYBE. An efficient 
method for constructing such solutions has been described by Drinfel'd 
\drinfeld, generalizing earlier work of Kulish et al. \krs. The basis of 
the method is an expansion of the $R$-matrix $R(u) : W_1 \otimes W_2 
\to W_1 \otimes W_2$ around $u = \infty$ in the form $R(u) = 1 + u^{-1} 
r + ...$ where $u^{-1} r$ is a solution of the classical Yang-Baxter
equation. The crucial difference for our superspin chains is that {\it 
no such classical limit exists}. Because of the alternating structure 
of the chain, the $R$-matrix is a map $R(u) : V \otimes V^* \to V^* 
\otimes V$ (Figure 8), which implies $R(u) \not= 1$ for all $u$. (One 
might think that the isomorphism ${\cal T} : V \to V^*$, $f \mapsto \bar 
f$ could be used instead of the identity. However, $R(\infty) = {\cal T}$ 
fails to meet the requirement of {\bf G}-invariance.)

In my opinion, the most promising line of further attack is to study 
superspin chains with $1/r^2$ exchange of the Haldane-Shastry type \hs\
and/or explore the possible relation to a gauged WZW model. (Gauging is 
necessary to combat the indefinite metric on the noncompact symmetry 
group {\bf G}.)\medskip

\noindent{\bf Acknowledgements.} My particular thanks go to N.~Read who, 
by kindly describing the modules $V_{\pm 1}$ to me, revitalized my dormant 
engagement in this difficult project. I also thank A.~H\"uffmann and 
M.~Jan\ss en for useful discussions and A.~Altland for much constructive
criticism of the manuscript. \bigskip

\noindent{\bigbf Appendix A: Proof of Formula \defmonHp}\medskip

\noindent Eq.~\defmonH\ expresses ${\cal L}^m$ by left-invariant vector 
fields. To switch to right-invariant vector fields we proceed as follows.
In Eq.~\deflapl\ we introduced a second-order differential operator
$\lambda_{\cal P}(\partial/\partial x) := \lambda(\partial/\partial x)
= \sum g^{ij} \partial^2/\partial x^j \partial x^i$ by inverting the
invariant quadratic form ${\rm B}$ on ${\cal P}$. Analogous differential 
operators $\lambda_{\cal G}$ and $\lambda_{\cal K}$ are defined by inverting
${\rm B}$ on ${\cal G} = {\rm Lie}({\bf G})$ and ${\cal K} = {\rm Lie}
({\bf K})$, respectively. These operators are related by the identity 
$\lambda_{\cal P} = \lambda_{\cal G} - \lambda_{\cal K}$, which we insert 
into Eq.~\defmonH. By definition of the projection $\pi : {\bf G} \to 
{\bf G/K}$, we have $f(\pi(g\exp{\scriptstyle\sum}x^i e_i)) = f(\pi(g))$ 
for ${\scriptstyle\sum}x^i e_i \in {\cal K}$. Moreover, $\mu(k(g\exp{
\scriptstyle\sum}x^i e_i)) = \mu(k(g)) \mu(\exp{\scriptstyle\sum}
x^i e_i)$ for ${\scriptstyle\sum}x^i e_i \in {\cal K}$, and
	$$
	\lambda_{\cal K}(\partial/\partial x) 
	\mu(k(\exp{\scriptstyle\sum} x^i e_i))^{-1} \Big|_{x=0} 
	= \lambda_{\cal K} (\partial/\partial x) \exp \left( - m \str 
	\pi_+ {\scriptstyle\sum} x^i e_i \right) \Big|_{x=0} = 0
	$$
by a simple calculation. Therefore, we may drop $\lambda_{\cal K}$. Doing 
so we arrive at a formula for ${\cal L}^m$ identical to \defmonH\ except
that $\lambda = \lambda_{\cal P}$ is now replaced by $\lambda_{\cal G}$. 
After this preparation we can move $\exp{\scriptstyle\sum}x^i e_i$ to
the left of $g$ by using the invariance of $\lambda_{\cal G}$ under
$\exp{\scriptstyle\sum}x^i e_i \to g^{-1} (\exp{\scriptstyle\sum}x^i 
e_i) g$. This proves Eq.~\defmonHp. \bigskip

\noindent{\bigbf Appendix B: Derivation of Eqs.~\symtraco, \calHcos\ and 
\Tgplus}\medskip

\noindent We shall derive the coordinate expression \symtraco\ of the
representation $g \mapsto T_g$, Eq.~\symtrans. Recall the definition
of $s : {\bf G/K} \to {\bf G}_0 = {\rm SU}(n,n|2n)$ by
	$$
	s(\pi( \left( \mymatrix{A &B\cr C &D\cr} \right) )) = 
	\pmatrix{1 &BD^{-1}\cr CA^{-1} &1 \cr} \pmatrix{
	(1-BD^{-1}CA^{-1})^{-1/2} &0\cr 0 &(1-CA^{-1}BD^{-1})^{-1/2}\cr}.
	$$
Solving the equation $g = s(\pi(g)) k(g)$ for $k$, we get
	$$
	k( \left( \mymatrix{A &B\cr C &D\cr} \right) ) = 
	\pmatrix{	(1-BD^{-1}CA^{-1})^{1/2} A	&0		\cr
			0		&(1-CA^{-1}BD^{-1})^{1/2} D	\cr}.
	\eqno({\rm B}.1)
	$$
{}From $\pi_+ = (1+\Lambda)/2$ we have $\mu(k) = \exp(m \str \pi_+ \ln k) 
= (\sdet k)^{m/2} \exp \left( {m\over 2} \str \Lambda \ln k\right)$.
For the second factor we obtain from (B.1) the simple formula
	$$
	\exp \bigl( {m\over 2} \str \Lambda \ln k ( \left(
	\mymatrix{A &B\cr C &D\cr} \right) ) \bigr) = \sdet(A/D)^{m/2} .
	$$
To express the factor $\sdet k(g)$, we take the superdeterminant of both
sides of the equation $g = s(\pi(g)) k(g)$, which gives $\sdet k(g) = 
\sdet g$, since $s(\pi(g)) \in {\bf G}_0$. Hence,
	$$
	\mu(k( \left( \mymatrix{A &B\cr C &D\cr} \right) )) = 
	\sdet \left( \mymatrix{A &B\cr C &D\cr} \right)^{m/2}
	\sdet (A/D)^{m/2} .
	\eqno({\rm B}.2)
	$$
{}From this we infer $\mu(k(s(Z,\tilde Z))) = 1$. Moreover, Eq.~(B.2)
in combination with
	$$
	\pmatrix{ A &B\cr C &D\cr}^{-1} = 
	\pmatrix{ 1 &-A^{-1}B\cr -D^{-1}C &1}
	\pmatrix{ (A-BD^{-1}C)^{-1} &0\cr 0 &(D-CA^{-1}B)^{-1}\cr}
	\eqno({\rm B}.3)
	$$
yields $\mu(k(x)^{-1}) = \mu(k(x^{-1}))$ for any $x \in {\bf G}$. 
Applying this identity to $x = g^{-1}h$ with $h = s(Z,\tilde Z)$,
and using Eqs.~\gABCD, \sectis\ and (B.2) to calculate $\mu(k(s(Z,
\tilde Z)^{-1}g))$, we arrive at
	$$
	\mu(k(h)k(g^{-1}h)^{-1}) = (\sdet g)^{m/2}
	\sdet \left( A-\tilde ZC \over D-ZB \right)^{m/2} ,
	$$
which proves Eq.~\symtraco.

The second calculation we do in this appendix is to work out 
the similarity transformation $T_g^+ = v_m^{-1} T_g v_m$ for 
$v_m(Z,\tilde Z) = \sdet (1-\tilde Z Z)^{m/2}$ $(m>0)$. The effect 
of this transformation is to make the multiplier in Eq.~\symtraco\ 
pick up an extra factor $v_m(Z,\tilde Z)^{-1} v_m(g^{-1}\cdot 
Z,g^{-1}\cdot\tilde Z)$. To calculate it, we use the matrix 
function $Q = g \Lambda g^{-1} = s \Lambda s^{-1}$. From 
Eq.~(D.1) it is seen that $1 - \tilde Z Z = \left( (1+Q)_{BB}/2 
\right)^{-1}$ where the indices $BB$ mean the upper left (or 
Boson-Boson) block. Writing $g = \left( \mymatrix{A &B\cr C &D\cr} 
\right)$, $g^{-1} = \left( \mymatrix{A' &B'\cr C' &D'\cr} \right)$
and doing matrix multiplications, we find
	$$
	1 - (g^{-1}\cdot \tilde Z)(g^{-1}\cdot Z) = 
	\left( {1\over 2} (1 + g^{-1} Q g)_{BB} \right)^{-1}
	= (A-\tilde ZC)^{-1}(1-\tilde ZZ)(A'+B'Z)^{-1} .
	$$
Now, from Eq.~(B.3) we read off
$A' = (A-BD^{-1}C)^{-1}$ and $B' = -A'BD^{-1}$, so taking the
superdeterminant we obtain
	$$
	v_m(Z,\tilde Z)^{-1} v_m(g^{-1}\cdot Z,g^{-1}\cdot\tilde Z) = 
	\sdet \left( {AD(1-CA^{-1}BD^{-1}) \over (A-\tilde ZC)
	(D-ZB) } \right)^{m/2} .
	$$
Multiplying this with the multiplier in Eq.~\symtraco\ and using
$\sdet g = \sdet(AD) \sdet(1-CA^{-1}BD^{-1})$ we get the multiplier
in Eq.~\Tgplus.

Finally, we derive the coordinate expression \calHcos\ of the monopole 
Hamiltonian ${\cal H}_s = - {\cal L}^m$, Eq.~\defmonH. The pair of 
matrices $X, \tilde X$ appearing in \calHcos\ is defined by the equation 
$\exp({\scriptstyle\sum}x^i e_i) = s(D^{-1}XA,A^{-1}\tilde XD)$. Within 
the accuracy required by setting $X = \tilde X = 0$ after differentiation, 
and because of the invariance of the differential operator $\partial_{
X,\tilde X}$, Eq.~\invdiff, under $X \mapsto D^{-1}XA$, $\tilde X \mapsto
A^{-1}\tilde XD$, we may replace $\lambda(\partial/\partial x)$
in Eq.~\defmonH\ by $\partial_{X,\tilde X}$. Again, let $g = \left( 
\mymatrix{A &B\cr C &D\cr} \right)$. Using Eqs.~\gactsZ\ we get
	$$\eqalign{
	g \cdot (D^{-1}XA) &= (Z+X)(1+\tilde ZX)^{-1},	\cr
	g \cdot (A^{-1}\tilde XD) &= 
	(\tilde Z+\tilde X)(1+Z\tilde X)^{-1},		\cr}
	$$
where $Z = CA^{-1}$ and $\tilde Z = BD^{-1}$ are the coordinates of 
the coset $\pi(g)$. This explains the argument of the function $f$ on 
the right-hand side of Eq.~\calHcos. We turn to the multiplier 
$\mu(k(g)k(g\exp{\scriptstyle\sum}x^i e_i)^{-1})$ in Eq.~\defmonH.
The first factor, $\mu(k(g))$, has been given in Eq.~(B.2). For the
second factor, we use
	$$
	\mu(k(gs(X,\tilde X))) = (\sdet g)^{m/2} \sdet \left(
	{A+BX \over D+C\tilde X} \right)^{m/2} ,
	$$
as follows easily from (B.2). Making the substitutions $X \mapsto 
D^{-1}XA$, $\tilde X \mapsto A^{-1}\tilde XD$ and putting the factors 
together, we get the multiplier in Eq.~\calHcos. \bigskip

\noindent{\bigbf Appendix C: Representation of Lie superalgebras on
superfunctions}\medskip

\noindent We denote the superparity on any ${\rm Z}_2$-graded vector space 
$V$ by $|\cdot |$, i.e. $|X| = 0$ if $X \in V$ is even and $|X| = 1$ if $X$ 
is odd. A complex Lie superalgebra ${\cal G}$ is a ${\rm Z}_2$-graded 
vector space over $\CN$ with a bracket operation satisfying the axioms
	$$\eqalign{
	&| [ X , Y ] | = |X| + |Y| \ {\rm mod} \ 2 ,		\cr
	&[ X , Y ] = (-)^{|X||Y|+1} [ Y , X ] ,			\cr
	&[X,[Y,Z]] = [[X,Y],Z] + (-)^{|X||Y|} [Y,[X,Z]]	.	\cr}
	\eqno({\rm C}.1)
	$$
An example is ${\rm gl}(p,q;\CN)$ defined by the bracket \comrel.
Let now some Grassmann algebra $\Lambda$ be given. Following Berezin
\berezin, we define the {\it Grassmann envelope} ${\cal G}(\Lambda)$ of 
${\cal G}$ as the even part of the tensor product $\Lambda \otimes {\cal 
G}$, that is to say, as the linear space of objects $aX$ with $a \in 
\Lambda$, $X \in {\cal G}$ and $|a| = |X|$. When ${\cal G} = 
{\rm gl}(p,q;\CN)$, an element of ${\cal G}(\Lambda)$ is what is 
commonly called a supermatrix of dimension $(p+q)\times(p+q)$. 
The Grassmann envelope ${\cal G}(\Lambda)$ can be turned \berezin\ into a
{\it Lie algebra with Grassmann structure} in either one of two ways:
	$$\eqalignno{
	[aX,bY]&:=ab [X,Y] ,		&(1^{\rm st} \ {\rm kind})\cr
	[aX,bY]&:=ab (-)^{|X||b|}[X,Y]. &(2^{\rm nd} \ {\rm kind})\cr}
	$$
For the definition of the second kind we imagine $X$ and $b$ to 
anticommute when both $X$ and $b$ are odd. The usual multiplication 
rules for supermatrices used in the physics literature imply $aXbY = 
abXY$ for $a, b \in \Lambda$ and $X, Y \in {\rm gl}(p,q;\CN)$. 
Therefore, the natural convention is to take the Grassmann envelope 
${\rm gl}(p,q|\Lambda)$ to be a Lie algebra with Grassmann structure 
of the {\it first kind}. This is the convention we adopt.

Now let ${\bf G}_{\Bbb C}$ be a Lie group which has ${\rm gl}(p,q|\Lambda)$
for its Lie algebra, and let $g \mapsto T_g$ be a representation of 
${\bf G}_{\Bbb C}$ on some space of superfunctions. (For example, for 
$p = q = 2n$, ${\bf G}_{\Bbb C}$ might be the complexified symmetry 
group of the quantum Hall model space {\bf G/K}, and $T$ might be the
representation defined by Eq.~\symtrans.) Taking the differential at the
group unit $(t \in \RN)$
	$$
	{\rm d}T(aX) f := {d \over dt} T_{\exp taX} f \Big|_{t = 0} \ ,
	$$
we get a representation of the Lie algebra:
	$$
	[{\rm d}T(aX),{\rm d}T(bY)] = {\rm d}T([aX,bY]) .
	\eqno({\rm C}.2)
	$$
${\rm d}T(X)$ for $X \in {\rm gl}(p,q;\CN)$ is defined by linearity, ${\rm 
d}T(aX) = a {\rm d}T(X)$. (This is the precise meaning of Eq.~\calBij.) 
Thus, ${\rm d}T$ assigns to $X$ a differential operator ${\rm d}T(X)$ 
acting on superfunctions $f$. Such an operator differentiates with 
respect to the commuting and anticommuting coordinates. If $X$ is 
odd, so is ${\rm d}T(X)$. Consistency now requires us to observe the 
graded commutation law ${\rm d}T(X)b = (-)^{|X||b|} b{\rm d}T(X)$ where 
$b \in \Lambda$ is some parameter. Hence, from (C.2) we deduce
	$$
	[{\rm d}T(X),{\rm d}T(Y)] = (-)^{|X||Y|} {\rm d}T([X,Y]) .
	$$
It is easy to see that this bracket obeys the axioms (C.1), i.e. it
defines a Lie superalgebra. Setting ${\cal E}_{ij} = {\rm d}T(E_{ij})$ with 
$E_{ij}$ the canonical generators of ${\rm gl}(p,q;\CN)$, we 
obtain the supercommutation relations $[{\cal E}_{ij},{\cal E}_{kl}] 
= (-)^{(|i|+|j|)(|k|+|l|)} \delta_{jk} {\cal E}_{il} - \delta_{li} 
{\cal E}_{kj}$, which are different from those for the $E_{ij}$. 
We thus need to distinguish between two Lie superalgebras. The first
one is generated by the $E_{ij}$, the second one by the ${\cal E}_{ij}$.
Our discussion clarifies how the two are connected: constructing from 
the former and the latter a Lie algebra with Grassmann structure of
the first and second kind, respectively, we get Lie algebras that match, 
i.e. the latter represents the former.

The sceptical reader might be reluctant to accept the violation of the 
representation property at the Lie superalgebra level. He is invited 
to convince himself that the only consistent way to fix the disease
is to modify the basic law of multiplication of supermatrices. 
Clearly, such a modification would be very undesirable. \bigskip

\noindent{\bigbf Appendix D: Calculation of a Reduced Matrix Element}\medskip

\noindent We calculate the value of the exchange coupling constant $J$ 
of the spin chain for $m > 0$. (Recall $m = 1$ for a spin-split and $m 
= 2$ for a spin-degenerate Landau level.) The matrix-valued function 
$Q(g) = g\Lambda g^{-1}$ is expressed by
	$$
	Q = \pmatrix{ 	1+\tilde Z Z 	&-2\tilde Z\cr
			2Z		&-1-Z\tilde Z\cr}
	    \pmatrix{ 	(1-\tilde Z Z)^{-1} 	&0\cr
			0		&(1-Z\tilde Z)^{-1}\cr}
	\eqno({\rm D}.1)
	$$
in the coordinates $Z, \tilde Z$. We denote the matrix element
$(2Z(1-\tilde Z Z)^{-1})_{11}$ by $Q_{11}^{21}$. From Sect.~5.1
we have the relation $J = c_0^2 \sigma_{xx}/a_1$ where $c_0$ is the
constant of proportionality in $P Q_{11}^{21} v = c_0 {\cal B}_{11} v$. 
($P$ projects onto $V$, and $v = v_m$.) 
Making use of the invariant quadratic form $(\cdot,\cdot)$
we get $c_0 = ({\cal B}_{11} v , Q_{11}^{21} v) / ({\cal B}_{11} v ,
{\cal B}_{11} v)$. From Eq.~\railow{\rm a} for $T_g^+ = v^{-1} T_g v$
we have $({\cal B}_{11} v)(Z,\tilde Z) = m Z_{11} v(Z,\tilde Z)$.
The invariant Berezin measure on {\bf G/K} is expressed in the
coordinates $Z,\tilde Z$ by the flat measure $dZ d\tilde Z$ [12], so
	$$
	({\cal B}_{11} v , Q_{11}^{21} v) = \int dZ d\tilde Z
	\ m \bar Z_{11} (2Z(1-\tilde Z Z)^{-1})_{11} v(Z,\tilde Z)^2 .
	\eqno({\rm D}.2)
	$$
To calculate this integral, we use the identity
	$$
	m ( Z(1-\tilde Z Z)^{-1} )_{11} v(Z,\tilde Z)^2
	= {d \over dt} v(Z,\tilde Z - tE_{11})^2 \Big|_{t=0}
	$$
which follows from $v(Z,\tilde Z)^2 = \sdet (1-\tilde Z Z)^m$.
By partial integration we get
	$$
	({\cal B}_{11} v , Q_{11}^{21} v ) = 2 \int dZ d\tilde Z
	\ v(Z,\tilde Z)^2 = 2 (v,v) = 2 .
	\eqno({\rm D}.3)
	$$
On the other hand, Eq.~\normform\ for $a_{m+2,-m} = {\cal B}_{11} v$ 
gives $({\cal B}_{11} v , {\cal B}_{11} v ) = m$. Hence, $c_0 = 2/m$
and $J = 4\sigma_{xx} / m^2 a_1$. 

The above calculation does not stand up to close scrutiny, for it
turns out that the invariant Berezin integral in the coordinates
$Z, \tilde Z$ is plagued by boundary ambiguities \rothstein\ in 
general. This means that the integrals in (D.2) and (D.3) must be
corrected by the addition of extra terms. (I have analyzed the
situation for $n = 1$ in some detail. In this case I find the
correction terms for (D.2) to be nonvanishing only for $m = 1,2$.
For $m = 2$, the correction term is precisely cancelled
by the boundary term resulting from partial integration, so 
the first equality sign in (D.3) is correct as it stands in this
case.) One can avoid dealing with this issue by switching to
Efetov's polar coordinates \efetov\ where the problem of boundary
ambiguities is well understood [52,53]. Doing so, I find the
final result to be correct for all $m \ge 1$ (and all $n$). \bigskip

\noindent{\bigbf Appendix E: Relation between the two-spin Hamiltonian
and the Laplacian for {\bf G/K}}\medskip

\noindent From Eq.~\defmonHp, the monopole Hamiltonian ${\cal H}_s = 
- {\cal L}^m$ is seen to have the expression ${\cal H}_s = \sum_{ij} 
(-)^{|i|+1} {\cal E}_{ij}{\cal E}_{ji}$, with the spin operators 
${\cal E}_{ij} = {\rm d}T(E_{ij})$ being defined by the differential 
of $T_g$, Eq.~\symtrans. Since ${\cal H}_s$ is zero on its ground-state 
module $V = V_m$, we may write the two-spin Hamiltonian $H : V \otimes 
V^* \to V \otimes V^*$ in the form
	$$
	H = \half\sum_{ij} (-)^{|i|+1} ({\cal E}_{ij} + {\cal E}^*_{ij})
	({\cal E}_{ji} + {\cal E}^*_{ji} ) .
	$$
Returning to the notation used in Eq.~\defmonHp\ we get
	$$\eqalign{
	&\left( H (f \otimes f^*) \right) (\pi(g),\pi(h)) = 
	- \half\lambda_{\cal G}(\partial/\partial x) 
	\mu(k(g)k(e^{\Sigma x^i e_i}g)^{-1})			\cr
	&\times \mu(k(h)^{-1}k(e^{\Sigma x^i e_i}h))
	\left( f \otimes f^* \right) (\pi(e^{\Sigma x^i e_i}g),
	\pi(e^{\Sigma x^i e_i}h))\Big|_{x=0} .			\cr}
	$$
Now, following Sect.~5.2, we use the trick of identifying $g \equiv h$.
The multipliers then cancel each other, and switching from right- to
left-invariant vector fields we arrive at the formula
	$$\eqalign{
	\left( H (f \otimes f^*) \right) (\pi(g)) &= - \half\lambda_{\cal G}
	(\partial/\partial x) \left( f \otimes f^* \right) 
	(\pi(e^{\Sigma x^i e_i}g))\Big|_{x=0} 				\cr
	&= - \half\lambda_{\cal P}(\partial/\partial x) \left( f \otimes 
	f^* \right) (\pi(g\exp{\scriptstyle\sum}x^i e_i))\Big|_{x=0} . 	\cr}
	$$
Comparison with formula \deflapl\ proves that $H$ coincides with minus one 
half times the Laplacian ${\cal L}$ on {\bf G/K}, after restriction to 
the subspace of functions which have the quantum numbers that are generated 
by the tensor product $V \otimes V^*$. \bigskip

\noindent {\bigbf Figure Captions}\medskip

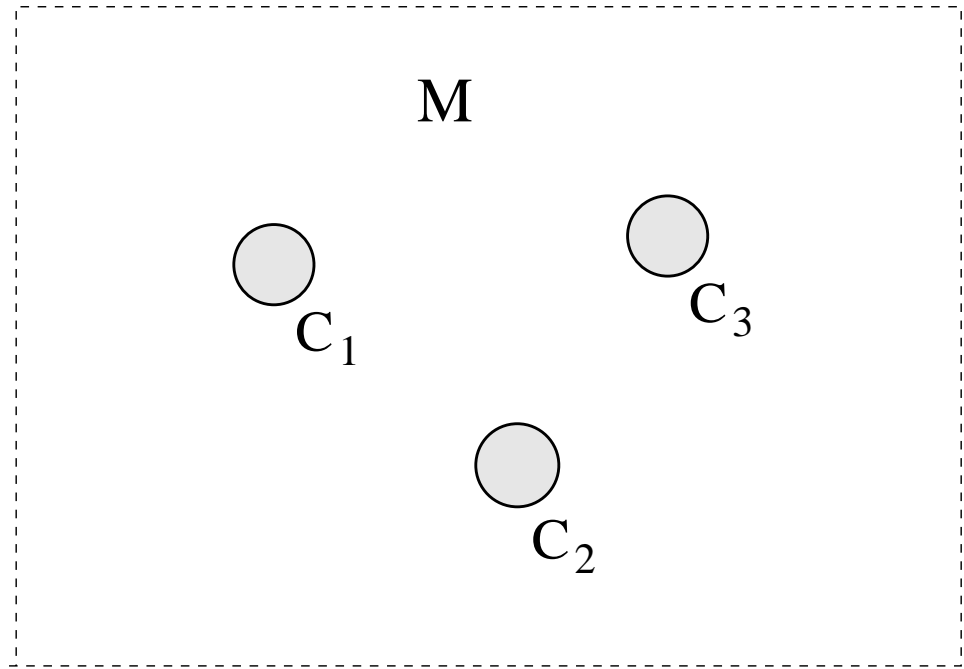
\noindent Fig.~1: Three-terminal device (schematic) with small contacts. 
The configuration space ${\cal M}$ of the $2d$ electron gas is finite 
or infinite. The contacts $C_i$ connect ${\cal M}$ with particle
reservoirs.

\smallskip\noindent
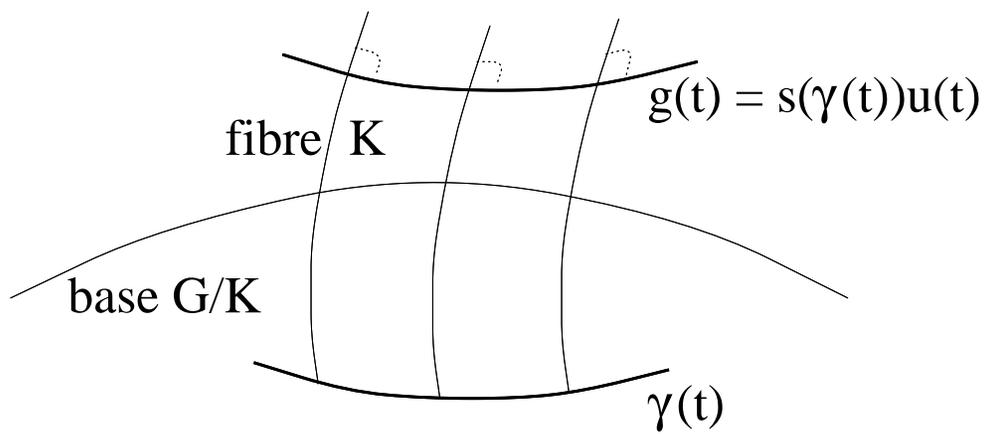
Fig.~2: Lifting of a curve $t \mapsto \gamma(t)$ on a principal fibre 
bundle ${\bf G} \to {\bf G/K}$ with gauge connection $(g^{-1} {\rm d}g)_
{\cal K}$.

\smallskip\noindent
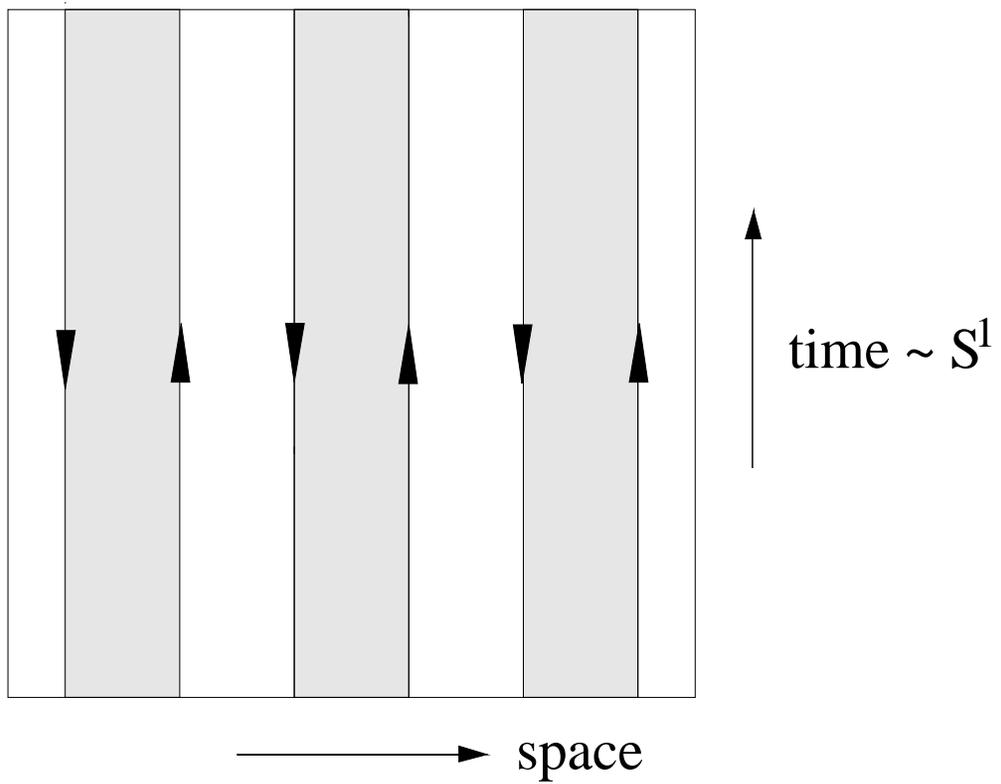
Fig.~3: Modification of the topological density. ${\cal M}_1$ is the
union of the shaded regions. The topological coupling constant is
doubled on ${\cal M}_1$ and set to zero on the empty regions.

\smallskip\noindent
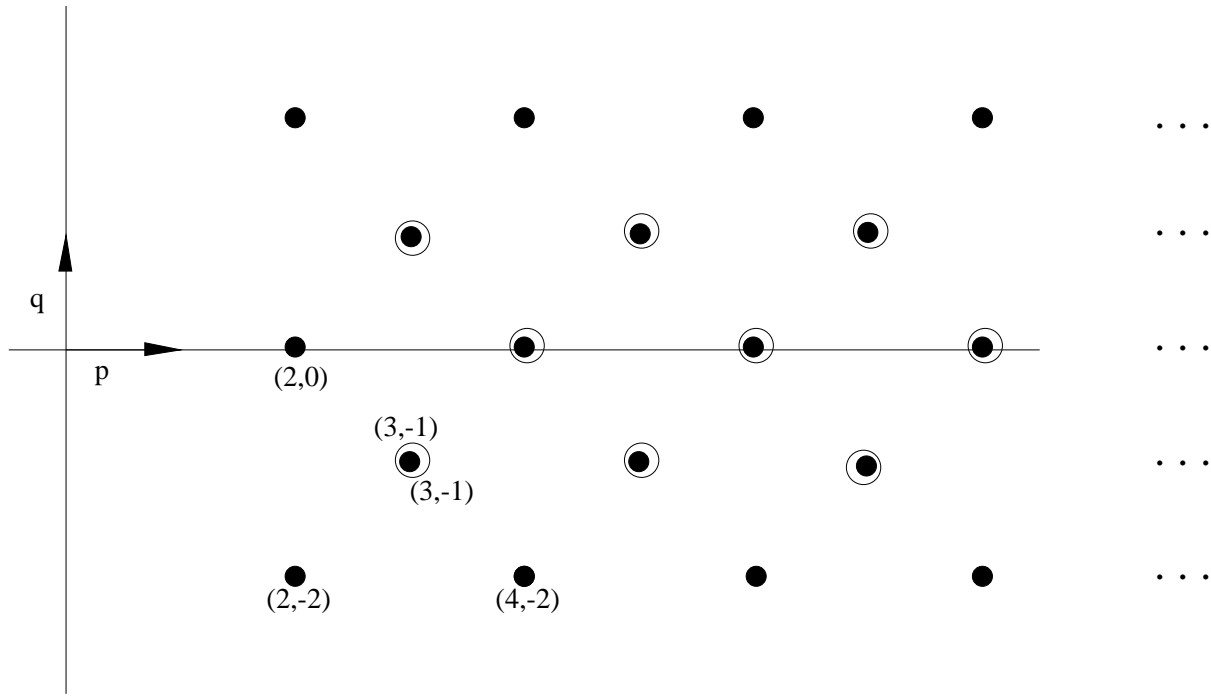
Fig.~4: Weight diagram of the lowest-weight module $V_2$. Weights
are marked by full black dots. Weights enclosed by an additional
circle are doubly degenerate. The lowest weight is $(2,-2)$.

\smallskip\noindent
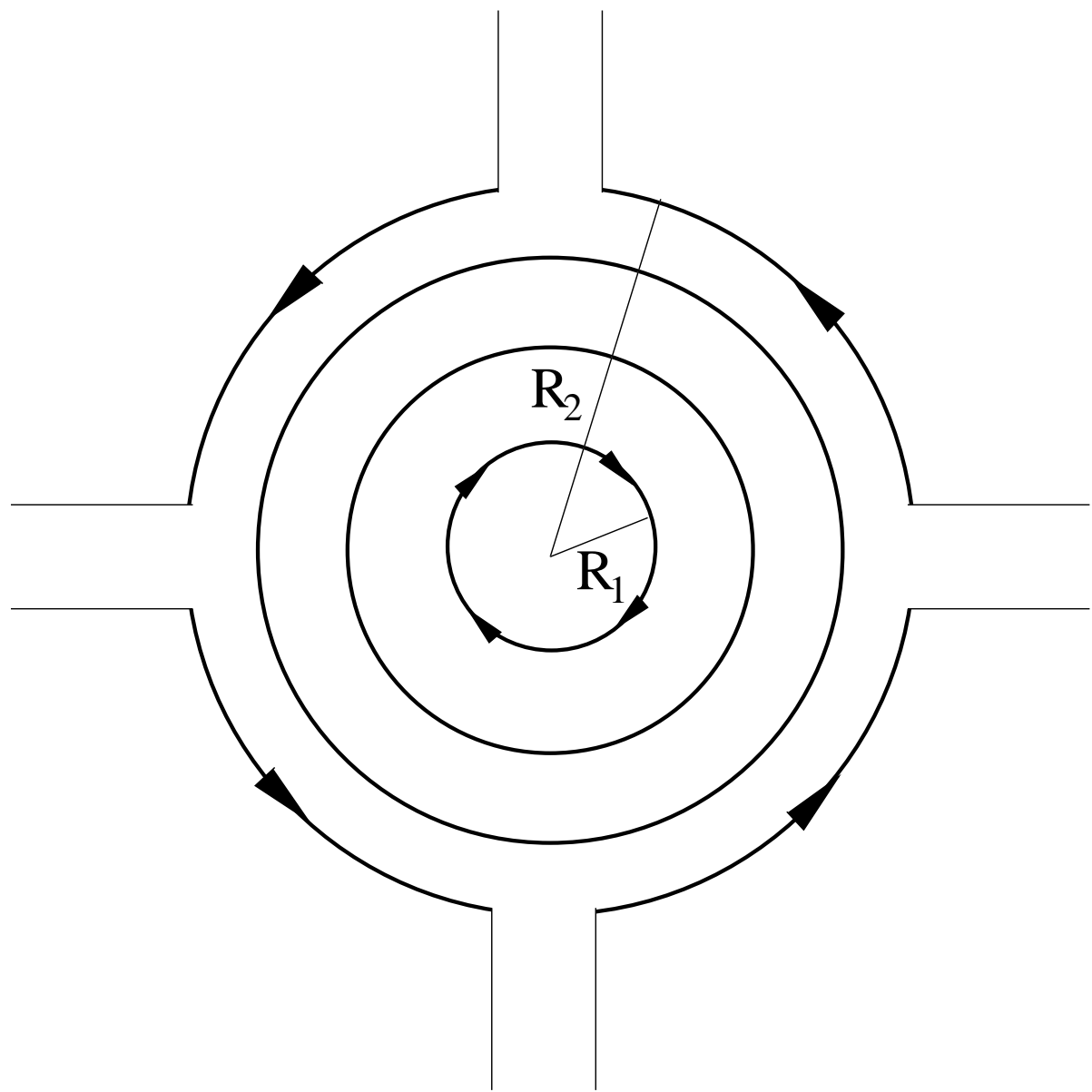
Fig.~5: Corbino disk geometry with four terminals. Fat black lines
symbolize {\bf G/K} superparticles propagating in imaginary time.
The particles at the outer and inner edges and in the bulk see 
monopole charges $+m$, $-m$, and 0, respectively. The sign of the
monopole charge at the edges is indicated by arrows. 

\smallskip\noindent
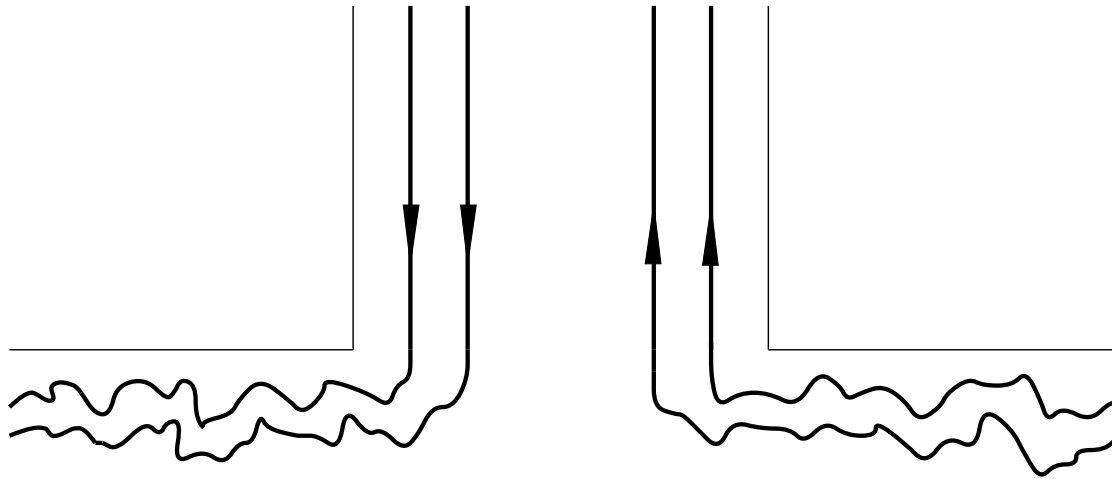
Fig.~6: Ideal lead connected to a disordered quantum Hall sample.
There are no direct transitions between incoming and outgoing
channels.

\smallskip\noindent
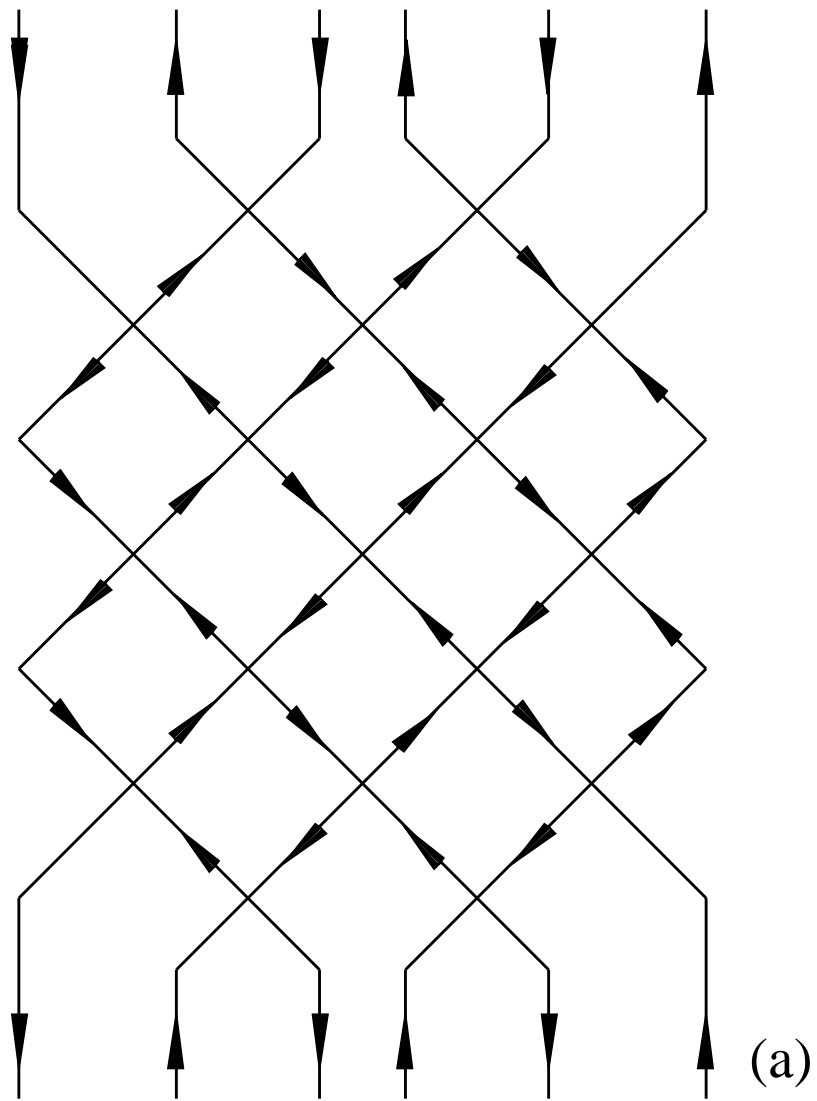
Fig.~7: (a) The network model of Chalker and Coddington. 
(b) Anisotropic limit of the network model. The dotted lines 
indicate tunnelling occurring with a small probability amplitude.

\smallskip\noindent
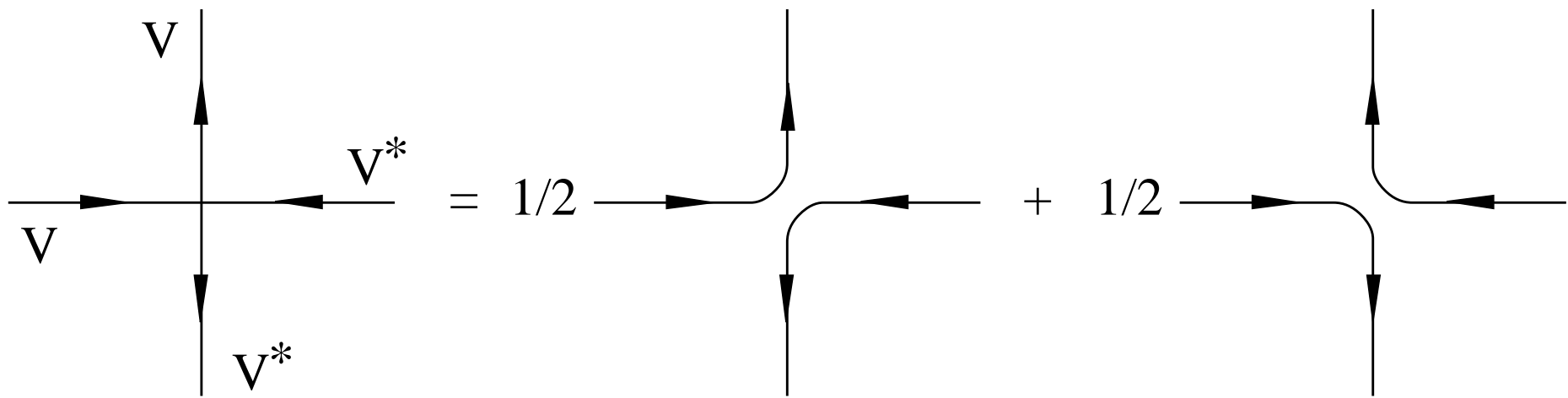
Fig.~8: $R$-matrix of the supersymmetric vertex model at the 
classical percolation transition.

\listrefs
\vfill\eject
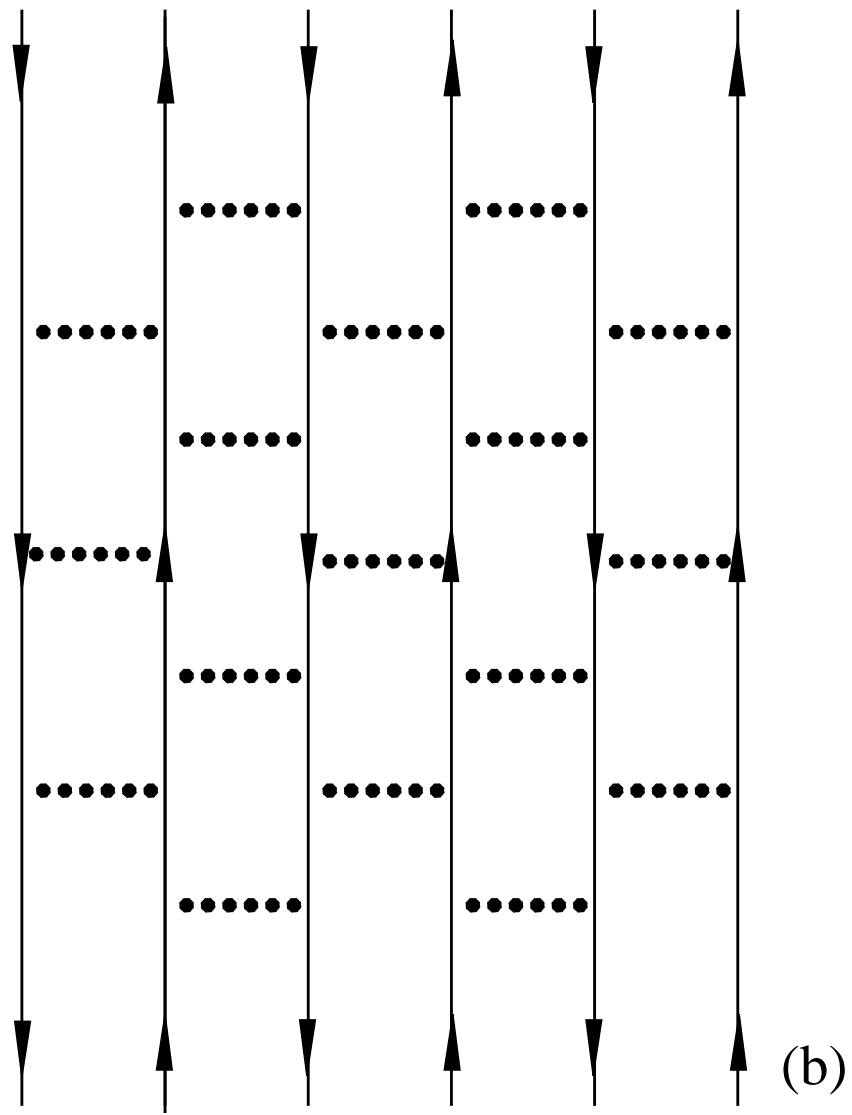
\end